\documentclass[10pt]{article}
\usepackage[margin=1in]{geometry}
\usepackage{amsfonts,amsmath,amssymb}
\usepackage[none]{hyphenat}
\usepackage{fancyhdr}
\usepackage{graphicx}
\usepackage{float}
\usepackage[nottoc,notlot,notlof]{tocbibind}
\usepackage{bigints}
\usepackage{amssymb,amsmath,tikz}
\usepackage{pgfplots} 
\usepackage{amsthm}
\usepackage{mathrsfs}
\usepackage{pifont}
\usepackage{slashed}
\usepackage{mathtools}
\usepackage{cancel}

\showoutput
\DeclareMathOperator*{\p}{p}

\DeclareMathOperator*{\q}{k}

\newcommand*{\ud}{\mathrm{\,d}} 

\renewcommand{\qedsymbol}{$\blacksquare$}

\theoremstyle{plain}

\newtheorem*{twr*}{Theorem}
\newtheorem*{lem*}{Lemma}
\newtheorem{twr}{Theorem}
\newtheorem{lem}{Lemma}
\newtheorem{defin}{Definition}
\newtheorem*{defin*}{Definition}

\newtheorem*{rem*}{Remark}
\newtheorem{cor}{Corollary}
\newtheorem{cor*}{Corollary}
\newtheorem{ex}{Example}

\newtheorem*{notn*}{Notation}
\newtheorem*{wiener-ito*}{Wiener-It\^o-Segal Decomposition}
\newtheorem*{prop*}{Proposition}

\DeclareMathAlphabet{\mathpzc}{OT1}{pzc}{m}{it}

\begin{document}

\begin{titlepage}
\begin{center}
\vspace*{1cm}
\large{\textbf{CAUSAL PERTURBATIVE QFT}}\\
\large{\textbf{AND WHITE NOISE}}\\
\vspace*{2cm}
\small{JAROS{\L}AW WAWRZYCKI}\\[1mm]
\tiny{Bogoliubov Labolatory of Theoretical Physics}
\\
\tiny{Joint Institute of Nuclear Research, 141980 Dubna, Russia}
\\
\vspace*{1cm}
\tiny{\today}\\
\vfill
\begin{abstract}
We present the Bogoliubov's causal perturbative QFT, which includes 
only one refinement: the creation-annihilation operators at a point, \emph{i.e.} 
for a specific momentum, are mathematically 
interpreted as the Hida operators from the white noise analysis. 
We leave the rest of the theory completely unchanged. This allows avoiding 
infrared-- and ultraviolet -- divergences in the transition 
to the adiabatic limit for interacting fields. We present here
the analysis of the causal axioms for the scattering operator with the Hida
operators as the creation-annihilation operators. 
\end{abstract}
\vspace*{0.5cm}
\tiny{{\bf Keywords}: scattering operator, causal perturbative method in QFT, \\
white noise, Hida operators, integral kernel operators, Fock expansion}
\end{center}
\vfill
\end{titlepage}

\tableofcontents
\thispagestyle{empty}
\clearpage

\setcounter{page}{1}

\section{The main idea and the main goal}

In \cite{Bogoliubov_Shirkov} it was recognized that the $n$-th order contribution $S_n$ to the scattering operator
$S$, with the scalar coefficients (Green functions) computed with the standard perturbative method with 
renormalization, give $S_n$ which respect four simple laws, called Bogolibov's causality axioms,
with $S_n$ regarded as a kind of generalized operator-valued distributions, with the coupling constant(s) replaced by space-time 
test function(s) $g$ and with $g^{\otimes n}$ regarded as the test function of the distribution $S_n$ -- a Wick polynomial with the Green 
functions (tempered translationally invariant distributions in $n$ space-time coordinates) as coefficients. 
The freedom in renormalization allows computing
$S_n$ up to similar expressions with the coefficients $\Lambda_n$ supported at the full diagonal, 
but as recognized in \cite{Bogoliubov_Shirkov}, in each case $S_n$ respect the said axioms. But concerning the rigorous 
reconstruction of $S_n$ based solely on the said axioms, it was only 
outlined shortly a proof of existence of $S_n$, which is in agreement with the $S_n$ with the Green functions obtained 
in the perturbation method and renormalization. The task of a rigorous computation of $S_n$ based solely on the Bogoliubov' 
axioms, outlined in \cite{Bogoliubov_Shirkov} was then undertaken in \cite{Epstein-Glaser}. It is obvious that the said 
axioms will acquire a concrete mathematical sense only after we fix a strict sense 
in which the operators of free fields, their Wick product, and $S_n$, are regarded as generalized operators. At this point 
\cite{Bogoliubov_Shirkov} leaves some freedom, indicating only that the class of generalized operators should include 
free fields, their Wick products, products of the Wick products of free fields, eventually, 
with translationally invariant tempered distributions as coefficients. In order to
put forward the idea of \cite{Bogoliubov_Shirkov}, Epstein and Glaser \cite{Epstein-Glaser} 
assumed that these generalized operators are operator valued distributions in the Wightman sense \cite{wig}. Having fixed
in this way the mathematical meaning of the Bogoliubov's axioms, Epstein and Glaser \cite{Epstein-Glaser} were able to convert the axioms, 
especially the causality axiom, into a computational inductive construction of the higher order contributions $S_n$. In this way 
Epstein and Glaser reduced the computation of $S_n$ to the problem of splitting of causal translationally invariant 
tempered distributions into the retarded and advanced parts, and have obtained mathematically consistent perturbative
QFT, eliminating altogether the ultraviolet (UV) divergences from the perturbative theory. The splitting can be computed
only up to a finite number of constants, depending on the singularity order of the splitted distribution -- freedom corresponding
to the renormalization group and non uniqueness of the ordinary renormalization procedure.

Nonetheless, some infrared divergences remained for those perturbative QFT with infinite range of interaction, like QED,
whenever we want to pass to the limit of the intensity of interaction function $g$ to the constant function $1$, performed
in order to get results with the physical interaction also in the remote part of space-time. 
This is the \emph{adiabatic limit problem}. In some problems of QFT, passing to this limit is unavoidable.

Here we propose to improve the Bogoliubov-Shirkov-Epstein-Glaser theory by reinterpreting the generalized operators
and regard them not as operator valued distributions in the Wightman sense, but as the integral kernel operators 
with vector valued kernels in the white noise calculus \cite{obataJFA}. 
Beside this reinterpretation, we leave the whole theory completely unchanged. This, among other things, allows us to solve
the adiabatic limit problem.

More specifically: we are using the Hida white noise operators 
(here $\boldsymbol{\p}$ subsumes the spatial momenta components and the corresponding discrete spin
components in order to simplify notation) $\partial_{\boldsymbol{\p}}^{*}, \,\,\, \partial_{\boldsymbol{\p}}$
which respect the canonical commutation or anticommutation relations 
$\big[\partial_{\boldsymbol{\p}}, \partial_{\boldsymbol{\q}}^{*}\big]_{{}_{\mp}} = \delta(\boldsymbol{\p}-\boldsymbol{\q})$,
as the creation, annihilation operators $a(\boldsymbol{\p})^{+}, \,\,\, a(\boldsymbol{\p})$,
of the free fields in the causal perturbative Bogoliubov, Epstein, Glaser QFT, \cite{Bogoliubov_Shirkov}, \cite{Epstein-Glaser}, 
leaving all the rest of the axioms of the theory completely unchanged.
\emph{I.e.}
we are using the standard Gelfand triple
\[
\left. \begin{array}{ccccc} 
E & \subset & \mathcal{H} & \subset & E^* 
\end{array}\right.,
\]
over the single particle Hilbert space $\mathcal{H}$ of the total system of free fields 
determined by the corresponding standard self-adjoint operator $A$ in $\mathcal{H}$ (with some positive power $A^r$ being nuclear), 
and its lifting to the standard Gelfand triple 
\[
\left. \begin{array}{ccccc} 
(E) & \subset & \Gamma(\mathcal{H}) & \subset & (E)^* 
\end{array}\right.,
\]
over the total Fock space $\Gamma(\mathcal{H})$ of the total system of free fields underlying the actual QFT, e.g. QED, 
which is naturally determined by the self-adjoint and standard operator $\Gamma(A)$, and
with the nuclear Hida test space $(E)$ and its strong dual $(E)^*$. The free field operators become sums of two (respectively positive and negative frequency) 
integral kernel operators with vector valued kernels in the sense of \cite{obataJFA}.  

Suppose we are given a generalized interaction Lagrangian with switching off test functions $g = (g_{{}_{0}}, \ldots, g_{{}_{k}})$
\begin{equation}\label{L}
\mathcal{L}(x) = \sum\limits_{j=0}^{k} g_{{}_{j}}(x) \mathcal{L}_{{}_{j}}(x) = g_{{}_{0}}(x) \mathcal{L}_{{}_{0}}(x) +
\sum\limits_{j=1}^{k} g_{{}_{j}}(x) W_{{}_{j}}(x), \,\,\,\, g_{{}_{j}} \in \mathcal{S}(\mathbb{R}^4; \mathbb{R})
\end{equation}
and with $\mathcal{L}_j(x)$ equal to any Wick products $W_{{}_{j}}(x)$ of free fields of even degree in Fermi fields,
with $\mathcal{L}_{{}_{0}}$ being the true interaction
Lagrangian, with $g_{{}_{0}}$ then eventually tending to the physical coupling constant 
(adiabatic limit probem), and with the other terms in $\mathcal{L}$ introduced after Schwinger
in order to compute the interacting counterparts $W_{{}_{j \, \textrm{int}}}$ of the free Wick products 
$W_{{}_{j}}$, $j>0$, or for the treatment of the external field problem.
Introduction of the Hida operators into the the Boglilubov, Epstein, Glaser construction of the scattering operator
converts the $n$-th order contributions $S_n(g^{\otimes \, n})$
and $W_{{}_{\textrm{int}}}^{(n)}(g_{{}_{0}}^{\otimes \, n},\phi)$ 
to the scattering operator  
\begin{equation}\label{PowerSeriesS}
S(g) = \boldsymbol{1} + \sum\limits_{n=1}^{\infty} {\textstyle\frac{1}{n!}} S_n(g^{\otimes \, n}),
\,\,\,\,
S(g)^{-1} = \boldsymbol{1} + \sum\limits_{n=1}^{\infty} {\textstyle\frac{1}{n!}} \overline{S_n}(g^{\otimes \, n}),
\end{equation}
and to the interacting Wick product fields 
\[
W_{{}_{j \,\, \textrm{int}}}(g_{{}_{0}},\phi) = \bigintsss \left[
S(g)^{-1}
{\textstyle\frac{i\delta S(g)}{\delta g_{{}_{j}}(x)}} 
\right]\Bigg|_{{}_{g_{{}_{i\neq0}}=0}} \phi(x) dx,
\]
into the finite sums of generalized integral kernel operators $\Xi(\kappa_{\mathpzc{l}\mathpzc{m}})$
\begin{multline}\label{Sn(g)}
S_n(g^{\otimes \, n}) = \sum\limits_{\mathpzc{l},\mathpzc{m}}
\int \kappa_{\mathpzc{l}\mathpzc{m}}\big(\boldsymbol{\p}_1, \ldots, \boldsymbol{\p}_\mathpzc{l},
\boldsymbol{\q}_1, \ldots, \boldsymbol{\q}_\mathpzc{m};  g^{\otimes \, n} \big) \,\,
\partial_{\boldsymbol{\p}_1}^{*} \ldots \partial_{\boldsymbol{\p}_\mathpzc{l}}^{*} 
\partial_{\boldsymbol{\q}_1} \ldots \partial_{\boldsymbol{\q}_\mathpzc{m}} 
d\boldsymbol{\p}_1 \ldots d\boldsymbol{\p}_\mathpzc{l}
d\boldsymbol{\q}_1 \ldots d\boldsymbol{\q}_\mathpzc{m}
\\
=
\sum\limits_{\mathpzc{l},\mathpzc{m}} \Xi\left((\kappa_{\mathpzc{l}\mathpzc{m}}(g^{\otimes \, n}) \right)
=
\sum\limits_{j_1,\ldots,j_n=0}^{k}\bigintsss \ud^4 x_1 \ldots \ud^4x_n \, S_n(j_1,x_1, \ldots, j_n, x_n) \, g_{{}_{j_1}}(x_1) \ldots g_{{}_{j_n}}(x_n),
\end{multline}
and
\begin{multline}\label{Wjint}
W_{{}_{j \,\, \textrm{int}}}^{(n)}(g_{{}_{0}}^{\otimes \, n},\phi) = \sum\limits_{\mathpzc{l},\mathpzc{m}}
\int \kappa_{\mathpzc{l}\mathpzc{m}}\big(\boldsymbol{\p}_1, \ldots, \boldsymbol{\p}_\mathpzc{l},
\boldsymbol{\q}_1, \ldots, \boldsymbol{\q}_\mathpzc{m};  g_{{}_{0}}^{\otimes \, n} \otimes \phi \big) \,\,
\partial_{\boldsymbol{\p}_1}^{*} \ldots \partial_{\boldsymbol{\p}_\mathpzc{l}}^{*} 
\partial_{\boldsymbol{\q}_1} \ldots \partial_{\boldsymbol{\q}_\mathpzc{m}} 
d\boldsymbol{\p}_1 \ldots d\boldsymbol{\p}_\mathpzc{l}
d\boldsymbol{\q}_1 \ldots d\boldsymbol{\q}_\mathpzc{m}
\\
=
\sum\limits_{\mathpzc{l},\mathpzc{m}} \Xi\left((\kappa_{\mathpzc{l}\mathpzc{m}}(g_{{}_{0}}^{\otimes \, n}\otimes \phi) \right)
=
\int \ud^4 x_1 \ldots \ud^4x_n \ud^4 x \, W_{{}_{j \,\, \textrm{int}}}^{(n)}(x_1, \ldots, x_n; x) \, g_{{}_{0}}(x_1) \ldots g_{{}_{0}}(x_n) \, \phi(x),
\end{multline}
with vector-valued distributional kernels $\kappa_{\mathpzc{l}\mathpzc{m}}$ in the sense of \cite{obataJFA}, with the values in the distributions
over the test nuclear space
\[
\left(\oplus_{0}^{k}\mathscr{E}\right)^{\otimes \, n} \ni g^{\otimes \, n} 
\,\,\,\,\,\,\,\,\,
\textrm{or, respectively,}
\,\,\,\,\,\,\,\,\,
\mathscr{E}^{\otimes \, n} \otimes (\oplus_{1}^{d}\mathscr{E}) \ni g_{{}_{0}}^{\otimes \, n} \otimes \phi
\]
with $\mathscr{E} = \mathcal{S}(\mathbb{R}^4)$.
Each of the $3$-dim Euclidean integration $d\boldsymbol{\p}_i$ with respect to the spatial momenta $\boldsymbol{\p}_i$ components
$\boldsymbol{\p}_{i1}, \boldsymbol{\p}_{i2}, \boldsymbol{\p}_{i3}$,
also includes here summation over the corresponding discrete spin components $s_i\in(1,2,\ldots)$ hidden under the symbol $\boldsymbol{\p}_i$. 

The class to which the operators $S_n$ and $W_{{}_{j \,\, \textrm{int}}}^{(n)}$ belong, expressed in terms of the Hida test space,
depend on the fact if there are massless free fields present in the interaction Lagrange density operator $\mathcal{L}$ or not.
Namely: 
\[
S_n \in
\begin{cases}
\mathscr{L}\left(\left(\oplus_{0}^{k}\mathscr{E}\right)^{\otimes \, n}, \, \mathscr{L}((E),(E))\right) \cong
\mathscr{L}\left((E), (E)\right) \otimes \mathscr{L}\left( \left(\oplus_{0}^{k}\mathscr{E}\right)^{\otimes \, n}, \mathbb{C} \right), 
& \text{if all fields in $\mathcal{L}$ are massive},\\
\mathscr{L}\left(\left(\oplus_{0}^{k}\mathscr{E}\right)^{\otimes \, n} , \, \mathscr{L}((E),(E)^*)\right)
\cong \mathscr{L}\left((E), (E)^* \right) \otimes \mathscr{L}\left( \left(\oplus_{0}^{k}\mathscr{E}\right)^{\otimes \, n}, \mathbb{C} \right), 
& \text{if massless fields are in $\mathcal{L}$}.\\
\end{cases}
\]
Because each skew-symmetric tempered distribution also is a continuous Grassmann-valued functional on the \emph{Grassmann test function space} 
(Subsection \ref{Grassmann}), then causal perturbative method makes rigorous sense also 
in case, some Wick products $W_{{}_{j}}$ are odd in Fermi fields, 
with the corresponding test components $g_{{}_{j}}$ replaced with Grassmann test functions. In this case
\[
S_{n} \in
\begin{cases}
\underset{r+p=n}{\oplus}\mathscr{L}\left((E), (E)\right) \otimes \mathscr{L}\left(\left(\oplus_{0}^{k}\mathscr{E}\right)^{\otimes \, r} 
\otimes \mathscr{E}^p, \mathcal{E}^{p \, *}\right), 
& \text{if all fields in $\mathcal{L}$ are massive},\\
\underset{r+p=n}{\oplus}\mathscr{L}\left((E), (E)^*\right) \otimes \mathscr{L}\left(\left(\oplus_{0}^{k}\mathscr{E}\right)^{\otimes \, r} 
\otimes \mathscr{E}^p, \mathcal{E}^{p \, *}\right), 
& \text{if there are massless fields in $\mathcal{L}$},\\
\end{cases}
\]
with $\mathcal{E}^{p \, *}$ being the subspace of grade $p$ of the \emph{abstract Grassmann algebra} $\oplus_p\mathcal{E}^{p \, *}$ 
\emph{with inner product and involution}
in the sense of \cite{Berezin}. $\mathcal{E}^p$ denotes the space of \emph{Grassmann-valued test functions} $g^p$ of grade $p$ 
due to \cite{Berezin}, defined in Section \ref{Grassmann}, and replacing ordinary test functions $g^{\otimes \, p}$.
Here $\mathscr{L}(E_1,E_2)$ denotes the linear space of linear continuous operators 
$E_1\longrightarrow E_2$ endowed with the natural topology of uniform
convergence on bounded sets. 

Wick products $W$ of free fields, with $\mathscr{L}\left(\mathscr{E}^{\otimes r}\otimes \mathscr{E}^{\otimes p},\mathbb{C}\right)$-vaued or resp.
$\mathscr{L}\left(\mathscr{E}^{\otimes r}\otimes \mathscr{E}^{p},\mathcal{E}^{p \, *}\right)$-valued kernels $\kappa_{\mathpzc{l},\mathpzc{m}}$,
containing massless field as a factor, are more singular generalized operators
$W \in  \mathscr{L}\left((E), (E)^* \right) \otimes \mathscr{L}\left(\mathscr{E}^{\otimes r}\otimes \mathscr{E}^{\otimes p},\mathbb{C}\right)$
or resp. $W \in  \mathscr{L}\left((E), (E)^* \right) \otimes \mathscr{L}\left(\mathscr{E}^{\otimes r}\otimes \mathscr{E}^{p},\mathcal{E}^{p \, *}\right)$. 
Any two such operators, evaluated at space-time test functions in $\mathscr{E}^{\otimes r}\otimes \mathscr{E}^{\otimes p}$ or, 
resp. in $\mathscr{E}^{\otimes r}\otimes \mathscr{E}^{p}$, 
belong to  $\mathscr{L}\left((E), (E)^* \right)$
or, resp., to $\mathscr{L}\left((E), (E)^* \right) \otimes \mathcal{E}^{p \, *}$, and 
cannot be composed in the sense of operator composition (on the first factor, with the ordnary 
Grassmann product on the second factor). But the contributions 
$S_n$ have the form of Wick products of free fields. Thus, it is not obvious if the axioms involving the products
\begin{equation}
S_n(j_1,x_1,\ldots, j_n, x_n)S_r(i_1,y_1,\ldots, i_r, y_r),
\end{equation}
which in the pure massive case are defined through the operator composition (and Grassmann product), 
are indeed meaningful in case the interaction Lagrangian density operator $\mathcal{L}$ contains massless fields, as e.g. in QED.
In this paper we show that, even in this massless case, the product operation for $S_n$, regarded as a finite sum of integral kernel
operators with the Hida operators as the creation-annihilation operators, is well-defined, and given as a limit in which
the massless free fields are replaced with the corresponding massive fields, and by passing to the zero mass limit. 
We give also the Wick theorem decomposition for such products and show that the causality axioms (I)-(V) for $S_n$ 
make sense if we are using the Hida operators as the creation-annihilation operators, and regard the free field operators, as well
as their Wick products $W_{{}_{j}}$, higher order contributions $S_n$ and 
$W_{{}_{j \,\, \textrm{int}}}^{(n)}$, as finite sums of integral kernel operators with vector valued kernels.

It is important, because the usage of Hida operators allows us to solve the adiabatic limit problem. Namely, in case of QED,
the limit $g_{{}_{0}}\rightarrow 1$ exists for the interacting fields $\psi_{{}_{\textrm{int}}}^{(n)}(g_{{}_{0}}^{\otimes \, n})$ 
and $A_{{}_{\textrm{int}}}^{(n)}(g_{{}_{0}}^{\otimes \, n})$ in QED, and is equal to a finite sum of integral kernel operators
with vector valued kernels in the sense of \cite{obataJFA}, and belong to
\begin{equation}\label{ClassIntFields}
\mathscr{L}\big(\oplus_{1}^{d}\mathscr{E}, \, \mathscr{L}((E),(E)^*)\big), \,\,\,\,\phi \in \oplus_{1}^{d}\mathscr{E}.
\end{equation}
Existence of the product operation in the \emph{whole} class 
$\mathscr{L}\left((E), (E)^* \right) \otimes \mathscr{L}\left(\mathscr{E}^{\otimes \, n},\mathbb{C}\right)$
or $\mathscr{L}\big((E), (E)^*\big) \otimes \mathscr{L}(\mathscr{E}^{\otimes \, (n-p)} \otimes \mathscr{E}^p, \mathcal{E}^{p \, *})$ of operators
is quite not obvious. But the higher order contributions $S_n$ to the scattering operator, which also define the interacting fields,
are of special class, and admit the operation of product defined by the above-mentioned limit operation.  
The adiabatic limit for interacting fields exists in QED only if the normalization in the Epstein-Glaser splitting is ``natural'', called
``on mass shell normalization'' (in which the natural equations of motion for the interacting fields are fulfilled), so the arbitrariness
in the splitting can thus be naturally eliminated by using the Hida operators \cite{wawrzycki}. In our previous
paper \cite{wawrzycki} 
we have presented 
a proof of existence of the limit $g_{{}_{0}}\rightarrow 1$ for $\psi_{{}_{\textrm{int}}}^{(n)}(g_{{}_{0}}^{\otimes \, n})$ 
and $A_{{}_{\textrm{int}}}^{(n)}(g_{{}_{0}}^{\otimes \, n})$ in (\ref{ClassIntFields}) in QED.  

Also, the adiabatic limit $g_{{}_{0}}\rightarrow 1$ for the higher order contributions $S_n(g_{{}_{0}}^{\otimes \, n})$ for the scattering operator itself 
exists in QED in $\mathscr{L}((E),(E)^*)$. This limit is not unique, but the non uniqueness is irrelevant for the computation of the effective 
cross-section, with the \emph{in} and \emph{out} states as the many particle (non-normalizable) plane wave states. 
Using Hida operators in the causal perturbative approach we  thus obtain mathematically consistent  QFT, in particular QED on the Minkowski space-time, 
without UV and IR infinities, applicable to the scattering processes with the \emph{in} and \emph{out} states as the 
many particle (non-normalizable) plane wave states. Also, the Schwinger Green functions can thus be computed without
encountering any infinities, but the whole bound state problem cannot be consequently treated within such QED on the Minkowski space-time, 
because the higher order contributions to the conserved Noether integrals for interacting fields are not ordinary self-adjoint
operators in the Fock space, but only generalized operators of the class  $\mathscr{L}((E),(E)^*)$.  

The adiabatic limit $g_{{}_{0}}\rightarrow 1$ problem for strong interactions on the Minkowski space-time seems to be devoid of any deeper physical sense, 
due to the phenomenon called ``asymptotic freedom'' which allows the perturbative method to be applicable in this case only in the high energy limit
of the scattering processes.

\section{White noise construction of the free fields and of their Wick products}\label{HidaFreeFieldsWick}

In the sequel, we are using the standard Gelfand triple $E\subset\mathcal{H}\subset E^*$ over the single particle Hilbert space
$\mathcal{H}$, which by construction is standard, \emph{i.e.} it has the form $\mathcal{S}_{A}(\sqcup \mathbb{R}^3; \mathbb{C})
\subset L^2(\sqcup \mathbb{R}^3;\mathbb{C})\subset\mathcal{S}_{A}(\sqcup \mathbb{R}^3; \mathbb{C})^*$ with standard operator $A$
on $L^2(\sqcup \mathbb{R}^3;\mathbb{C})$ (in terminology and notation of \cite{hida}, \cite{obata}, \cite{obataJFA}) which 
respects the conditions: each element of $E$, regared as a class modulo equality a.e. in $L^2$ has unique
continuous (even smooth) representant. The Dirac delta functional $\delta_{s,\boldsymbol{\p}}$ defined on $E$ regarded as function space,
belongs to $E^*$, and the map $(s,\boldsymbol{\p}) \mapsto \delta_{s,\boldsymbol{\p}} \in E^*$ is continuous 
(convergence in $E$ implies uniform convergence of the elements
of $E$ regarded as functions). All topological vector spaces we are using are standard countably Hilbert nuclear, their strong
duals (which are also nuclear), or separable Hilbert spaces. All tensor products $\otimes$ of nuclear spaces we are using, are equal to the projective
tensor products, and which are essentially unique for nuclear spaces and equal to equicontinuous tensor products. Tensor products of Hilbert spaces
are understood as the Hilbert space tensor products (as in construction of Bose or Fermi Fock liftings $\Gamma$ or their mixtures).
The kernel theorem, universally valid for nuclear spaces, we are using without explicit mention. 
Let us recall that the Fock lifting of the single particle Hilbert spaces of any two free fields (Bose or Fermi) is equal to the tensor product
of the Bose and/or Fermi Fock liftings. In the sequel we will use the standard construction of Gelfand triples over Fock spaces
which are constructed through the Fock liftings of the single particle Gelfand triples -- a construction due to Hida.
Then we construct the free fields, their Wick products, and products of the Wick products, as the finite sums of the so-called integral
kernel operators.  Finally, we  use the characterization of the so-called integral kernel operators
through the properties of the corresponding kernels: Thms. 3.6, 3.9, 3.13 of \cite{obataJFA}. Said theorems have been proved for the Bose
case only, but the construction of the Fock lifting of the single particle Gelfand triple makes sense also for the Fermi Fock lifting,
and gives another Gelfand triple rigged with the standard nuclear Hida test space and its strong dual. The proof remains the same (it is even simpler)
in the Fermi Fock space, as the Fermi lifting $\Gamma(A)$ of a standard operator is again standard, and the estimations of the contractions
$\otimes_{k}$ remain the same, so that the said theorems remain true also in the Fermi case. It is clear that the said theorems are true
also in the generalized Fock space (without any symmetrizations or antisymmetriations). The norm estimations of the contractions
as given in \cite{obataJFA} are valid for the non-symmetrized case. The operation of alternation (performed 
separately within the set of variables corresponding to one an the same Fermi field) and the operation of symmetrization
(performed separately within each set of variables corresponding to one and the same Bose field) commute with each other and 
commute with the action of the tensored standard operator $A^{\otimes n}$, and do not affect the norm estimations of \cite{obataJFA}, 
therefore Thms. 3.6, 3.9, 3.13 of \cite{obataJFA} remain true
for the integral kernel operators acting in the mixed Bose-Fermi Fock space. We therefore will refer to these theorems, and regard them as 
valid for integral kernel operators acting on the Hida test space obtained by the Fock lifting $\Gamma$ of the direct sum of single particle 
Gelfand triples of the Fermi and Bose fields, which is equal to the tensor product of the respective Fermi and Bose Fock liftings
of the respective Fermi and Bose single particle spaces.
We use the same definitions and notation as in \cite{obataJFA} as closely as possible and will not repeat the formulation
of the said theorems here. However, some minimal repetitions are unavoidable, as we have to recall the definition of the Hida operators and integral
kernel operators.

Thus, we construct the Gelfand triple over the full single particle Hilbert space 
$\mathcal{H} = \mathcal{H}_1 \oplus \ldots \oplus \mathcal{H}_N$ of all $N$ free fields underlying the
considered QFT
\[
\left. \begin{array}{ccccc} 
E & \subset & \mathcal{H} & \subset & E^* \\
\parallel & & \parallel & & \parallel \\
E_1 \oplus \ldots \oplus E_N & & \mathcal{H}_1 \oplus \ldots \oplus \mathcal{H}_N & & E_{1}^{*} \oplus \ldots \oplus E_{N}^{*}
\end{array}\right.,
\]
and then its natural standard realization, \emph{i.e.} a unitary isomorphism $U$
which is likewise continuous with respect to the nuclear topologies on $E$ and its strong (nuclear) dual $E^*$ (when regarded as acting in
$E^*$ as the linear transpose of $U$ acting in $E$).
Single particle Hilbert space $\mathcal{H}_i$, $i=1, \ldots, N$, of each field becomes identified with the standard Hilbert space
$\mathcal{H}_i \cong L^2(\mathbb{R}^3;\mathbb{C}^n) = L^2(\sqcup \mathbb{R}^3;\mathbb{C})$ and similarly
for the full single particle Hilbert space  $\mathcal{H} = \oplus_{i}^{N} \mathcal{H}_i \cong L^2(\sqcup \mathbb{R}^3;\mathbb{C})$.
\[
\left. \begin{array}{ccccc} \mathcal{S}_{A}(\sqcup \mathbb{R}^3; \mathbb{C}) & \subset & L^2(\sqcup \mathbb{R}^3;\mathbb{C}) & \subset & \mathcal{S}_{A}(\sqcup \mathbb{R}^3; \mathbb{C})^* \\
\downarrow \uparrow & & \downarrow \uparrow & & \downarrow \uparrow \\
E & \subset & \mathcal{H} & \subset & E^*
\end{array}\right..
\]

This means that the space $E$ consisting of direct sums of restrictions of the Fourier transforms of space-time test functions
to the respective orbits in momenta defining the representation of $T_4 \ltimes SL(2, \mathbb{C})$ in the Mackey's classification
(acting in the single particle Hilbert space for each corresponding free field) has to be given the standard realization with the help of a standard operator
$A$ in $L^2(\sqcup \mathbb{R}^3; \mathbb{C}) \cong \mathcal{H}$ (with standard $A$, \emph{i.e.} 
self-adjoint positive, with some negative power of which being nuclear, or trace-class,
and with the minimal spectral value greater than $1$). It is equal to the 
the direct sum $\oplus_{i=1}^{N} A_i$ of the standard operators $A_i$ corresponding to the single particle Hilbert space
of the $i$-th free field. Thus, first we need to construct the standard $A_i$ and the standard Gelfand triples for each of the free fields
of the theory.  $A= \oplus_{i=1}^{N} A_i$ is also standard.  It serves for the construction of the standard Gelfand
triple over the full single particle Hilbert space $L^2(\sqcup \mathbb{R}^3; \mathbb{C}) \cong \mathcal{H}$. 
Namely, with the help of the operator $A$ we construct $E$ as the following projective limit
\[
E = \underset{k\in \mathbb{N}}{\bigcap} \textrm{Dom} \, A^k
\]
with a standard realization of $E$ as the countably Hilbert nuclear test space and its strong (also
nuclear) dual as the inductive limit
\[
E^* = \underset{k\in \mathbb{N}}{\bigcup} \overline{\textrm{Dom} A^{-k}}
\]
with the Hilbertian defining norms
\[
\big| \cdot \big|_{{}_{k}} \overset{\textrm{df}}{=} \big|A^k \cdot \big|_{{}_{L^2}}, \,\,\,\,\,\,
\big| \cdot \big|_{{}_{-k}} \overset{\textrm{df}}{=} \big|A^{-k} \cdot \big|_{{}_{L^2}}, \,\,\, k = 0,1,2,3, \ldots,
\]
and with the above said closulre $\overline{\textrm{Dom} A^{-k}}$ of $\textrm{Dom} A^{-k} {:} = \mathcal{H}$ with respect to 
the norm $\big| \cdot \big|_{{}_{-k}}$.

We construct the standard Gelfand triple with the joining isomorphism $\Gamma(U)$ 
\[
\left. \begin{array}{ccccc} \big(\mathcal{S}_{A}(\sqcup \mathbb{R}^3; \mathbb{C})\big) & \subset & \Gamma\big(L^2(\sqcup \mathbb{R}^3;\mathbb{C})\big) & \subset & \big(\mathcal{S}_{A}(\sqcup \mathbb{R}^3; \mathbb{C}))^* \\
\downarrow \uparrow & & \downarrow \uparrow & & \downarrow \uparrow \\
(E) & \subset & \Gamma(\mathcal{H}) & \subset & (E)^* \\
\parallel & & \parallel & & \parallel \\
(E_1)\otimes \ldots \otimes (E_N)& \subset & \Gamma\big(\mathcal{H}_1\big) \otimes \ldots \otimes \Gamma\big(\mathcal{H}_N\big) &
\subset & (E_1)^*\otimes \ldots \otimes (E_N)^*
\end{array}\right.,
\]
over the full Fock space $\Gamma(\mathcal{H})$ of all free fields underlying the considered QFT theory,
replacing $A$ with $\Gamma(A)$, which is also standard. 
The tensor products, 
symmetrized in Bose variables and antisymmetrized in Fermi
variables, will be denoted by $\widehat{\otimes}$, as for example in expressions 
\begin{gather*}
E_{l_1} \widehat{\otimes} \ldots \widehat{\otimes}E_{l_l} \widehat{\otimes} E_{m_1} \widehat{\otimes} \ldots \widehat{\otimes}E_{m_m}
\subset E^{\widehat{\otimes} \, (l+m)}, 
\\
E_{l_1} \widehat{\otimes} \ldots \widehat{\otimes}E_{l_l} \otimes E_{m_1}^* \widehat{\otimes} \ldots \widehat{\otimes}E_{m_m}^* \subset 
E^{\widehat{\otimes}l} \otimes E^{*\widehat{\otimes}\, m}.
\end{gather*} 
\emph{e.t.c}.

Let us recall the definition of the right contraction $\otimes_{m}$ and its symmetrized/antisymetrized version. To this end let
$e^{{}^{i}}_{{}_{j}}$ be the $j$-th eigenvector of the standard operator $A_i$ acting in the single particle Hilbert space of the
$i$-th field. 
The pairings $\langle \cdot, \cdot \rangle$,
are given by the inner product followed by the complex conjugation over the Hermitian variable of the inner product, 
so that they are always bilinear, as in \cite{obataJFA}. Let $\kappa \in E^{\otimes (l+m)}$, $f \in E^{\otimes (n+m)}$, or more precisely
\begin{gather*}
\kappa \in E_{l_1} \otimes \ldots \otimes E_{l_l} \otimes E_{m_1} \otimes \ldots \otimes E_{m_m}
\subset E^{\otimes \, (l+m)}
\\
f \in E_{n_1} \otimes \ldots \otimes E_{n_n} \otimes E_{m_1} \otimes \ldots \otimes E_{m_m}
\subset E^{\otimes \, (l+m)},
\end{gather*}
and let
\[
e(l') = e^{{}^{l_1}}_{{}_{l'_1}} \otimes \ldots \otimes e^{{}^{l_l}}_{{}_{l'_l}},
\,\,\,
e(m') = e^{{}^{m_1}}_{{}_{m'_1}} \otimes \ldots \otimes e^{{}^{m_m}}_{{}_{m'_m}},
\,\,\,
e(n') = e^{{}^{n_1}}_{{}_{n'_1}} \otimes \ldots \otimes e^{{}^{n_n}}_{{}_{n'_n}},
\]    
Let $\kappa$ and $g$ have the following Fourier decompositions (summation with $l',m',n'$ understood as multiidices)  
\[
\kappa = \sum\limits_{l',m'} \big\langle \kappa, e(l') \otimes e(m') \big\rangle \, e(l') \otimes e(m'),
\,\,\,
f = \sum\limits_{n',m'} \big\langle f, e(n') \otimes e(m') \big\rangle \, e(n') \otimes e(m'). 
\]
For such $\kappa,f$ we define the right $m$-contraction
\[
\kappa \otimes_m f {:}= \sum\limits_{l',n'} \left( \sum\limits_{m'} \big\langle \kappa, e(l') \otimes e(m') \big\rangle 
\big\langle f, e(n') \otimes e(m') \big\rangle \right)
 \, e(l') \otimes e(n').
\]
This right $m$-contraction defines uniquely a bilinear continuous map $\otimes_m: E^{\otimes (l+m)} \times E^{\otimes (n+m)} \rightarrow E^{\otimes (l+n)}$, 
which can be extended to a bilinear continuous map $\otimes_m: (E^{\otimes (l+m)})^* \times E^{\otimes (n+m)} \rightarrow (E^{\otimes (l+n)})^*$
by the norm estimation \cite{obataJFA}, Lemma 3.1. Next we define the symmetrized-antisymmetrized contraction, induced uniquely by 
\[
\kappa \widehat{\otimes_m} f {:}= \textrm{Alt} \circ \textrm{Sym} \left(\kappa \otimes_m f \right) \in E^{* \widehat{\otimes} (l+n)},
\]
where $\textrm{Sym}$ denotes symmetrization separately within each set of variables corresponding to one and the same Bose field,
and $\textrm{Alt}$ denotes the alternation performed separately within each set of variables corresponding to one and the same Fermi field.
$\widehat{\otimes_m}$ extends to a bilinear continuous map  
$\widehat{\otimes_m}: E^{* \otimes (l+m)} \times E^{\otimes (n+m)} \rightarrow E^{*\widehat{\otimes} (l+n)}$, and thus to
a bilinear continuous map $\widehat{\otimes_m}: E^{* \otimes (l+m)} \times E^{\widehat{\otimes} (n+m)} \rightarrow E^{*\widehat{\otimes} (l+n)}$.

For any $\Phi$ in $(E)$ or in $(E)^*$ let
\begin{equation}\label{PhiDecomposition}
\Phi = \sum\limits_{n=0}^{\infty} \Phi_{n} \,\,\,\,
\textrm{with} \,\,\, \Phi_n \in E^{\widehat{\otimes} \, n} \,\, \textrm{or, respectively}, \,\,\, \Phi_n \in E^{*\widehat{\otimes} \, n}
\end{equation}
be its decomposition into $n$-particle states of an element $\Phi$ of the test Hida space $(E)$
or in its strong dual $(E)^*$, convergent, respectively, in $(E)$ or in $(E)^*$. We define
\[
\begin{split}
a(w) \Phi_0 = 0, \,\,\,\, a(w) \Phi_n = n \, \overline{w} \widehat{\otimes}_1 \Phi_n
\\
a(w)^+ \Phi_n = w \widehat{\otimes} \Phi_n, \,\,\,\, \textrm{for each fixed} \,\, w \in E^*.
\end{split}
\]

\begin{defin}
 The Hida operators are obtained when we put here the Dirac delta functional for $w= \delta_{{}_{s,\boldsymbol{\p}}}$
\[
\partial_{{}_{s,\boldsymbol{\p}}} = a_s(\boldsymbol{\p}) = a(\delta_{{}_{s,\boldsymbol{\p}}}),
\,\,\,\,
\partial_{{}_{s,\boldsymbol{\p}}}^+ = a_s(\boldsymbol{\p})^+ =  a(\delta_{{}_{s,\boldsymbol{\p}}})^+
\]
\label{HidaOperators}
\end{defin}

For each fixed spin-momentum point $(s,\boldsymbol{\p})$ the Hida operators are well-defined
(generalized) operators 
\[
\begin{split}
a_s(\boldsymbol{\p}) \in \mathscr{L}\big((E), (E)\big) \subset \mathscr{L}\big((E), (E)^*\big),
\\
a_s(\boldsymbol{\p})^+ \in \mathscr{L}\big((E)^*, (E)^*\big) \subset \mathscr{L}\big((E), (E)^*\big),
\end{split}
\]
with the natural topological inclusions induced by the topological natural inclusion $(E) \subset (E)^*$. 
Let $\phi \in \mathscr{E}$ (here $\mathscr{E}$ is the space-time test standard countably Hilbert nuclear space $\mathcal{S}$ or $\mathcal{S}^{00}$, 
with definition given further) and let $\kappa_{l, m}$ be any $\mathscr{L}(\mathscr{E}, \mathbb{C}) = \mathscr{E}^*$-valued distribution
\[
\kappa_{l, m} \in  \mathscr{L}(E^{\otimes (l+m)}, \mathscr{E}^*)  
= \mathscr{L}(\mathscr{E}, E^{*\otimes (l+m)}) = E^{* \otimes (l+m)} \otimes \mathscr{E}^*. 
\]
\begin{defin}
Then, for $\Phi \in (E)$ with decomposition (\ref{PhiDecomposition}), we put
\[
\left. \begin{array}{ccc} 
\Xi_{{}_{l,m}}(\kappa_{{}_{l,m}})(\Phi \otimes \phi) & \overset{\textrm{df}}{=}  &
\sum\limits_{n=0}^{\infty} \kappa_{l,m} \widehat{\otimes_m} (\Phi_{n+m} \otimes \phi)  \\
\parallel & & \parallel   \\
\Xi_{{}_{l,m}}\big(\kappa_{{}_{l,m}}(\phi)\big) \, \Phi &  &  \sum\limits_{n=0}^{\infty} \kappa_{l,m}(\phi) \widehat{\otimes_m} \Phi_{n+m}
\end{array}\right..
\]
which for any $\mathscr{E}^*$-valued distribution $\kappa_{l,m}$ is a well-defined (generalized) operator 
\[
\Xi_{{}_{l,m}}(\kappa_{{}_{l,m}}) \in \mathscr{L}\big((E)\otimes \mathscr{E}, (E)^* \big) \cong
\mathscr{L}\big(\mathscr{E}, \mathscr{L}((E),(E)^*)\big).
\]
\label{XiDefinition}
\end{defin}
If $\kappa_{l,m}\in E_{l_1}^* \otimes \ldots \otimes E_{l_1}^* 
\otimes E_{m_1}^* \widehat{\otimes} \ldots \otimes E_{m_m}^* \otimes \mathscr{E}^*$ in definition \ref{XiDefinition}, then 
only the last $m$ factors $E_{m_i}^*$ of $\kappa_{l,m}$ are \emph{all} contracted with the corresponding 
factors of the component $\Phi_{n+m}$. Here $\Phi_{n+m}$ denote the components in $E^{\widehat{\otimes} (n+m)}$ 
which contain all the $m$ fators $E_{m_1} \otimes \ldots \otimes E_{m_m}$ which are dual to the last $m$ factors $E_{m_i}^*$  of 
$\kappa_{l,m}$, and which can thus be contracted with each other. This is consistent with the definition of the vacuum: 
any $\kappa_{l,m}(\phi) \widehat{\otimes_m} \Phi_{n'}$ with $n'=(n'_1, \ldots, n'_{n'})$
not containing any variable contained in $m = (m_1, \ldots, m_m)$ is identically zero, as the contraction operation 
$\kappa_{l,m}(\phi) \widehat{\otimes_m} \Phi_{n'}$ 
acts as the annihilation of the states lying in the tensor product $E_{m_1} \widehat{\otimes} \ldots \widehat{\otimes} E_{m_{m}}$
followed by the creation of the states lying in the tensor product $E_{l_1} \widehat{\otimes} \ldots \widehat{\otimes} E_{l_{l}}$. 
But the annihilation of the states lying in the tensor product $E_{m_1} \widehat{\otimes} \ldots \widehat{\otimes} E_{m_{m}}$
by assumption, commutes (or anticommutes) with the creation operators of the states lying in the tensor product 
$E_{n'_1} \widehat{\otimes} \ldots \widehat{\otimes} E_{n'_{n'}}$ and annihilation operators in action on the vacuum are always equal to zero.

$\Xi_{{}_{l,m}}(\kappa_{{}_{l,m}})$ defines integral kernel operator $\Xi_{{}_{l,m}}(\kappa_{{}_{l,m}})$ 
which is uniquely determined by the condition
\cite{obataJFA}:
\[
\langle\langle \Xi_{{}_{l,m}}(\kappa_{{}_{l,m}})(\Phi \otimes \phi), \Psi \rangle\rangle
= \langle \kappa_{{}_{l,m}}(\eta_{{}_{\Phi,\Psi}}), \phi \rangle,
\,\,\,
\Phi, \Psi \in (E), \phi \in \mathscr{E}
\] 
or, respectively,
\[
\langle\langle \Xi_{{}_{l,m}}(\kappa_{{}_{l,m}})(\Phi \otimes \phi), \Psi \rangle\rangle
= \langle \kappa_{{}_{l,m}}(\phi), \eta_{{}_{\Phi,\Psi}}\rangle,
\,\,\,
\Phi, \Psi \in (E), \phi \in \mathscr{E},
\] 
depending on $\kappa_{{}_{l,m}}$ is regarded as element of
\[
\mathscr{L}(E^{\hat{\otimes} (l+m)}, \mathscr{E}^*)  
\,\,\,
\textrm{or, respectively, of} \,\,\,
\mathscr{L}(\mathscr{E}, E^{*\hat{\otimes} (l+m)}).
\]
Here 
\[
\eta_{{}_{\Phi,\Psi}}(s_1, \boldsymbol{\p}_1, \ldots, s_l, \boldsymbol{\p}_l,
s_{l+1}, \boldsymbol{\p}_{l+1}, \ldots, s_{l+m}, \boldsymbol{\p}_{l+m})
\overset{\textrm{df}}{=}
\langle\langle a_{s_1}(\boldsymbol{\p}_{1})^+ \ldots a_{s_l}(\boldsymbol{\p}_{l})^+ a_{s_{l+1}}(\boldsymbol{\p}_{l+1})
\ldots a_{s_{l+m}}(\boldsymbol{\p}_{l+m})\Phi, \Psi \rangle\rangle
\]
is the function which always belongs to $E^{\hat{\otimes}(l+m)}$ (compare \cite{hida}, with the same proof valid for the Fermi Fock space
and general Fock space).

\begin{ex}
Free fields $\mathbb{A}$ are sums of two integral kernel operators
\[
\mathbb{A}(\phi) = \mathbb{A}^{(-)}(\phi) + \mathbb{A}^{(+)}(\phi) = \Xi(\kappa_{0,1}(\phi)) +
\Xi(\kappa_{1,0}(\phi))
\]
with the integral kernels $\kappa_{l,m}$ represented by ordinary functions:
\[
\begin{split}
\kappa_{0,1}(\nu, \boldsymbol{\p}; \mu, x) =
{\textstyle\frac{\delta_{\nu \mu}}{(2\pi)^{3/2}\sqrt{2p_0(\boldsymbol{\p})}}}
e^{-ip\cdot x}, \,\,\,\,\,\,
p = (|p_0(\boldsymbol{\p})|, \boldsymbol{\p}), \, p\cdot p=0, p\neq 0, \\
\kappa_{1,0}(\nu, \boldsymbol{\p}; \mu, x) =
{\textstyle\frac{-g_{\nu \mu}}{(2\pi)^{3/2}\sqrt{2p_0(\boldsymbol{\p})}}}
e^{ip\cdot x},
\,\,\,\,\,\,
p \cdot p = 0, p \neq 0,
\end{split}
\]
for the free e.m.potential field $\mathbb{A}=A$ (in the Gupta-Bleuler gauge) and
\[
\kappa_{0,1}(s, \boldsymbol{\p}; a,x) = \left\{ \begin{array}{ll}
(2\pi)^{-3/2}u_{s}^{a}(\boldsymbol{\p})e^{-ip\cdot x}, \,\,\, 
\textrm{$p = (|p_0(\boldsymbol{\p})|, \boldsymbol{\p}), \, p \cdot p = \mathfrak{m}^2$} & \textrm{if $s=1,2$}
\\
0 & \textrm{if $s=3,4$}
\end{array} \right.,
\]
\[
\kappa_{1,0}(s, \boldsymbol{\p}; a,x) = \left\{ \begin{array}{ll}
0 & \textrm{if $s=1,2$}
\\
(2\pi)^{-3/2} v_{s-2}^{a}(\boldsymbol{\p})e^{ip\cdot x}, \,\,\, \textrm{$p \cdot p = \mathfrak{m}^2$} & \textrm{if $s=3,4$}
\end{array} \right.
\]
for the free Dirac spinor field $\mathbb{A}=\boldsymbol{\psi}$ of mass $\mathfrak{m}$, and
which are in fact the respective plane wave solutions of d'Alembert and of Dirac equation, 
which span the corresponding generalized eigen-solution sub spaces. 
Here $g_{\nu\mu}$ are the components of the space-time Minkowski metric tensor with signature $(1,-1,-1,-1)$, and 
\[
u_s(\boldsymbol{\p}) =  \frac{1}{\sqrt{2}} \sqrt{\frac{|p_0(\boldsymbol{\p})| + \mathfrak{m}}{2 |p_0(\boldsymbol{\p})|}}
\left( \begin{array}{c}   \chi_s + \frac{\boldsymbol{\p} \cdot \boldsymbol{\sigma}}{|p_0(\boldsymbol{\p})| + \mathpzc{m}} \chi_s
\\                                           
              \chi_s - \frac{\boldsymbol{\p} \cdot \boldsymbol{\sigma}}{|p_0(\boldsymbol{\p})| + \mathfrak{m}} \chi_s                         \end{array}\right),
\,\,\,\,  
v_s(\boldsymbol{\p}) =  \frac{1}{\sqrt{2}} \sqrt{\frac{|p_0(\boldsymbol{\p})| + \mathfrak{m}}{2 |p_0(\boldsymbol{\p})|}}
\left( \begin{array}{c}   \chi_s + \frac{\boldsymbol{\p} \cdot \boldsymbol{\sigma}}{|p_0(\boldsymbol{\p})| + \mathfrak{m}} \chi_s
\\                                           
              -\big(\chi_s - \frac{\boldsymbol{\p} \cdot \boldsymbol{\sigma}}{|p_0(\boldsymbol{\p})| + \mathfrak{m}}\chi_s \big)                          \end{array}\right),
\]
where
\[
\chi_1 = \left( \begin{array}{c} 1  \\
                                                  0 \end{array}\right), \,\,\,\,\,
\chi_2 = \left( \begin{array}{c} 0  \\
                                                  1 \end{array}\right),
\]
are the Fourier transforms of the complete system of the free Dirac equation in the chiral represenation in which 
\[
\gamma^0 = \left( \begin{array}{cc}   0 &  \bold{1}_2  \\
                                           
                                                   \bold{1}_2              & 0 \end{array}\right), \,\,\,\,
\gamma^k = \left( \begin{array}{cc}   0 &  -\sigma_k  \\
                                           
                                                   \sigma_k             & 0 \end{array}\right),
\]
with the Pauli matrices
\[
\boldsymbol{\sigma} = (\sigma_1,\sigma_2,\sigma_3) = 
\left( \,\, \left( \begin{array}{cc} 0 & 1 \\

1 & 0 \end{array}\right), 
\,\,\,\,\,
\left( \begin{array}{cc} 0 & -i \\

i & 0 \end{array}\right),
\,\,\,
\left( \begin{array}{cc} 1 & 0 \\

0 & -1 \end{array}\right)
\,\,
\right).
\]
\label{KernelsForAandPsi}
\end{ex}

\begin{lem}
The free e.m. potential and spinor field operators $\mathbb{A}=A,\boldsymbol{\psi}$ are operator-valued distributions, \emph{i.e.} belong 
to $\mathscr{L}\big(\mathscr{E}, \mathscr{L}((E),(E))\big)$, \emph{i.e.}
\[
\kappa_{0,1}, \kappa_{1,0} \in \mathscr{L}(E^*, \mathscr{E}^{*}) = \mathscr{L}(\mathscr{E}, E) \subset \mathscr{L}(E, \mathscr{E}^{*})
\cong E^{*} \otimes \mathscr{E}^{*},
\]
if and only if
\[
\begin{split}
E= \mathcal{S}^{0}(\mathbb{R}^3; \mathbb{C}^{4}) = \mathcal{S}_{A_{(3)}}(\mathbb{R}^3;
\mathbb{C}^4), \,\, \mathscr{E} = \mathcal{S}^{00}(\mathbb{R}^4; \mathbb{C}^4) = \mathcal{S}_{\widetilde{A}_{(4)}}(\mathbb{R}^4;
\mathbb{C}^4)
\,\,\,\, \textrm{for} \,\,\, A
\\
E= \mathcal{S}(\mathbb{R}^3; \mathbb{C}^4) = \mathcal{S}_{\widetilde{H}_{(3)}}(\mathbb{R}^3;
\mathbb{C}^4), \,\, \mathscr{E} = \mathcal{S}(\mathbb{R}^4; \mathbb{C}^4) = \mathcal{S}_{H_{(4)}}(\mathbb{R}^4; \mathbb{C}^4)
\,\,\,\, \textrm{for} \,\,\, \boldsymbol{\psi}.
\end{split}
\]
Moreover
\begin{equation}\label{kappa(0,1)(xi)inOM}
\kappa_{0,1}, \kappa_{1,0} \in \mathscr{L}(E, \mathcal{O}_M)
\end{equation}
and for each $\xi \in E$, $\kappa_{0,1}(\xi), \kappa_{1,0}(\xi)$ are smooth having all derivatives bounded.
\qed
\label{Cont.free.field.kernels}
\end{lem}

Here $\mathcal{S}^{0}(\mathbb{R}^n;\mathbb{C})=\mathcal{S}_{A_{(n)}}(\mathbb{R}^n;\mathbb{C})$
is the closed subspace of the Schwartz space $\mathcal{S}(\mathbb{R}^n;\mathbb{C})$ of all functions
whose all derivatives vanish at zero. $\mathcal{S}^{00}(\mathbb{R}^n;\mathbb{C})$ is the Fourier transform
inverse image of $\mathcal{S}^{0}(\mathbb{R}^n;\mathbb{C})$. The space $\mathcal{S}^{0}(\mathbb{R}^n;\mathbb{C})$ can be  
realized as a countably Hilbert nuclear space $\mathcal{S}_{A_{(n)}}(\mathbb{R}^n;\mathbb{C})$ associated, in the sense of \cite{GelfandIV},
\cite{obata}, with a positive self adjoint operator $A_{(n)}$ in $L^2(\mathbb{R}^n;\ud^np)$, with $\textrm{Inf Spec} \, A_{(n)} >1$
whose some negative power $\big[A_{(n)} \big]^{-r}$, $r>0$, is of Hilbert-Schmidt class. In order to  construct an example of a series
of operators $A_{(n)}$, $n=2,3, \ldots$, let us consider the harmonic oscillator Hamiltonian operators 
\[
H_{(n)} = - \Delta_{{}_{\mathbb{R}^n}} + r^2 +1,
\,\,\,\,\,\,\,\, r^2= (p_{1})^2 + \ldots +(p_{n})^2,
\,\,\,\,\,\,\,
\textrm{in}
\,\,
L^2(\mathbb{R}^n, \ud^np),
\] 
and the following $L^2(\mathbb{R}\times \mathbb{S}^{n-1}; dt\times d\mu_{{}_{\mathbb{S}^{n-1}}})$,
$L^2(\mathbb{R}\times \mathbb{S}^{n-1}; \nu_{{}_{n}}(t) dt\times d\mu_{{}_{\mathbb{S}^{n-1}}})$ -spaces
on $\mathbb{R}\times \mathbb{S}^{n-1}$ with the weights
\[
\nu_{{}_{n}}(t) = {\textstyle\frac{(t+\sqrt{t^2+4})^{n-1}}{2^{n-2}(t^2+4-t\sqrt{t^2+4})}},
\] 
and the following unitary operators $U_2: L^2(\mathbb{R}\times \mathbb{S}^{n-1}; \nu_{{}_{n}}(t) dt\times d\mu_{{}_{\mathbb{S}^{n-1}}})
\rightarrow L^2(\mathbb{R}^n; \ud^n p)$, $U_1: L^2(\mathbb{R}\times \mathbb{S}^{n-1}; dt\times d\mu_{{}_{\mathbb{S}^{n-1}}})
\rightarrow L^2(\mathbb{R}\times \mathbb{S}^{n-1}; \nu_{{}_{n}}(t) dt\times d\mu_{{}_{\mathbb{S}^{n-1}}})$,
$U= U_2U_1: L^2(\mathbb{R}\times \mathbb{S}^{n-1}; dt\times d\mu_{{}_{\mathbb{S}^{n-1}}}) \rightarrow L^2(\mathbb{R}^n; \ud^n p)$
given by the following formulas
\begin{gather*}
U_1 f(t,\omega) = {\textstyle\frac{1}{\sqrt{\nu_{{}_{n}}(t)}}}f(t,\omega), 
\,\,\,\,\,\,\,\,\,\,\, f \in L^2(\mathbb{R}\times \mathbb{S}^{n-1}; dt\times d\mu_{{}_{\mathbb{S}^{n-1}}}),
\\
U_2 f(r,\omega) = f(t(r),\omega), 
\,\,\,\,\,\,\,\,\,
t(r) = r-r^{-1},
\,\,\,\,\,\,\,\,\,
f\in L^2(\mathbb{R}\times \mathbb{S}^{n-1}; \nu_{{}_{n}}(t) dt \times d\mu_{{}_{\mathbb{S}^{n-1}}}),
\end{gather*}
where $U_2 f$ is expressed in spherical coordinates $(r,\omega)$ in $\mathbb{R}^n$. We can put
\[
A_{(n)} = U \big(H_{(1)}\otimes \boldsymbol{1} + \boldsymbol{1} \otimes \Delta_{{}_{\mathbb{S}^{n-1}}} \big) U^{-1}.
\]
Here $\Delta_{{}_{\mathbb{R}^n}}, \Delta_{{}_{\mathbb{S}^{n-1}}}$ are the standard Laplace operators on $\mathbb{R}^n$
and $\mathbb{S}^{n-1}$, with the standard invariant measures $\ud^np, d\mu_{{}_{\mathbb{S}^{n-1}}}$. In particular
\[
A_{(3)} = - {\textstyle\frac{r^2}{r^2+1}} \partial_{r}^{2} - {\textstyle\frac{r^3(r^2+4)}{(r^2+1)^3}} \partial_{r}
+ \Big[{\textstyle\frac{r^2(r^2+4)(r^2-2)}{4(r^2+1)^4}} + r^2+r^{-2} \Big]  
\]
in spherical coordinates. 

Let us consider Wick products of free fields
\begin{equation}\label{FreeFields}
\mathbb{A}^{{}^{i}}(\phi) = \mathbb{A}^{{}^{i \,\, (-)}}(\phi) + \mathbb{A}^{{}^{i \,\, (+)}}(\phi) = 
\sum_{a_i}\mathbb{A}^{{}^{i \,\, (-)}}_{{}_{a_i}}(\phi^{{}^{a_i}}) + \sum_{a_i}\mathbb{A}^{{}^{i \,\, (+)}}_{{}_{a_i}}(\phi^{{}^{a_i}}) 
=\Xi_{{}_{0,1}}(\kappa_{{}_{0,1}}^{{}^{i}}(\phi)) +
\Xi_{{}_{1,0}}(\kappa_{{}_{0,1}}^{{}^{i}}(\phi))
\end{equation}
with the integral kernels 
\begin{equation}\label{FreeFieldKernels}
\kappa_{{}_{0,1}}^{{}^{i}}(s_i, \boldsymbol{\p}_i; a_i,x) = u_{{}_{s_i \,\,\,a_i}}^{{}^{i}}(\boldsymbol{\p}_i)e^{-ip_i\cdot x},
\,\,\,
\kappa_{{}_{1,0}}^{{}^{i}}(s_i, \boldsymbol{\p}_i; a_i,x) = v_{{}_{s_i \,\,\,a_i}}^{{}^{i}}(\boldsymbol{\p}_i)e^{ip_i\cdot x},
\,\,\,
p_i = (\boldsymbol{\p}_i, p_{0i}(\boldsymbol{\p}_i)) = (\boldsymbol{\p}_i, \sqrt{|\boldsymbol{\p}_i|^2+ \mathfrak{m}_i^2})
\end{equation}
where $\mathfrak{m}_i$ are the masses of the fields $\mathbb{A}^{{}^{i}}$ (and which are always represented by ordinary functions for
all free fields -- the corresponding basic plane wave solutions) and with $u,v$ being always multipliers of the corresponding single particle
nuclear test spaces $E_i\subset \mathcal{H}_i \subset E_i^*$. Let the respective space-time test space of the field $\mathbb{A}^{{}^{i}}$ be equal 
$\mathscr{E}_i = \mathscr{E}\otimes \mathbb{C}^{d_i}$, where $\mathscr{E} = \mathcal{S}(\mathbb{R}^4; \mathbb{C})$ in massive case 
or $\mathscr{E} = \mathcal{S}^{00}(\mathbb{R}^4; \mathbb{C})$ in massless case. From now on (except lemma \ref{MasslessWickProdS00}) 
we restrict the whole theory to the case 
$\mathscr{E} = \mathcal{S}(\mathbb{R}^4; \mathbb{C})$, both in massive and massless case, with the massless fields lying 
in the more singular class $\mathscr{L}\big(\mathscr{E}, \mathscr{L}((E),(E)^*)\big)$. This is because we have in view application to causal
perturbative method which requires as space-time test space $\mathscr{E}$ which includes all smooth functions of compact support, and which is true
for $\mathscr{E} = \mathcal{S}(\mathbb{R}^4; \mathbb{C})$ but it is not the case for $\mathscr{E} = \mathcal{S}^{00}(\mathbb{R}^4; \mathbb{C})$.
$E_i= \mathcal{S}^{0}(\mathbb{R}^3; \mathbb{C}^{d_i})$ in the massless case and $E_i= \mathcal{S}(\mathbb{R}^3; \mathbb{C}^{d_i})$ in the massive case.
In the following lemmas the tensor product $\mathscr{E}_1 \otimes \ldots \mathscr{E}_n$ and its strong dual we simply denote 
by $\mathscr{E}^{\otimes n}$ and  $\mathscr{E}^{* \otimes n}$.

\begin{lem}
The functions $u^{{}^{i}}, u^{{}^{i}}$ in (\ref{FreeFieldKernels}) are multipliers of the single particle nuclear spaces $E_i$
of the corresponding free fields $\mathbb{A}^{{}^{i}}$.
\qed
\label{FreeFieldMultipliers}
\end{lem}

\begin{lem}
The Wick product ${:}\mathbb{A}^{{}^{1}}(x_1) \ldots \mathbb{A}^{{}^{n}}(x_n) {:}$ of the free fields 
$\mathbb{A}^{{}^{1}}(x_1), \ldots, \mathbb{A}^{{}^{n}}(x_n)$, belongs
to $\mathscr{L}\big(\mathscr{E}^{\otimes n}, \mathscr{L}((E),(E))\big)$, \emph{i.e.} it is  an operator valued distribution
in the white-noise sense, and its kernels $\kappa_{l,m}$ are equal to the  tensor products
\begin{gather*}
\kappa_{l,m} = \kappa_{{}_{l_1m_1}}^{{}^{1}} \otimes \ldots \otimes \kappa_{{}_{l_nm_n}}^{{}^{n}}
\in \mathscr{L}(E^{*\otimes (l+m)}, \mathscr{E}^{*\otimes n}) = \mathscr{L}(\mathscr{E}^{\otimes n}, E^{\otimes (l+m)})
\subset \mathscr{L}(E^{\otimes (l+m)}, \mathscr{E}^{* \otimes n})
\cong E^{*\otimes (l+m)} \otimes  \mathscr{E}^{* \otimes n},
\\
l=l_1 + \ldots + l_n, \, m = m_1 + \ldots + m_n,
\end{gather*}
which can be extended to $E^{*}$ in each tensor product factor in $E^{\otimes (l+m)}$, \emph{i.e.}
$\kappa_{l,m} \in  E^{\otimes l} \otimes E^{\otimes m} \otimes  \mathscr{E}^{* \otimes n}$
(by Lemma \ref{Cont.free.field.kernels} and by the extension theorem 3.13 of \cite{obataJFA} or, eventually, its generalization to the Fermi case).
\qed
\label{MasslessWickProdS00}
\end{lem}

\begin{lem}
The Wick product ${:}\mathbb{A}^{{}^{1}}_{{}_{a_1}}(x) \ldots \mathbb{A}^{{}^{n}}_{{}_{a_n}}(x) {:}$ of the free fields 
$\mathbb{A}^{{}^{1}}_{{}_{a_1}}(x_1), \ldots, \mathbb{A}^{{}^{n}}_{{}_{a_n}}(x_n)$, belongs
to $\mathscr{L}\big(\mathscr{E}^{\dot{\otimes} n}, \mathscr{L}((E),(E)^*)\big)$ but not to $\mathscr{L}\big(\mathscr{E}^{\dot{\otimes} n}, \mathscr{L}((E),(E))\big)$, 
\emph{i.e.} it is not equal to any operator valued distribution
in the white-noise sense, if among the free fields $\mathbb{A}^{{}^{1}}, \ldots, \mathbb{A}^{{}^{n}}$ there are massless fields,
and its kernels $\kappa_{{}_{l,m}}(s_1,\boldsymbol{\p}_1, \ldots, \boldsymbol{\p}_n, s_n; a_1, \ldots, a_n, x)$
are equal to the tensor pointwise products 
\begin{multline*}
\kappa_{{}_{l_1m_1}}^{{}^{1}} \dot{\otimes} \ldots \dot{\otimes} \kappa_{{}_{l_nm_n}}^{{}^{n}}(s_1,\boldsymbol{\p}_1, \ldots, s_n,\boldsymbol{\p}_n; a_1, \ldots, a_n,x)
\\
=\kappa_{{}_{l_1m_1}}^{{}^{1}}(s_1\boldsymbol{\p}_1; a_1,x)
\ldots
 \kappa_{{}_{l_n,m_n}}^{{}^{n}}(s_n\boldsymbol{\p}_n; a_n,x) \,\,\,
l=l_1 + \ldots + l_n, \, m = m_1 + \ldots + m_n,
\\
\textrm{and} \,\, \kappa_{{}_{l,m}} \in \mathscr{L}(E^{\otimes (l+m)}, \mathscr{E}^{*})
\cong E^{* \otimes(l+m)} \otimes \mathscr{E}^{* \dot{\otimes} n}.
\end{multline*}
Here $\mathscr{E}^{\dot{\otimes} n} = \mathscr{E} \otimes \mathbb{C}^{d_1} \otimes \ldots \otimes \mathbb{C}^{d_n}$
and $\mathscr{E}^{* \dot{\otimes} n} = \mathscr{E}^* \otimes \mathbb{C}^{d_1} \otimes \ldots \otimes \mathbb{C}^{d_n}$.
\qed
\label{MasslessWickPointwiseProd}
\end{lem}

\begin{lem}
The Wick product ${:}\mathbb{A}^{{}^{1}}_{{}_{a_1}}(x) \ldots \mathbb{A}^{{}^{n}}_{{}_{a_n}}(x) {:}$ of massive free fields belongs
to $\mathscr{L}\big(\mathscr{E}^{\dot{\otimes} n}, \mathscr{L}((E),(E))\big)$, \emph{i.e.} it is an operator valued distribution,
and its kernels $\kappa_{{}_{l,m}}(s_1,\boldsymbol{\p}_1, \ldots, \boldsymbol{\p}_n, s_n; a_1, \ldots, a_n,x)$
are equal to the pointwise products
\begin{multline*}
\kappa_{{}_{l_1m_1}}^{{}^{1}} \dot{\otimes} \ldots \dot{\otimes} 
\kappa_{{}_{l_nm_n}}^{{}^{n}}(s_1,\boldsymbol{\p}_1, \ldots, s_n,\boldsymbol{\p}_n; a_1, \ldots, a_n,x)
\\
=\kappa_{{}_{l_1m_1}}^{{}^{1}}(s_1,\boldsymbol{\p}_1; a_1,x)
\ldots
 \kappa_{{}_{l_n,m_n}}^{{}^{n}}(s_n,\boldsymbol{\p}_n; a_n,  x) \,\,\,
l=l_1 + \ldots + l_n, \, m = m_1 + \ldots + m_n,
\\
\textrm{and} \,\, \kappa_{{}_{l,m}} \in \mathscr{L}(E^{\otimes (l+m)}, \mathscr{E}^{*})
\cong E^{*\otimes (l+m)} \otimes \mathscr{E}^{* \dot{\otimes} n},
\end{multline*}
which can be extended to $E^{*}$ in each tensor product factor in $E^{\otimes (l+m)}$ which corresponds to an annihilation operator,
\emph{i.e.} $\kappa_{l,m} \in E^{\otimes l} \otimes E^{* \otimes m}\otimes \mathscr{E}^{* \dot{\otimes} n}$.
\label{MassiveWickPointwiseProd}
\end{lem}
\begin{lem}
The kernels $\kappa_{{}_{l,m}}$ of the Wick product ${:}\mathbb{A}^{{}^{1}}(x) \ldots \mathbb{A}^{{}^{n}}(x) {:}$ of free fields
map continuously
\[
E^{\otimes n } \ni \xi = \xi_1 \otimes \ldots \otimes \xi_n \longmapsto
\kappa_{{}_{l,m}}(\xi) =
\kappa_{{}_{l_1,m_1}}^{{}^{1}}(\xi_1)
\ldots
 \kappa_{{}_{l_n,m_n}}^{{}^{n}}(\xi_n)  \in \mathcal{O}_{M}(\mathbb{R}^4)
\]
and each space-time derivative of $\kappa_{{}_{l,m}}(\xi)$ is bounded.
\label{DistrProdWickProd}
\end{lem}
\qedsymbol \,
Lemma \ref{DistrProdWickProd} immediately follows from (\ref{kappa(0,1)(xi)inOM}) of lemma \ref{Cont.free.field.kernels}
and from the product formula
\[
\kappa^{{}^{1}}_{l_1, m_1} \dot{\otimes} \ldots \dot{\otimes} \kappa^{{}^{n}}_{l_n, m_n}(\xi_1 \otimes \ldots \xi_m)
= \kappa^{{}^{1}}_{l_1, m_1}(\xi_1) \cdots \kappa^{{}^{n}}_{l_n, m_n}(\xi_n), 
\,\,\,\,
(l_i,m_i) = (0,1) \,\, \textrm{or} \,\, (l_i,m_i) = (1,0), \xi_i \in E_i. 
\qed
\] 
In particular the Wick product ${:}\mathbb{A}^{{}^{1}}(x) \ldots \mathbb{A}^{{}^{n}}(x) {:}$ can be multiplied
by any tempered distribution $t(x)$ and gives a well-defined generalized operator. This is because the value $\kappa_{{}_{l,m}}(\xi)$
of each of its kernels at $\xi\in E^{\hat{\otimes}n}$ can be multiplied by the distribution $t(x)$, because the value
of this product at $\xi\in E^{\hat{\otimes}n}$ is well-defined
and continuously depends on $\xi$ (as $\kappa_{{}_{l,m}}(\xi) \in \mathcal{O}_M$ continuously depends on $\xi$ by Lemma \ref{DistrProdWickProd}).

The proof of lemma \ref{Cont.free.field.kernels} is rather an easy verification of the definition of continuity
which can be checked on using explicit form of the plane-wave kernels $\kappa_{{}_{l_im_i}}^{{}^{i}}(s_i,\boldsymbol{\p}_i; a_i, x)$ of free fields.
As already said,
lemmas \ref{MasslessWickProdS00} and \ref{DistrProdWickProd} are immediate consequences of lemma \ref{Cont.free.field.kernels}
and theorems 3.9, 3.13 of \cite{obataJFA} (or eventually its extension to the Fermi case).
Lemma \ref{MasslessWickPointwiseProd} immediately follows from lemma \ref{DistrProdWickProd} and theorem 3.9 of \cite{obataJFA}.
The proof of lemma \ref{MassiveWickPointwiseProd} we are giving only later, compare lemma \ref{rett(x,y):W'(x)W''(y):InSxS->L((E),(E))Contq1>2}, 
the case of kernel (\ref{x1(x2)}) in which we put the distribution $t=\kappa_q$ identically $1$. 

Here we only emphasize the essential point which makes difference between the lemmas \ref{MassiveWickPointwiseProd}
and \ref{MasslessWickPointwiseProd}, in which the presence of the massless free field is essential. 
To understand the difference between the massive and massless cases let $\kappa_{1,0}^{{}^{1}}, \ldots, \kappa_{1,0}^{{}^{n}}$ 
be kernels of free fields, among which at least one, say $\kappa_{1,0}^{{}^{1}}$, 
is massless. Let us show that in this case the assertion of lemma \ref{MassiveWickPointwiseProd} cannot hold. 
The pointwise product $\kappa_{1,0}^{{}^{1}} \overset{\cdot}{\otimes} \ldots \overset{\cdot}{\otimes} \kappa_{1,0}^{{}^{n}}$
we can treat as an element of
\[
\mathscr{L}(\mathscr{E}^{\dot{\otimes} n}, \, E_{{}_{1}}^{*}
\otimes \ldots \otimes E_{{}_{n}}^{*}) \cong
\mathscr{L}(E_{{}_{1}} \otimes \ldots \otimes E_{{}_{n}}, \,\,
\mathscr{E}^{*\dot{\otimes} n}).
\]
Assertion of the lemma \ref{MassiveWickPointwiseProd} would be proved iff we show that (Thm 3.13 of \cite{obataJFA})
\[
\kappa_{1,0}^{{}^{1}} \overset{\cdot}{\otimes} \ldots \overset{\cdot}{\otimes} \kappa_{1,0}^{{}^{n}}
\in
\mathscr{L}(\mathscr{E}^{\dot{\otimes} n}, \, E_{{}_{1}}^{*}
\otimes \ldots \otimes E_{{}_{n}}^{*})
\]
actually belongs to $\mathscr{L}(\mathscr{E}^{\dot{\otimes} n}, \, E_{{}_{1}}
\otimes \ldots \otimes E_{{}_{n}})$.
In this case
\[
\kappa_{1,0}^{{}^{1}} \overset{\cdot}{\otimes} \ldots \overset{\cdot}{\otimes} \kappa_{1,0}^{{}^{n}}
\in
\mathscr{L}(\mathscr{E}, \, E_{{}_{1}}^{*}
\otimes \ldots \otimes E_{{}_{n}}^{*}) \cong
\mathscr{L}(E_{{}_{1}} \otimes \ldots \otimes E_{{}_{n}}, \,\,
\mathscr{E}^{*}).
\]
would be extendible to an element of (compare Thm 3.13 of \cite{obataJFA} and kernel theorem)
\[
\mathscr{L}(E_{{}_{1}}^{*} \otimes \ldots \otimes E_{{}_{n}}^{*}, \,\,
\mathscr{E}^{*\dot{\otimes} n}) \cong
\mathscr{L}(\mathscr{E}^{\dot{\otimes} n}, \, E_{{}_{1}}
\otimes \ldots \otimes E_{{}_{n}})
\]
iff $\kappa_{1,0}^{{}^{1}} \overset{\cdot}{\otimes} \ldots \overset{\cdot}{\otimes} \kappa_{1,0}^{{}^{n}}$
actually belongs to $\mathscr{L}(\mathscr{E}^{\dot{\otimes} n}, \, E_{{}_{1}}
\otimes \ldots \otimes E_{{}_{n}})$.
This however is impossible if the kernel $\kappa_{1,0}^{{}^{1}}$ is associated to a free 
massless field (or its derivative). Indeed, easy computation shows that
$\kappa_{1,0}^{{}^{1}} \overset{\cdot}{\otimes} \ldots \overset{\cdot}{\otimes} \kappa_{1,0}^{{}^{n}}(\phi)$,
$\phi \in \mathscr{E}^{\dot{\otimes} n}$, has the following general form (with $a$ regarded as multiindex $(a_1, \dots, a_n)$)
\begin{multline*}
\kappa_{1,0}^{{}^{1}} \overset{\cdot}{\otimes} \ldots \overset{\cdot}{\otimes} \kappa_{1,0}^{{}^{n}}(\phi)
(s_1,\boldsymbol{\p}_1, \ldots, s_n, \boldsymbol{\p}_n) =
v_{{}_{s_1}}^{{}^{1}}(\boldsymbol{\p}_1) \ldots v_{{}_{s_n}}^{{}^{n}}(\boldsymbol{\p}_n) \widetilde{\phi}(\boldsymbol{\p}_1 + \ldots + \boldsymbol{\p}_n,
p_{01}(\boldsymbol{\p}_1) + \ldots + p_{0n}(\boldsymbol{\p}_n))
\\
= \sum\limits_{a} v_{{}_{s_1 \,\, a_1}}^{{}^{1}}(\boldsymbol{\p}_1) \ldots v_{{}_{s_n \,\, a_n}}^{{}^{n}}(\boldsymbol{\p}_n) 
\widetilde{\phi^a}(\boldsymbol{\p}_1 + \ldots + \boldsymbol{\p}_n,
p_{01}(\boldsymbol{\p}_1) + \ldots + p_{0n}(\boldsymbol{\p}_n)),
\end{multline*}
where $v^{{}^{i}}$ is a multiplier of $E_{{}_{i}} = \mathcal{S}_{{}_{A_{i}}}(\mathbb{R}^3)$ and, by assumption, $p_{01}(\boldsymbol{\p}_1) = |\boldsymbol{\p}_1|$,
and generally $p_{0i}(\boldsymbol{\p}_i) = |\boldsymbol{\p}_i|$ if $i$-th field is massless.
We can now see that $\kappa_{1,0}^{{}^{1}} \overset{\cdot}{\otimes} \ldots \overset{\cdot}{\otimes}\kappa_{1,0}^{{}^{n}}(\phi)$
cannot even be extended to elements of $\mathscr{C}^{\infty}(\mathbb{R}^3 \times \ldots \times \mathbb{R}^3)$, with their derivations blowing up to infinity 
along the subset $|\boldsymbol{\p}_1|=0$ and, generally, along $|\boldsymbol{\p}_i|=0$, corresponding to the $i$-th massless field. 
So all the more it cannot belong to $\mathcal{S}(\mathbb{R}^3) \otimes \ldots \otimes \mathcal{S}(\mathbb{R}^3) \otimes \ldots$
or to $\mathcal{S}^{0}(\mathbb{R}^3) \otimes \ldots \otimes \mathcal{S}(\mathbb{R}^3) \otimes \ldots 
= E_1 \otimes \ldots \otimes E_n$ (where $E_{{}_{1}} = \mathcal{S}^{0}(\mathbb{R}^3)\otimes \mathbb{C}^{d_1}$ for  $\mathbb{A}^{{}^{1}}$ and generally 
$\mathcal{S}^{0}(\mathbb{R}^3)\otimes \mathbb{C}^{d_i}$ for massless fields $\mathbb{A}^{{}^{i}}$
and $E_{{}_{i}} = \mathcal{S}(\mathbb{R}^3)\otimes \mathbb{C}^{d_i}$ for massive fields $\mathbb{A}^{{}^{i}}$).
In particular
$\phi \longmapsto
\kappa_{1,0}^{{}^{1}} \overset{\cdot}{\otimes} \ldots \overset{\cdot}{\otimes} \kappa_{1,0}^{{}^{n}}(\phi)$
cannot be continuous as a map
$\mathscr{E}^{\dot{\otimes} n} \longmapsto E_{{}_{1}}
\otimes \ldots \otimes E_{{}_{n}}$.
From this it follows that $\kappa_{1,0}^{{}^{1}} \overset{\cdot}{\otimes} \ldots \overset{\cdot}{\otimes} \kappa_{1,0}^{{}^{n}}$
cannot be extended to an element of $\mathscr{L}(E_{{}_{1}}^{*} \otimes \ldots \otimes E_{{}_{n}}^{*}, \,\,
\mathscr{E}^{* \dot{\otimes} n})$.

\section{Product operation. Wick theorem for products}

Let us consider Wick theorem for ``products'' of Wick polynomials of free fields. In the intermediate stage of the computations
of the scattering operator and interacting fields the so-called Wick theorem (\cite{Bogoliubov_Shirkov}, \S 17.2) is used for
decomposition of the ``product''
\begin{equation}\label{Wick(x)Wick(y)}
\boldsymbol{{:}} \mathbb{A'}^{{}_{1}}_{{}_{a_1}}(x) \ldots \mathbb{A'}^{{}_{n}}_{{}_{a_n}}(x) \boldsymbol{{:}} \,
\boldsymbol{{:}} \mathbb{A''}^{{}_{1}}_{b_{1}}(y) \ldots \mathbb{A''}^{{}_{q}}_{b_{q}}(y) \boldsymbol{{:}}
\end{equation}
of Wick product monomials
\begin{equation}\label{WickMonomialsInGeneralFreeFields}
\boldsymbol{{:}} \mathbb{A'}^{{}_{1}}_{{}_{a_1}}(x) \ldots \mathbb{A'}^{{}_{n}}_{{}_{a_n}}(x) \boldsymbol{{:}}
\,\,\,
\textrm{and}
\,\,\,
\boldsymbol{{:}} \mathbb{A''}^{{}_{1}}_{b_{1}}(y) \ldots \mathbb{A''}^{{}_{q}}_{b_{q}}(y) \boldsymbol{{:}}
\end{equation}
in fixed components $\mathbb{A'}^{{}_{i}}_{{}_{a_i}}$, $\mathbb{A''}^{{}_{j}}_{b_{j}}$ of free fields
$\mathbb{A'}^{{}_{i}}$, $\mathbb{A''}^{{}_{j}}$, each separately evaluated at the same space-time point $x$ or respectively, $y$,
into the sum of Wick monomials (each in the so called ``normal order'').

The point lies in the correct definition of such ``product'', because in general (lemma \ref{MasslessWickPointwiseProd}) each factor evaluated respectively at $x$ or $y$, represents a generalized integral kernel operator transforming continuously the Hida space $(E)$ into its strong dual
$(E)^*$, so that the product cannot be understood as ordinary operator composition, and therefore a correct definition is here required.

As the first step in definition of the product we note that the free fields and their Wick products define (finite sums of) integral kernel operators with vector-valued
kernels in the sense of \cite{obataJFA}, as we have explained above, and the ''product'' can be given as a distributional
kernel operator. Indeed, from lemma \ref{MassiveWickPointwiseProd} it follows that each factor (\ref{WickMonomialsInGeneralFreeFields}) separately
represents an integral kernel operator which belongs to
\[
\mathscr{L}\big(\mathscr{E}_{i}, \, \mathscr{L}((E), \, (E)) \big), \,\,\, i =1 \, \textrm{or} \, 2,
\]
if among the factors $\mathbb{A'}^{{}_{i}}$, $\mathbb{A''}^{{}_{j}}$, there are no massless fields (or their derivatives).
This means that the first factor in (\ref{WickMonomialsInGeneralFreeFields}) defines the corresponding continuous map
\[
\mathscr{E}_{1} \ni \phi \longmapsto \Xi'(\phi) =
\sum\limits_{\substack{\ell',m' \\ \ell'+m'=n}}
\Xi'_{\ell',m'}\big(\kappa'_{\ell',m'}(\phi)\big)
\in \mathscr{L}\big( (E), (E)\big), 
\]
and similarly the second factor in (\ref{WickMonomialsInGeneralFreeFields}) defines continuous map
\[
\mathscr{E}_{2} \ni \varphi \longmapsto \Xi''(\varphi) =
\sum\limits_{\substack{\ell'',m'' \\ \ell''+m''=q}}
\Xi''_{\ell'',m'''}\big({\kappa''}_{\ell'',m''}(\varphi)\big)
\in \mathscr{L}\big( (E), (E)\big),
\]
where $\mathscr{E}_{i} = \mathcal{S}(\mathbb{R}^4; \mathbb{C}^{d_i})$
$i= 1,2$. Both factors $\Xi'$ and $ \Xi'''$
are equal to finite sums of integral kernel operators with $\mathscr{E}_{1}^{*}$- or $\mathscr{E}_{2}^{*}$-valued distributional kernels
$\kappa'_{\ell',m'}, \kappa''_{\ell'',m''}$. In this case both factors $\Xi'(\phi)$ and $ \Xi''(\varphi)$, when evaluated at the test functions $\phi, \varphi$,
are ordinary operators on the Fock space transforming continuously the Hida space $(E)$ into itself,
and thus can be composed $\Xi'(\phi) \circ \Xi''(\varphi)$ as operators, giving the composition operator
\[
\Xi'(\phi) \circ \Xi''(\varphi) \in \mathscr{L}\big( (E), (E)\big),
\]
defining the map
\[
\mathscr{E}_{1} \otimes \mathscr{E}_{2} \ni \phi \otimes \varphi \longmapsto \Xi'(\phi) \circ \Xi''(\varphi)
\in \mathscr{L}\big( (E), (E)\big),
\]
which by construction is separately continuous in the arguments $\phi \in \mathscr{E}_{1}$ and $\varphi \in \mathscr{E}_{2}$. Because
$\mathscr{E}_{1}, \mathscr{E}_{2}$ are complete Fr\'echet spaces, then by Thm 2.17 of \cite{Rudin} (or Proposition 1.3.11 of \cite{obataJFA}) there exist
the corresponding operator-valued continuous map (say operator-valued distribution with $a,b$ understood as multiindices)
\begin{multline}\label{Xi(phi)circXi(varphi)distribution}
\phi \otimes \varphi \longmapsto
\Xi(\phi \otimes \varphi) =
\sum\limits_{a,b}
\int\limits_{\big[\mathbb{R}^4\big]^{\times \, 2}}
\boldsymbol{{:}} \mathbb{A'}^{{}_{1}}_{{}_{a_1}}(x) \ldots \mathbb{A'}^{{}_{n}}_{{}_{a_n}}(x) \boldsymbol{{:}} \,
\boldsymbol{{:}} \mathbb{A''}^{{}_{1}}_{{}_{b_1}}(y) \ldots \mathbb{A''}^{{}_{q}}_{{}_{b_q}}(y) \boldsymbol{{:}} \,
\phi_a \otimes \varphi_b(x,y) \, \ud^4x \, \ud^4 y
\\
= \Xi'(\phi) \circ \Xi''(\varphi).
\end{multline}
This gives us the Wick theorem for the ``product'' (\ref{Wick(x)Wick(y)}) in case in which all factors
$\mathbb{A'}^{{}_{i}}$, $\mathbb{A''}^{{}_{j}}$, are massive free fields or their derivatives.

Namely, if the free fields 
in (\ref{Wick(x)Wick(y)}) are massive, then 
\begin{eqnarray*}
\Xi'(\phi) = \sum \limits_{a} \int\limits_{\mathbb{R}^4}
\boldsymbol{{:}} \mathbb{A'}^{{}_{1}}_{{}_{a_1}}(x) \ldots \mathbb{A'}^{{}_{n}}_{{}_{a_n}}(x) \boldsymbol{{:}} \, \phi^a(x) \,\,\,\, \ud^4x
\in \mathscr{L}((E),(E)),
\\
\Xi''(\varphi) = \sum \limits_{b} \int\limits_{\mathbb{R}^4}
\boldsymbol{{:}} \mathbb{A''}^{{}_{1}}_{{}_{b_1}}(y) \ldots \mathbb{A''}^{{}_{b_q}}_{{}_{b_q}}(y) \boldsymbol{{:}} \,\,\,\, \varphi^b(y) \, \ud^4y
\in \mathscr{L}((E),(E)),
\\
\Xi(\phi \otimes \varphi) =
\sum \limits_{a,b}
\int\limits_{\big[\mathbb{R}^4\big]^{\times \, 2}}
\boldsymbol{{:}} \mathbb{A'}^{{}_{1}}_{{}_{a_1}}(x) \ldots \mathbb{A}^{{}_{n}}_{{}_{a_n}}(x) \boldsymbol{{:}} \,\,
\boldsymbol{{:}} \mathbb{A''}^{{}_{1}}_{{}_{b_1}}(y) \ldots \mathbb{A''}^{{}_{q}}_{{}_{b_q}}(y) \boldsymbol{{:}} \,\, \times
\\ \times \,\,\,\,\,\,\,
\phi^a \otimes \varphi^b(x,y) \, \ud^4x \ud^4y \,\,\, \in \mathscr{L}((E),(E)),
\end{eqnarray*}
and moreover
\begin{eqnarray*}
\Xi' \in \mathscr{L}(\mathscr{E}_{1}, \mathscr{L}((E),(E))\big),
\\
\Xi''\in \mathscr{L}(\mathscr{E}_{2}, \mathscr{L}((E),(E))\big),
\\
\Xi\in \mathscr{L}(\mathscr{E}_{1} \otimes \mathscr{E}_{2}, \mathscr{L}((E),(E))\big),
\end{eqnarray*}
and the following Wick theorem for massive fields holds
\begin{twr}
\[
\Xi =
\sum_{\substack{\kappa'_{\ell',m'} \\ \kappa''_{\ell'',m''}}} \sum \limits_{0\leq q \leq \textrm{min} \{m',\ell''\}} \,
\Xi_{\ell'+\ell''-q,m'+m''-q}\big( \kappa'_{\ell',m'} \otimes_{q} \, {\kappa''}_{\ell'',m''}\big),
\]
where
\begin{multline}\label{ContractionIntegral}
\kappa'_{\ell',m'} \otimes_{q} \, {\kappa''}_{\ell'',m''}(\phi\otimes\varphi) = \kappa'_{\ell',m'}(\phi) \otimes_{q} \, {\kappa''}_{\ell'',m''}(\varphi)
\\
=
\sum\limits_{\substack{ \ldots, s'_{\ell'+i}, s''_{j}  \ldots}}
\int
\kappa'_{\ell',m'}(\phi)( \ldots, s'_{\ell'+i}, \boldsymbol{\p}'_{\ell'+i}, \ldots )
\kappa''_{\ell'',m''}(\varphi)( \ldots, s''_{j}, \boldsymbol{\p}''_{j}, \ldots,s''_1, \boldsymbol{\p}''_{\ell''+1}, \ldots) \,\,\,
\times
\\
\times \,\,\, \cdots \delta_{{}_{s'_{\ell'+i} \,\, s''_{j}}} \,\, \delta(\boldsymbol{\p}'_{\ell'+i}-\boldsymbol{\p}''_{j}) 
\cdots \ud^3 \boldsymbol{\p}'_{\ell'+i} \ud^3 \boldsymbol{\p}''_{j} \cdots,
\end{multline}
and where the kernels $\kappa'_{\ell',m'}, \kappa''_{\ell'',m''}$ range, respectively, over the kernels of finite Fock expansions
of the operators $\Xi', \Xi''$. The $q$-contractions $\kappa_{\ell,m} = \kappa'_{\ell',m'} \otimes_{q} \, {\kappa''}_{\ell'',m''}$ are performed
upon all $q$ pairs of variables $(s'_{\ell'+i}, \boldsymbol{\p}'_{\ell'+i}), (s''_{j}, \boldsymbol{\p}''_{j})$ 
in which the first variable $(s'_{\ell'+i}, \boldsymbol{\p}'_{\ell'+i})$ of the contracted pair lies among the last $m'$ variables of the kernel
$\kappa'_{\ell',m'}(\phi)$ and the second variable $(s''_{j}, \boldsymbol{\p}''_{j})$ 
of the contracted pair lies among the first $l''$ variables of the kernel $\kappa''_{\ell'',m''}(\varphi)$, and to both variables
of the contracted pair correspond respectively annihilation and creation operator of one and the same free field. The contractions
$\kappa_{\ell,m} = \kappa'_{\ell',m'} \otimes_{q} \, {\kappa''}_{\ell'',m''}$, $\ell=\ell'+\ell''-q, m=m'+m''-q$ 
should be symmetrized in Bose variables and antisymmetrized 
in Fermi variables in order to keep one-to-one correspondence between the kernels and the corresponding integral kernel operator.
The symmetrization or alternation should be performed within each set of variables corresponding to one and the same free field,
and, moreover, which lie among the first $\ell$ variables and separately within the variables lying
among the last $m$ variables of the contraction $\kappa_{\ell,m}$. 
\label{WickThmForMassiveFields}
\end{twr}

\qedsymbol \,  For the proof we apply definition \ref{XiDefinition} to the first and then to the second of the composed operators 
\[
\sum_{\substack{\kappa'_{\ell',m'},  \kappa''_{\ell'',m''}}} 
\Xi'_{\ell',m'}\big(\kappa'_{\ell',m'}(\phi)\big) \circ \Xi''_{\ell'',m''}\big({\kappa''}_{\ell'',m''}(\varphi)\big).
\] 
That the contraction integral (\ref{ContractionIntegral}) is absolutely convergent follows from
lemma \ref{MassiveWickPointwiseProd}, becuse in the considered massive case 
$\kappa'_{\ell',m'} \in E^{\otimes \ell'} \otimes E^{* \otimes m'}\otimes \mathscr{E}_{1}^{*}$
and $\kappa''_{\ell'',m'} \in E^{\otimes \ell''} \otimes E^{* \otimes m''}\otimes \mathscr{E}_{2}^{*}$ ad the contraction (dual pairing)
is well-defined. \qed

Note that the one-contraction $\kappa^{{}^{i}}_{0,1}\otimes_1 \kappa^{{}^{j}}_{1,0}$ is nothing else but the pairing 
$\quad \underbracket{\mathbb{A}^{{}^{i}} \mathbb{A}^{{}^{j}}}$ of the fields $\mathbb{A}^{{}^{i}},\mathbb{A}^{{}^{j}}$ in the notation of \cite{Bogoliubov_Shirkov}.
And more generally the $\otimes_q$-contractions (\ref{masslesskappa01.masslesskappa10...q'-contraction...masslesskappa10.masslesskappa10}),
(\ref{(x)q}) are nothing else but what is called product $\quad \underbracket{\mathbb{A}^{{}^{q_1}} \mathbb{A}^{{}^{q_1}}} \ldots
\quad \underbracket{\mathbb{A}^{{}^{q_q}} \mathbb{A}^{{}^{q_q}}}$ of $q$ pairings 
$\quad \underbracket{\mathbb{A}^{{}^{q_i}} \mathbb{A}^{{}^{q_i}}}$ of the fields $\mathbb{A}^{{}^{q_i}}$ in \cite{Bogoliubov_Shirkov}.
That  $\kappa_{\ell,m}$ are indeed well-defined vector-valued distributions is not visible in the conventional formulation of the (formal/symbolic version of the) Wick theorem, where regularization is required.
The point is that kernels $\kappa_{\ell,m}$ of the Fock expansion (\ref{Xi(phi)circXi(varphi)distribution}) can be obtained in well-defined
mathematical terms by the operations of pointed tensor product $\dot{\otimes}$, ordinary tensor product $\otimes$, $q$-contraction
$\otimes_q$, symmetrization and antisymmetrization,
from the kernels $\kappa_{0,1}, \kappa_{1,0}$ defining the free fields $\mathbb{A}_{{}_{k}}$, but the kernels
$\kappa_{\ell,m}$ of the Fock expansion (\ref{Xi(phi)circXi(varphi)distribution}) cannot be obtained from the
pairing distributions of the fields $\mathbb{A}_{{}_{k}}$ without the intermediate step, in which we are 
about to multiply products of pairings. Using the white noise construction of the free fields and the kernels
$\kappa^{{}^{i}}_{0,1}, \kappa^{{}^{i}}_{1,0}$ defining the free fields $\mathbb{A}^{{}^{i}}$, we can
forget about any ''regularization'', at least for massive fields $\mathbb{A}^{{}^{i}}$. Indeed, because
contraction of the kernels is well-defined, then so is the product of pairings, which can be defined through it. So we have rigorously
shown, using white noise, that the pairings are exceptional, and their products are well-defined
distributions. The fact that ``regularization'' can be avoided in the computation of the product of pairings in the Wick theorem
was also noticed in \cite{Scharf}, but in a rather informal manner. Here we give a mathematical justification for the 
practical computations of the products of pairings, as given in \cite{Scharf}.

This form of Wick theorem, \emph{i.e.} Theorem \ref{WickThmForMassiveFields}, is however
insufficient in realistic QFT, such as QED, because in the causal construction of the scattering operator or causal construction of interacting
fields from the scattering operator, the  Wick factors (\ref{WickMonomialsInGeneralFreeFields}) necessary include the Lagrange interaction density
\[
\mathcal{L}(x) = \boldsymbol{{:}} \boldsymbol{\psi}(x)^{+}\gamma_{0} \gamma^\mu \boldsymbol{\psi}(x) A_\mu(x) \boldsymbol{{:}} 
\]
and necessary include the massless electromagnetic potential field $A$ as one of the factors $\mathbb{A}_{{}_{k}}$
in the Wick products which have to be considered. In particular, in the second inductive computational step (second order contribution) 
in the causal perturbative construction of the 
scattering operator we need to consider the``product'' (compare \cite{Scharf}, \cite{Bogoliubov_Shirkov}, \cite{Epstein-Glaser})
\begin{equation}\label{L(x)L(y)}
\mathcal{L}(x) \mathcal{L}(y)
\end{equation}
and apply Wick theorem of \cite{Bogoliubov_Shirkov} in order to write it in the form of ``normally ordered''
operators. 

But Wick Theorem \ref{WickThmForMassiveFields} for massive fields can be generalized over to the case
of product (\ref{Wick(x)Wick(y)}) in which each of the normal factors (\ref{WickMonomialsInGeneralFreeFields})
contains massless fields $\mathbb{A}_{{}_{k}}$.

In this situation, when among the factors $\mathbb{A}_{{}_{k}}$ in (\ref{Wick(x)Wick(y)}) there are present mass-less 
fields (or their derivatives) in (\ref{Wick(x)Wick(y)}), as e.g. in (\ref{L(x)L(y)}) of QED -- a particular case of (\ref{Wick(x)Wick(y)}) -- 
then we replace the $\mathscr{E}_{2}^{*}$-valued kernels $\kappa_{0,1}, \kappa_{1,0}$ defining the massless factors by their massive counterparts.
In practice, we just replace the zero mass energy functions
\[
p_{0}(\boldsymbol{\p}) = |\boldsymbol{\p}|
\]
in the massless kernels $\kappa_{0,1}, \kappa_{1,0}$ by the massive energy functions
\[
p_{0}(\boldsymbol{\p}) = \sqrt{|\boldsymbol{\p}|^2 + \epsilon^2}
\]
and obtain in this manner the kernels $\kappa_{\epsilon \,\, 0,1}, \kappa_{\epsilon \,\, 1,0}$ with exponents $e^{\mp ip\cdot x}$
with $p$ ranging over the positive energy massive hyperboloid $\mathscr{O}_{{}_{\epsilon,0,0,0}}$, the nuclear massless single particle spaces
$E_i$ corresponding to the massless fields we keep unchanged, but the space-time test spaces
\[
\mathscr{E}_i =  \mathcal{S}^{00}(\mathbb{R}^4)\subset  \mathcal{S}(\mathbb{R}^4),
\]
corresponding to these fields, we can, and we do, enlarge to  $\mathscr{E}_i =  \mathcal{S}(\mathbb{R}^4)$, although the regularity of the corresponding
field operator will have to be weakened. In the limit $\epsilon \rightarrow 0$, the replacement of the space-time test space
does not make any difference, as the regularity of pointwise Wick product operators, in both cases,  is weakened and in both cases
the pointwise Wick products transform continuously the space-time test space into operators transforming $(E)$ into $(E)^*$.  
Taking into account application to causal QFT, the space-time test space should be enlarged to the Schwartz test space $\mathcal{S}(\mathbb{R}^4)$ 
as it contains all smooth functions of compact support, which is needed for the implementation of causality. 

Then we construct the analogue  $\Xi_{\epsilon}$ of the
product (\ref{Wick(x)Wick(y)}) in which all the massless free field kernels are replaced with massive ``counterparts'' 
$\kappa_{\epsilon \,\, 0,1}, \kappa_{\epsilon \,\, 1,0}$ (more precisely:
we replace $|\boldsymbol{\p}|$ with $(|\boldsymbol{\p}|^2+\epsilon^2)^{1/2}$ in the massless kernles $\kappa_{0,1}, \kappa_{1,0}$). Next, we compute
the Fock expansion of $\Xi_{\epsilon}$:
\begin{equation}\label{FockExpansionXiepsilon}
\Xi_{\epsilon} = 
\sum\limits_{\ell, m} \Xi_{\epsilon \,\, \ell, m}({\kappa}_{\epsilon \,\,  \ell, m})
\,\,\,\,\,\,\,
\in \,\,\,\,\,\,\, \mathscr{L}((E) \otimes \mathscr{E}_{1} \otimes \mathscr{E}_{2}, \, (E)), \,\,\, \ell+ m \leq n+q,
\end{equation}
into integral kernel operators
with $\mathscr{E}_{1}^{*} \otimes \mathscr{E}_{2}^{*} = \mathscr{L}(\mathscr{E}_{1} \otimes \mathscr{E}_{2}, \mathbb{C})$-valued
kernels ${\kappa}_{\epsilon \,\, \ell, m}$, exactly as above for the massive fields, with $\mathscr{E}_{1}$ and  $\mathscr{E}_{2}$ 
being the ordinary Schwartz spaces, as was for the purely massive case.

Next we observe that
${\kappa}_{\epsilon \,\, \ell, m} \rightarrow {\kappa}_{\ell, m}, {\kappa}_{\epsilon \,\, 0,0}
\rightarrow {\kappa}_{0,0}$ converge in
\[
\mathscr{L}(E^{\otimes (\ell+m)}; \mathscr{E}_{1}^{*} \otimes \mathscr{E}_{2}^{*}), \,\,\, \ell+ m \leq n+q,
\]
and, respectively (for the scalar contractions), in $\mathscr{E}_{1}^{*} \otimes \mathscr{E}_{2}^{*}$,
when $\epsilon \rightarrow 0$, with $\mathscr{E}_{1}$, $\mathscr{E}_{2}$, equal to the ordinary Schwartz
space-time test spaces $\mathcal{S}(\mathbb{R}^4;\mathbb{C}^{d_1})$, $\mathcal{S}(\mathbb{R}^4;\mathbb{C}^{d_2}), \ldots$, 
depending on the fields in (\ref{Wick(x)Wick(y)}), in which the spaces
$\mathcal{S}^{00}(\mathbb{R}^4; \mathbb{C}^{d_i})$ corresponding to the massless fields 
are replaced with $\mathcal{S}(\mathbb{R}^4)$.  
By Thm. 3.9 and Theorem 4.8 of \cite{obataJFA} (or, respectively, by their Fermi-Fock analogues) the operator ''product'' $\Xi_{\epsilon}$
converges to an operator
\begin{equation}\label{XimasslessProduct}
\Xi \in \mathscr{L}(\,\, (E) \otimes \mathscr{E}_{1} \otimes \mathscr{E}_{2} \,\, , \,\,\, (E)^* \,\,)
\end{equation}
when $\epsilon \rightarrow 0$, which in general does not belong to
\[
\mathscr{L}(\,\,(E) \otimes \mathscr{E}_{1} \otimes \mathscr{E}_{2} \,\, , \,\,\, (E) \,\,).
\]
This operator, when evaluated at a fixed element $\phi \otimes \varphi \in \mathscr{E}_{1} \otimes \mathscr{E}_{2}$,
gives an operator in $\mathscr{L}((E), \, (E)^*)$, 
but this time, its value cannot be written as operator composition.
Again, by Thms. 3.9 and 4.8 of \cite{obataJFA} or their fermionic analogues,
which are applicable to general operators  $\Xi$ of the class (\ref{XimasslessProduct}),
possesses unique (here finite, because we include the scalar term $\Xi_{0,0}$)
Fock expansion
\[
\Xi =
\sum\limits_{\ell, m} \Xi_{\ell, m}({\kappa}_{\ell, m})
\,\,\,\,
\in \,\,\,\, \mathscr{L}((E) \otimes \mathscr{E}_{1} \otimes \mathscr{E}_{2}, \, (E)^*), \,\,\, \ell+ m \leq n+q,
\]
which in fact gives the rigorous version of the Wick theorem stated in
\cite{Bogoliubov_Shirkov}, \S 17, as ''The Wick's Theorem for Ordinary Products''\footnote{In the English Edition we read there: ''Wick Theorem for
Normal Products'', but ''The Wick's Theorem for Ordinary Products'' would be a better translation of the Russian original.}.
Generalization of this theorem to the ''products''
(\ref{Wick(x)Wick(y)}) containing a greater number of normally ordered Wick product factors (\ref{WickMonomialsInGeneralFreeFields}),
is obvious. This gives the mathematical justification for the Wick theorem stated
in \cite{Bogoliubov_Shirkov}, and shows that indeed the decomposition, or Fock expansion, can be
computed through the kernels $\kappa_{0,1}^{{}^{k}}, \kappa_{1,0}^{{}^{k}}$ defining the free field factors $\mathbb{A}_{{}_{k}}$,
because indeed they can be effectively computed through the operator
products (with massless factors replaced with the massive counterparts) and by the observation that in the zero-mass limit
of the massive pairings we indeed get the kernels of the massless fields. We thus have a generalization
of the Wick Theorem \ref{WickThmForMassiveFields} to the case of (\ref{Wick(x)Wick(y)}) in which some free fields
$\mathbb{A}_{{}_{k}}$, or even all, are massless. Although in this case, convergence of the contractions $\otimes_q$
is not automatic and does not follow from the very construction. Nonetheless, 
we can generalize the notion of contraction by the indicated above limit process, replacing the massless
kernels by the massive counterparts and passing to the limit. 
Namely, for the kernels $\kappa'_{\ell',m'}, \kappa''_{\ell'',m''}$ of the Fock
expansions, respectively, of the operators $\Xi', \Xi''$, we define the \emph{limit contraction} $\otimes|_{{}_{q}}$
\begin{equation}\label{kappaepsilonconvergence}
\kappa'_{\ell',m'} \otimes|_{{}_{q}} \, \kappa''_{\ell'',m''} \overset{\textrm{df}}{=}
\textrm{lim}_{\epsilon \rightarrow 0} \,\,\,
\kappa'_{\epsilon \,\, \ell',m''} \otimes_q \, \kappa''_{\epsilon \,\, \ell'',m''}
\end{equation}
which exists in
\[
\mathscr{L}\big(\mathscr{E}_{1}\otimes \mathscr{E}_{2} , E^{* \otimes (\ell'+\ell''-q + m'+m''-q)} \big)
\cong
\mathscr{L}\big(E^{\otimes (\ell'+\ell''-q + m'+m''-q)}, \, \mathscr{E}_{1}^{*}\otimes \mathscr{E}_{2}^{*} \big).
\]

We can thus generalize  the Wick Theorem \ref{WickThmForMassiveFields} over to massless fields $\mathbb{A}_{{}_{k}}$, by replacing the
ordinary contractions $\otimes_q$ with the \emph{limit contractions} $\otimes|_{{}_{q}}$ in it.
\begin{twr}
In case there are massless fields $\mathbb{A}_{{}_{k}}$ in each of the normal factors (\ref{WickMonomialsInGeneralFreeFields})
of the ``product'' (\ref{Wick(x)Wick(y)}) we obtain the following Fock expansion, or Wick theorem, 
for the generalized ``product'' operator $\Xi = \textrm{lim}_{{}_{\epsilon \rightarrow 0}} \, \Xi_\epsilon$.

The operators
\begin{eqnarray*}
\Xi'(\phi) = \sum \limits_{a} \int\limits_{\mathbb{R}^4}  
\boldsymbol{{:}} \mathbb{A}_{{}_{1}}^{a_1}(x) \ldots \mathbb{A}_{{}_{N}}^{a_N}(x) \boldsymbol{{:}} \, \phi^a(x) \,\,\,\, \ud^4x
\in \mathscr{L}((E),(E)^*),
\\
\Xi''(\varphi) = \sum \limits_{b} \int\limits_{\mathbb{R}^4}  
\boldsymbol{{:}} \mathbb{A}_{{}_{N+1}}^{b_{N+1}}(y) \ldots \mathbb{A}_{{}_{M}}^{b_M}(y)  \boldsymbol{{:}}  \,\,\,\, \varphi^b(y) \, \ud^4y
\in \mathscr{L}((E),(E)^*),
\\
\Xi(\phi \otimes \varphi) = 
\sum \limits_{a,b}
\int\limits_{\big[\mathbb{R}^4\big]^{\times \, 2}} 
\boldsymbol{{:}} \mathbb{A}_{{}_{1}}^{a_1}(x) \ldots \mathbb{A}_{{}_{N}}^{a_N}(x) \boldsymbol{{:}} \,\,
\boldsymbol{{:}} \mathbb{A}_{{}_{N+1}}^{b_{N+1}}(y) \ldots \mathbb{A}_{{}_{M}}^{b_M}(y)  \boldsymbol{{:}} \,\, \times
\\ \times \,\,\,\,\,\,\,
\phi^a \otimes \varphi^b(x,y) \, \ud^4x \ud^4y \,\,\, \in \mathscr{L}((E),(E)^*),
\end{eqnarray*} 
and moreover
\begin{eqnarray*}
\Xi' \in \mathscr{L}(\mathscr{E}_{1}, \mathscr{L}((E),(E)^*)\big),
\\
\Xi''\in \mathscr{L}(\mathscr{E}_{2}, \mathscr{L}((E),(E)^*)\big),
\\
\Xi\in \mathscr{L}(\mathscr{E}_{1} \otimes \mathscr{E}_{2}, \mathscr{L}((E),(E)^*)\big),
\end{eqnarray*}
and the following Wick theorem holds
\[
\Xi =
\sum_{\substack{\kappa'_{\ell',m'} \\ \kappa''_{\ell'',m''}}} \sum \limits_{0\leq q \leq \textrm{min} \, \{m',\ell''\}}  \,
\Xi_{\ell'+\ell''-q,m'+m''-q}\big( \kappa'_{\ell',m'} \otimes|_{{}_{q}} \, {\kappa''}_{\ell'',m''}\big),
\]
where the kernels $\kappa'_{\ell',m'}, \kappa''_{\ell'',m''}$ range, respectively, over the  kernels of finite Fock expansions 
of the operators $\Xi', \Xi''$. The limit $q$-contractions $\otimes|_{{}_{q}}$ are performed
upon the pairs of variables in which the first element of the contracted pair lies among the last $m'$ variables of the kernel 
$\kappa'_{\ell',m'}$ and the second variable
of the contracted pair lies among the first $l''$ variables of the kernel $\kappa''_{\ell'',m''}$, and to both variables
of the contracted pair correspond respectively annihilation and creation operator of one and the same free field. The limit contraction 
$\kappa_{l,m}= \kappa'_{\ell',m'} \otimes|_{{}_{q}} \, {\kappa''}_{\ell'',m''}$ should be symmetrized in Bose variables and antisymmetrized in Fermi variables
in order to achieve one-to-ne correspondence between kernels and corresponding integral kernel operators and this should be done separately
for the first variables $l$ and separately for the last $m$ variables. 
Here $\mathscr{E}_{1}$ and $\mathscr{E}_{2}$, are the  ordinary Schwartz spaces 
$\mathcal{S}(\mathbb{R}^4; \mathbb{C}^{d_1})$, $\mathcal{S}(\mathbb{R}^4; \mathbb{C}^{d_2})$,  depending on
the free fields in the Wick product. 
\label{WickThmForMasslessFields}
\end{twr}

\begin{rem*}
We should also emphasize here that although in principle the \emph{limit contraction} $\otimes|_{{}_{q}}$ 
is computed as the limit $\textrm{lim}_{\epsilon \rightarrow 0}$
of the contraction with the massless kernels $\kappa_{0,1}^{{}^{k}},\kappa_{1,0}^{{}^{k}}$ replaced by the massive counterparts
$\kappa_{\epsilon \,\, 0,1}^{{}^{k}},\kappa_{\epsilon \,\, 1,0}^{{}^{k}}$, the integral representing ordinary contraction with
$\kappa_{\epsilon \,\, 0,1}^{{}^{k}},\kappa_{\epsilon \,\, 1,0}^{{}^{k}}$ in it remains convergent even if
we put $\epsilon$ simply equal zero in it, and represents the \emph{limit contraction}.
This we have proved below
and, as we will see, the proof 
remains valid if the number of basic plane wave kernels is arbitrary large.
Thus, the \emph{limit contraction} can be expressed by ordinary formulas for contraction integrals
expressed through the original kernels $\kappa_{0,1}^{{}^{k}},\kappa_{1,0}^{{}^{k}}$ of the free fields $\mathbb{A}^{{}^{k}}$.
Therefore, in the sequel we will denote the limit contraction  $\otimes|_{{}_{q}}$ simply by $\otimes{{}_{q}}$.
\qed
\end{rem*}

{\bf Proof of theorem \ref{WickThmForMasslessFields}}.
For the proof it is sufficient to show the convergence (\ref{kappaepsilonconvergence}). Without loosing generality, we assume that all 
considered free field kernels are massless. We start with a proof of the remark, \emph{i.e.}
we show absolute convergence of the integral representing  $\otimes_{q}$-contractions, staring with scalar contractions 
$\kappa_{\epsilon \,\, 0,0}$, in which all spin-momentum variables are contracted, 
with $\epsilon$ simply put equal zero in it, giving a norm estimation for the absolute value of this integral:
\begin{multline}\label{masslesskappa01.masslesskappa10...q'-contraction...masslesskappa10.masslesskappa10}
\kappa^{{}^{q_1}}_{0,1} \dot{\otimes} \kappa^{{}^{q_2}}_{0,1} \ldots \dot{\otimes} \kappa^{{}^{q_q}}_{0,1} (\phi)
\, \otimes|_{{}_{q}}  \,\, \kappa^{{}^{q_1}}_{1,0} \dot{\otimes} \kappa^{{}^{q_2}}_{1,0} \ldots \dot{\otimes} \kappa^{{}^{q_q}}_{1,0} (\varphi) 
= \overset{q}{\underset{i=1}{\dot{\otimes}}} \kappa^{{}^{q_i}}_{0,1}(\phi)
\, \otimes|_{{}_{q}} \,\,
\overset{q}{\underset{i=1}{\dot{\otimes}}} \kappa^{{}^{q_i}}_{1,0}(\varphi)
\\
=\sum \limits_{s_1, \ldots, s_{q'}}
\int
u^{{}^{q_1}}_{{}_{s_{q_1}}}(\boldsymbol{\p}_{{}_{q_1}}) \ldots u^{{}^{q_q}}_{{}_{s_{q_q}}}(\boldsymbol{\p}_{q_q}) 
\widetilde{\phi}\big(-\boldsymbol{\p}_{{}_{q_1}}- \ldots - \boldsymbol{\p}_{{}_{q_q}}, \, -|\boldsymbol{\p}_{{}_{q_1}}|
\ldots - |\boldsymbol{\p}_{{}_{q_q}}| \big)
\,\, \times
\\
\times \,\,\,
v^{{}^{q_1}}_{{}_{s_{q_1}}}(\boldsymbol{\p}_{{}_{q_1}}) \ldots v^{{}^{q_q}}_{{}_{s_{q_q}}}(\boldsymbol{\p}_{{}_{q_q}}) 
\widetilde{\varphi}\big(\boldsymbol{\p}_{{}_{q_1}}+ \ldots + \boldsymbol{\p}_{{}_{q_q}}, |\boldsymbol{\p}_{{}_{q_1}}|
\ldots +|\boldsymbol{\p}_{{}_{q_q}}| \big)
\,\, \ud^3 \boldsymbol{\p}_{{}_{q_1}} \ldots  \ud^3 \boldsymbol{\p}_{{}_{q_q}}.
\end{multline}
Indeed, because for the Schwartz function $\phi \in \mathscr{E}_1$
its Fourier transform $\widetilde{\phi}$ also is a Schwatrz function, ranging over a bounded set, whenever so does $\phi$, 
we can use this fact to estimate 
(\ref{masslesskappa01.masslesskappa10...q'-contraction...masslesskappa10.masslesskappa10}). Namely the function
\[
V_{(4)}(\boldsymbol{\p}, p_0) = 1+ p_{0}^2 + |\boldsymbol{\p}|^2  
\]
is a multiplier of  the Schwartz space  $\widetilde{\mathscr{E}_1}$. 
For each natural $n$, and $\phi$ ranging over a bounded set $B$ in the Schwartz space, or $\widetilde{\phi}$
ranging over  a bounded set $\widetilde{B}$ in the Schwartz space $\widetilde{\mathscr{E}_1}$,
the set of functions $V_{(4)}^{n} \cdot \widetilde{\phi}$
ranges over a bounded set in the Schwartz space $\widetilde{\mathscr{E}_1}$, depending only on $B$ and $n$, 
on which the countable set of norms
$| \cdot |_1$, $|\cdot |_2$, $\ldots$, 
defining  $\mathscr{E}_1$ are finite. For any fixed number $q$ of functions $u^{{}^{q_i}},v^{{}^{q_i}}$ determining kernels of free (say massless) 
fields there exists a natural $n$ such that
\begin{equation}\label{V(4)}
V_{(4)}^{-n}(\boldsymbol{\p}_{{}_{q_1}} + \ldots + \boldsymbol{\p}_{{}_{q_q}}, \, |\boldsymbol{\p}_{{}_{q_1}}|+ \ldots 
+ |\boldsymbol{\p}_{{}_{q_q}}|) u^{{}^{q_1}}(\boldsymbol{\p}_{{}_{q_1}})
v^{{}^{q_1}}(\boldsymbol{\p}_{{}_{q_1}}) \ldots u^{{}^{q_q}}(\boldsymbol{\p}_{{}_{q_q}})v^{{}^{q_q}}(\boldsymbol{\p}_{{}_{q_q}}) 
\end{equation}
is absolutely integrable regarded as the function of $\boldsymbol{\p}_{{}_{q_1}}, \ldots, \boldsymbol{\p}_{{}_{q_q}}$. Denoting the $L_1$-norm
of the function (\ref{V(4)}) by $c$, we have the following inequality for the absolute value of (\ref{masslesskappa01.masslesskappa10...q'-contraction...masslesskappa10.masslesskappa10})
\begin{equation}\label{AbsoluteConvergenceKappa00}
\Big|\overset{q}{\underset{i=1}{\dot{\otimes}}} \kappa^{{}^{q_i}}_{0,1}(\phi)
\, \otimes|_{{}_{q}} \,\,
\overset{q}{\underset{i=1}{\dot{\otimes}}} \kappa^{{}^{q_i}}_{1,0}(\varphi) \Big| \,\, \leq 
\,\,\, c \, \underset{p\in \mathbb{R}^4}{\textrm{sup}} \big|\widetilde{\phi}(p) \big| \,\,\,
\,\,\ \underset{p\in \mathbb{R}^4}{\textrm{sup}} \big|V_{(4)}^{n}. \widetilde{\varphi}(p) \big|
\leq
c \, \big|\widetilde{\phi}\big|_{{}_{k}} \,\,  \big|\widetilde{\varphi}\big|_{{}_{n}}
=c \, \big|\phi\big|_{{}_{k}} \, \big|\varphi\big|_{{}_{n}},
\end{equation}
for all sufficiently large natural $k,n$, with the last norms $|\cdot|_{{}_{k}} = |\widetilde{H_{{}_{(4)}}}^k \cdot|_{{}_{L^2}}$, 
$|\cdot|_{{}_{k}} = |H_{{}_{(4)}}^k \cdot|_{{}_{L^2}}$, on the Fourier image $\widetilde{\mathscr{E}_i}$, $i=1,2$, of the Schwartz space $\mathscr{E}_i$ and on 
the Schwartz space $\mathscr{E}_i$, coming from the system of Hilbertian norms on $\mathscr{E}_i$.
Note that the estimation (\ref{AbsoluteConvergenceKappa00}) is valid for the integral 
(\ref{masslesskappa01.masslesskappa10...q'-contraction...masslesskappa10.masslesskappa10}) in which the integrand is replaced
by its absolute value, so that (\ref{masslesskappa01.masslesskappa10...q'-contraction...masslesskappa10.masslesskappa10})
is absolutely convergent.

Let us pass now to the general $\otimes_{q}$-contraction. Let, for simplicity of notation,
the last $q$ spin-momentum variables $s_{{}_{q_1}},\boldsymbol{\p}_{{}_{q_1}}, \ldots, s_{{}_{q_q}},\boldsymbol{\p}_{{}_{q_q}}$ 
in the first kernel $\kappa'_{l, m+q}(\phi) = \overset{l}{\underset{i=1}{\dot{\otimes}}} \kappa^{{}^{l_i}}_{1,0} 
\overset{m}{\underset{i=1}{\dot{\otimes}}} \kappa^{{}^{m_i}}_{0,1}
\overset{q}{\underset{i=1}{\dot{\otimes}}} \kappa^{{}^{q_i}}_{0,1}(\phi)$  be contracted with the
first $q$ spin-momentum variables $s_{{}_{q_1}},\boldsymbol{\p}_{{}_{q_1}}, \ldots, s_{{}_{q_q}},\boldsymbol{\p}_{{}_{q_q}}$  
in the second kernel $\kappa''_{q+\ell, \mathpzc{m}}(\varphi) =\overset{q}{\underset{i=1}{\dot{\otimes}}} \kappa^{{}^{q_i}}_{1,0}
\overset{\ell}{\underset{i=1}{\dot{\otimes}}} \kappa^{{}^{\ell_i}}_{1,0}
\overset{\mathpzc{m}}{\underset{i=1}{\dot{\otimes}}} \kappa^{{}^{\mathpzc{m}_i}}_{0,1}(\varphi)$:
\begin{multline}\label{(l+m+q)q(q+ell+em)Contraction}
\overset{l}{\underset{i=1}{\dot{\otimes}}} \kappa^{{}^{l_i}}_{1,0} 
\overset{m}{\underset{i=1}{\dot{\otimes}}} \kappa^{{}^{m_i}}_{0,1}
\overset{q}{\underset{i=1}{\dot{\otimes}}} \kappa^{{}^{q_i}}_{0,1}(\phi)
\, \otimes|_{{}_{q}} \,\,
\overset{q}{\underset{i=1}{\dot{\otimes}}} \kappa^{{}^{q_i}}_{1,0}
\overset{\ell}{\underset{i=1}{\dot{\otimes}}} \kappa^{{}^{\ell_i}}_{1,0}
\overset{\mathpzc{m}}{\underset{i=1}{\dot{\otimes}}} \kappa^{{}^{\mathpzc{m}_i}}_{0,1}(\varphi)
\\
=\sum\limits_{s_{{}_{q_1}}, \ldots, s_{{}_{q_q}}} \bigintsss \Bigg[
\\
\prod\limits_{i=1}^{l} v^{{}^{l_i}}_{{}_{s_{{}_{l_i}}}}(\boldsymbol{\p}_{{}_{l_i}})
\prod\limits_{i=1}^{m} u^{{}^{m_i}}_{{}_{s_{{}_{m_i}}}}(\boldsymbol{\p}_{{}_{m_i}})
\prod\limits_{i=1}^{q} u^{{}^{q_i}}_{{}_{s_{{}_{q_i}}}}(\boldsymbol{\p}_{{}_{q_i}})
\widetilde{\phi}\left(\sum\limits_{i=1}^{l}\boldsymbol{\p}_{{}_{l_i}}- \sum\limits_{i=1}^{m}\boldsymbol{\p}_{{}_{m_i}}
- \sum\limits_{i=1}^{q} \boldsymbol{\p}_{{}_{q_i}}, \sum\limits_{i=1}^{l}|\boldsymbol{\p}_{{}_{l_i}}|- \sum\limits_{i=1}^{m}|\boldsymbol{\p}_{{}_{m_i}}|
- \sum\limits_{i=1}^{q} |\boldsymbol{\p}_{{}_{q_i}}|\right) \,\, \times
\\
\times \,\, 
\prod\limits_{i=1}^{q} v^{{}^{q_i}}_{{}_{s_{{}_{q_i}}}}(\boldsymbol{\p}_{{}_{q_i}})
\prod\limits_{i=1}^{\ell} v^{{}^{\ell_i}}_{{}_{s_{{}_{m_i}}}}(\boldsymbol{\p}_{{}_{\ell_i}})
\prod\limits_{q=1}^{\mathpzc{m}} u^{{}^{\mathpzc{m}_i}}_{{}_{s_{{}_{\mathpzc{m}_i}}}}(\boldsymbol{\p}_{{}_{\mathpzc{m}_i}})
\widetilde{\varphi}\left(\sum\limits_{i=1}^{q}\boldsymbol{\p}_{{}_{q_i}} + \sum\limits_{i=1}^{\ell}\boldsymbol{\p}_{{}_{\ell_i}}
-\sum\limits_{i=1}^{\mathpzc{m}} \boldsymbol{\p}_{{}_{\mathpzc{m}_i}}, \sum\limits_{i}^{q} |\boldsymbol{\p}_{{}_{q_i}}|+ 
\sum\limits_{i}^{\ell}|\boldsymbol{\p}_{{}_{\ell_i}}|
-\sum\limits_{i}^{\mathpzc{m}} |\boldsymbol{\p}_{{}_{\mathpzc{m}_i}}|\right) 
\\
\Bigg] \prod\limits_{i=1}^{q} \ud^3 \boldsymbol{\p}_{{}_{q_i}} 
\end{multline}
being the generalized function in $\overset{l}{\underset{i=1}{\otimes}} E_{{}_{l_i}}^*
\overset{m}{\underset{i=1}{\otimes}} E_{{}_{m_i}}^*
\overset{\ell}{\underset{i=1}{\otimes}} E_{{}_{\ell_i}}^*
\overset{\mathpzc{m}}{\underset{i=1}{\otimes}} E_{{}_{\mathpzc{m}_i}}^*$ of the non contracted spin-momenta variables
\begin{equation}\label{NonContractedVariables}
(\ldots, s_{{}_{l_i}},\boldsymbol{\p}_{{}_{l_i}}, \ldots, s_{{}_{m_i}},\boldsymbol{\p}_{{}_{m_i}}, \ldots, 
s_{{}_{\ell_i}},\boldsymbol{\p}_{{}_{\ell_i}}, \ldots, 
s_{{}_{\mathpzc{m}_i}},\boldsymbol{\p}_{{}_{\mathpzc{m}_i}}, \ldots),
\end{equation}
and with all free field kernels being massless. Replacement of $\widetilde{\phi}, \widetilde{\varphi}$ with their translations
$T_{{}_{w}}\widetilde{\phi} = \widetilde{e_w\phi}, T_{{}_{\upsilon}}\widetilde{\varphi} = \widetilde{e_{\upsilon}\varphi}$, 
with $w,\upsilon, e_{w}, e_{\upsilon}$ given below, in the proof of (\ref{AbsoluteConvergenceKappa00}),
gives a proof of the absolute convergence of the integral (\ref{(l+m+q)q(q+ell+em)Contraction}), majorized at infinity by a fixed 
polynomial in the non-contracted variables and depending only on the norms of $\phi,\varphi$, compare our analysis 
of the limit (\ref{kappaepsilonconvergence}) given below. We pass now to the analysis of the limit (\ref{kappaepsilonconvergence}).

Replacing $|\boldsymbol{\p}_i|$ by $|\boldsymbol{\p}_i|_{{}_{\epsilon}} = \sqrt{|\boldsymbol{\p}_i|^2+\epsilon^2}$ in
(\ref{(l+m+q)q(q+ell+em)Contraction}) and  subtracting (\ref{(l+m+q)q(q+ell+em)Contraction})  we obtain the following estimation
for the absolute value of such obtained difference 
\begin{multline}\label{q-q'-contractionEstimation}
\Big|
\overset{l}{\underset{i=1}{\dot{\otimes}}} \kappa^{{}^{l_i}}_{\epsilon \, 1,0} 
\overset{m}{\underset{i=1}{\dot{\otimes}}} \kappa^{{}^{m_i}}_{\epsilon \, 0,1}
\overset{q}{\underset{i=1}{\dot{\otimes}}} \kappa^{{}^{q_i}}_{\epsilon \, 0,1}(\phi)
\, \otimes|_{{}_{q}} \,\,
\overset{q}{\underset{i=1}{\dot{\otimes}}} \kappa^{{}^{q_i}}_{\epsilon \, 1,0}
\overset{\ell}{\underset{i=1}{\dot{\otimes}}} \kappa^{{}^{\ell_i}}_{\epsilon \, 1,0}
\overset{\mathpzc{m}}{\underset{i=1}{\dot{\otimes}}} \kappa^{{}^{\mathpzc{m}_i}}_{\epsilon \, 0,1}(\varphi)
\\
-
\overset{l}{\underset{i=1}{\dot{\otimes}}} \kappa^{{}^{l_i}}_{1,0} 
\overset{m}{\underset{i=1}{\dot{\otimes}}} \kappa^{{}^{m_i}}_{0,1}
\overset{q}{\underset{i=1}{\dot{\otimes}}} \kappa^{{}^{q_i}}_{0,1}(\phi)
\, \otimes|_{{}_{q}} 
\overset{q}{\underset{i=1}{\dot{\otimes}}} \kappa^{{}^{q_i}}_{1,0}
\overset{\ell}{\underset{i=1}{\dot{\otimes}}} \kappa^{{}^{\ell_i}}_{1,0}
\overset{\mathpzc{m}}{\underset{i=1}{\dot{\otimes}}} \kappa^{{}^{\mathpzc{m}_i}}_{0,1}(\varphi)
\Big|
\\
\leq
\epsilon \, c \,
\left(\sum\limits_{0=|\gamma|\leq |\alpha|\leq k} |w^\gamma|{\alpha \choose \gamma}P_{{}_{l,m}} \cdot \, \big|\phi\big|_{{}_{k}}\right)
\left(\sum\limits_{0=|\gamma|\leq |\alpha| \leq n}|\upsilon^\gamma|{\alpha \choose \gamma}P_{{}_{\ell,\mathpzc{m}}} \cdot \, \big|\varphi\big|_{{}_{n}}\right),
\end{multline}
for all $\phi \in \mathscr{E}_1$, $\varphi \in \mathscr{E}_2$ and with $k,n \in \mathbb{N}$ and finite $c>0$, independent of $\phi,\varphi$. 
Here
\begin{gather*}
w = 
\Bigg( \overbrace{\sum\limits_{i=1}^{l}\boldsymbol{\p}_{{}_{l_i}}- \sum\limits_{i=1}^{m}\boldsymbol{\p}_{{}_{m_i}}}^{\boldsymbol{w}} \,\,\,\, , 
\,\,\,\,\,\,\,\,\,\,\,
\overbrace{\sum\limits_{i=1}^{l}|\boldsymbol{\p}_{{}_{l_i}}|- \sum\limits_{i=1}^{m}|\boldsymbol{\p}_{{}_{m_i}}|}^{w_0}  \Bigg) 
= (\boldsymbol{w},w_0) = (w_1,w_2,w_3,w_0),
\\
\upsilon = 
\Big( \underbrace{\sum\limits_{i=1}^{\ell}\boldsymbol{\p}_{{}_{\ell_i}}
-\sum\limits_{i=1}^{\mathpzc{m}} \boldsymbol{\p}_{{}_{\mathpzc{m}_i}}}_{\boldsymbol{\upsilon}} \,\,\,\, , 
\,\,\,\,\,\,\,\,\,\,\,
\underbrace{\sum\limits_{i}^{\ell}|\boldsymbol{\p}_{{}_{\ell_i}}|
-\sum\limits_{i=1}^{\mathpzc{m}} |\boldsymbol{\p}_{{}_{\mathpzc{m}_i}}|}_{\upsilon_0}  \Big)
= (\boldsymbol{\upsilon},\upsilon_0) = (\upsilon_1,\upsilon_2,\upsilon_3,\upsilon_0),
\end{gather*}
and $\alpha,\beta, \gamma$, are $4$-component multiindices, for example $\gamma= (\gamma_0,\gamma_1, \gamma_2, \gamma_3)$ and
\[
w^{\gamma} = {w}_{0}^{\gamma_0}{w}_{1}^{\gamma_1} {w}_{2}^{\gamma_2} {w}_{3}^{\gamma_3}, 
\,\,\,\,\,
\upsilon^{\gamma} = {\upsilon}_{0}^{\gamma_0}{\upsilon}_{1}^{\gamma_1} {\upsilon}_{2}^{\gamma_2} {\upsilon}_{3}^{\gamma_3},
\,\,\,\,\,
{\alpha \choose \gamma} = {\alpha_0 \choose \gamma_0} {\alpha_1 \choose \gamma_1}{\alpha_2 \choose \gamma_2}{\alpha_3 \choose \gamma_3}.   
\]
\[
P_{{}_{l,m}} = \prod\limits_{i=1}^{l} \left| v^{{}^{l_i}}_{{}_{s_{{}_{l_i}}}}(\boldsymbol{\p}_{{}_{l_i}})\right| \,\,
\prod\limits_{i=1}^{m} \left| u^{{}^{m_i}}_{{}_{s_{{}_{m_i}}}}(\boldsymbol{\p}_{{}_{m_i}}) \right|,
\,\,\,
P_{{}_{\ell,\mathpzc{m}}} = \prod\limits_{i=1}^{\ell} \left| v^{{}^{\ell_i}}_{{}_{s_{{}_{\ell_i}}}}(\boldsymbol{\p}_{{}_{\ell_i}}) \right| \,\,
\prod\limits_{i=1}^{\mathpzc{m}} \left| u^{{}^{\mathpzc{m}_i}}_{{}_{s_{{}_{\mathpzc{m}_i}}}}(\boldsymbol{\p}_{{}_{\mathpzc{m}_i}}) \right|.
\]
In case of scalar $\otimes_q$-contraction, \emph{i.e.} when $|l|= |m| = |\ell| =|\mathpzc{m}|=0$ with all spin-momentum variables contracted,
$P_{{}_{l,m}},P_{{}_{\ell,\mathpzc{m}}}, w^\gamma, \upsilon^\gamma$ for $\gamma =0$ degenerate to the constant equal $1$,
and  $w^\gamma, \upsilon^\gamma =0$ for $\gamma >0$  in (\ref{q-q'-contractionEstimation}).

Indeed, let $e_{w}(x) = e^{iw\cdot x}$. 
Then $\widetilde{e_{w}\phi}(\boldsymbol{\p}, p_0 \big) = \widetilde{\phi}\big(\boldsymbol{\p}+ \boldsymbol{w}, p_0 + w_0 \big)$.
We then use the functions $e_{w}\phi$, $e_{\upsilon}\varphi$, expressing the left-hand-side of 
(\ref{q-q'-contractionEstimation}) as follows 
\begin{multline*}
\Bigg| \sum\limits_{\ldots} \int \cdots u^{{}^{q_i}}_{{}_{\epsilon}} \cdots \widetilde{e_{w}\phi}(-\ldots, -\ldots + \Delta) 
\cdots v^{{}^{q_i}}_{{}_{\epsilon}} \cdots \widetilde{e_{\upsilon}\varphi}(\ldots,\ldots + \Delta)
\\
-  \sum\limits_{\ldots} \int \cdots u^{{}^{q_i}} \cdots \widetilde{e_{w}\phi}(\ldots,\ldots) 
\cdots v^{{}^{q_i}} \cdots \widetilde{e_{\upsilon}\varphi}(\ldots,\ldots)\Bigg|
\end{multline*}
\begin{multline}\label{MeanValue1}
\leq
\Bigg| 
\sum\limits_{\ldots} \int \cdots u^{{}^{q_i}}_{{}_{\epsilon}} \cdots \widetilde{e_{w}\phi}(-\ldots,-\ldots + \Delta) 
\cdots v^{{}^{q_i}}_{{}_{\epsilon}} \cdots \widetilde{e_{\upsilon}\varphi}(\ldots,\ldots + \Delta)
\\
-  \sum\limits_{\ldots} \int \cdots u^{{}^{i}} \cdots \widetilde{e_{w}\phi}(-\ldots,-\ldots+\Delta) 
\cdots v^{{}^{q_i}} \cdots \widetilde{e_{\upsilon}\varphi}(\ldots,\ldots+\Delta)
\Bigg|
\end{multline}
\begin{multline}\label{MeanValue2}
\,\,\,\, +
\Bigg| \sum\limits_{\ldots} \int \cdots u^{{}^{q_i}} \cdots \widetilde{e_{w}\phi}(-\ldots,-\ldots + \Delta) 
\cdots v^{{}^{q_i}} \cdots \widetilde{e_{\upsilon}\varphi}(\ldots,\ldots + \Delta)
\\
-  \sum\limits_{\ldots} \int \cdots u^{{}^{q_i}} \cdots \widetilde{e_{w}\phi}(\ldots,\ldots) 
\cdots v^{{}^{q_i}} \cdots \widetilde{e_{\upsilon}\varphi}(\ldots,\ldots)\Bigg|,
\end{multline}  
where the first dots $(\ldots,\ldots)$ in the arguments of the functions $\widetilde{e_{w}\phi},\widetilde{e_{\upsilon}\varphi}$ 
denote the sum of all the spatial components of the contracted momenta, \emph{i.e.} integrated, and, the second dots
in $(\ldots,\ldots)$ representing the sum of zero components of contracted momenta;  
$u^{{}^{i}}_{{}_{\epsilon}}, v^{{}^{i}}_{{}_{\epsilon}}$ denote the multipliers 
$u^{{}^{i}}, v^{{}^{i}}$  defining the kernels of the massless fields
(\ref{FreeFields}) in which $|\boldsymbol{\p}_i|$ are replaced with $|\boldsymbol{\p}_i|_{{}_{\epsilon}}$. Here
\[
\Delta = \sum\limits_{i=1}^{l}\Delta_{{}_{l_i}}-\sum\limits_{i=1}^{m}\Delta_{{}_{m_i}}-\sum\limits_{i=1}^{q}\Delta_{{}_{q_i}},
\Delta = \sum\limits_{i=1}^{q}\Delta_{{}_{q_i}}+\sum\limits_{i=1}^{\ell}\Delta_{{}_{\ell_i}}-\sum\limits_{i=1}^{\mathpzc{m}}\Delta_{{}_{\mathpzc{m}_i}},
\,\, \textrm{resp. in} \,\, \widetilde{\phi}(-\ldots,-\ldots+\Delta), \widetilde{\varphi}(\ldots,\ldots+\Delta),
\]
with
\[
0 \leq |\boldsymbol{\p}_i|_{{}_{\epsilon}} - |\boldsymbol{\p}_i| 
= \Delta_i = {\textstyle\frac{\epsilon^2}{\sqrt{|\boldsymbol{\p}_i|^2 +\epsilon^2} + |\boldsymbol{\p}_i|}} \leq \epsilon,
\]
denotes the total increment of the zero component. Next we use the mean value theorem 
\begin{gather*}
\widetilde{e_{w}\phi}(-\ldots,-\ldots + \Delta) = \widetilde{e_{w}\phi}(-\ldots,-\ldots) 
+ \partial_{{}_{p_0}}\widetilde{e_{w}\phi}(-\ldots,-\ldots + \lambda_1 \Delta)\Delta,
\\
\widetilde{e_{\upsilon}\varphi}(\ldots,\ldots + \Delta) = \widetilde{e_{w}\phi}(\ldots,\ldots) 
+ \partial_{{}_{p_0}}\widetilde{e_{w}\phi}(\ldots,\ldots + \lambda_2 \Delta)\Delta, 
\end{gather*}
with smooth functions $\lambda_k$ of $\boldsymbol{\p}_i$, such that $0 \leq \lambda_k \leq 1$
(and divide the range $\mathbb{R}^3$ of integration in each variable $\boldsymbol{\p}_{{}_{q_i}}$, into two subsets 
$|\boldsymbol{\p}_{{}_{q_i}}| \leq 1$, $|\boldsymbol{\p}_{{}_{q_i}}| > 1$) in (\ref{MeanValue1}), (\ref{MeanValue2}).
Then, we insert under the integration the multiplier $V_{(4)}^{n}(\ldots,\ldots)V_{(4)}^{-n}(\ldots,\ldots)$ given by (\ref{V(4)}) 
and proceeding like in (\ref{AbsoluteConvergenceKappa00}) we obtain
\begin{multline*}
\big| \kappa'_{\epsilon \, l, m+q}(\phi)  \, \otimes|_{{}_{q}}  \,\, \kappa''_{\epsilon \, q+\ell, \mathpzc{m}}(\varphi)
- \kappa'_{l, m+q}(\phi)  \, \otimes|_{{}_{q}}  \,\, \kappa''_{q+\ell, \mathpzc{m}}(\varphi) \big|
\\
\leq c \, P_{{}_{l,m}} \,\underset{p\in \mathbb{R}^4}{\textrm{sup}}\big|\widetilde{e_{w}\phi}(p) \big| 
\,\,P_{{}_{\ell,\mathpzc{m}}} \, \underset{p\in \mathbb{R}^4}{\textrm{sup}} \big|V_{(4)}^{n}\cdot \widetilde{e_{\upsilon}\varphi}(p) \big|
\leq
c \, P_{{}_{l,m}} \cdot \big|\widetilde{e_{w}\phi}\big|_{{}_{k'}} \,P_{{}_{\ell,\mathpzc{m}}} \cdot  \big|\widetilde{e_{\upsilon}\varphi}\big|_{{}_{n'}}
=c \, P_{{}_{l,m}} \cdot \big|e_{w}\phi\big|_{{}_{k'}} \, P_{{}_{\ell,\mathpzc{m}}} \cdot \big|e_{\upsilon}\varphi\big|_{{}_{n'}}.
\end{multline*}
Using the equivalent system of norms 
\[
| \phi |_{{}_{k}} = \underset{|\alpha|,|\beta| \leq k, x\in\mathbb{R}^4}{\textrm{sup}} \, |x^\beta \partial^\alpha \phi(x)|,
\,\,\,\,\,\,
| \varphi |_{{}_{n}} = \underset{|\alpha|,|\beta| \leq n, x\in\mathbb{R}^4}{\textrm{sup}} \, |x^\beta \partial^\alpha \varphi(x)|,
\]
in this estimation, we obtain (\ref{q-q'-contractionEstimation}). 

From (\ref{q-q'-contractionEstimation}) it follows that the absolute value of
\begin{equation}\label{F(p^q',p^(q'+1),...)/|phi||varphi|epsilon}
{\textstyle\frac{1}{\big|\phi \big|_{{}_{k}} \big|\varphi \big|_{{}_{n}} \epsilon }}
\Big[
\kappa'_{\epsilon \, l, m+q}(\phi)  \, \otimes|_{{}_{q}}  \,\, \kappa''_{\epsilon \, q+\ell, \mathpzc{m}}(\varphi)
- \kappa'_{l, m+q}(\phi)  \, \otimes|_{{}_{q}}  \,\, \kappa''_{q+\ell, \mathpzc{m}}(\varphi)
\Big],
\end{equation}
regarded as the function of the non-contracted variables (\ref{NonContractedVariables}),
is bounded by the absolute value of a fixed multiplier (lemma \ref{FreeFieldMultipliers}) of the nuclear space (\ref{E^(q'+1)x ...xE^q}), 
independent of non-zero $\phi,\varphi$, $\epsilon$, 
bounded at infinity by a fixed polynomial in these variables. Therefore,  (\ref{F(p^q',p^(q'+1),...)/|phi||varphi|epsilon})
is a multiplier of the nuclear space
\begin{equation}\label{E^(q'+1)x ...xE^q}
\overset{l}{\underset{i=1}{\otimes}} E_{{}_{l_i}}
\overset{m}{\underset{i=1}{\otimes}} E_{{}_{m_i}}
\overset{\ell}{\underset{i=1}{\otimes}} E_{{}_{\ell_i}}
\overset{\mathpzc{m}}{\underset{i=1}{\otimes}} E_{{}_{\mathpzc{m}_i}},
\end{equation}
so that there exists a norm $|\cdot|_i$, among the norms $|\cdot|_1, |\cdot|_2, \ldots$, 
defining (\ref{E^(q'+1)x ...xE^q}), and a finite constant $c$, both independent of $\phi, \varphi, \xi$,  
such that 
\begin{equation}\label{BasicInequalityForWickEpsilon->0}
\Big|\Big\langle
\Big[
\kappa'_{\epsilon \, l, m+q}(\phi)  \, \otimes|_{{}_{q}}  \,\, \kappa''_{\epsilon \, q+\ell, \mathpzc{m}}(\varphi)
- \kappa'_{l, m+q}(\phi)  \, \otimes|_{{}_{q}}  \,\, \kappa''_{q+\ell, \mathpzc{m}}(\varphi)
\Big],
\, \xi
\Big\rangle
\Big|
\leq \epsilon \, c \, \big|\phi \big|_{{}_{k}} \big|\varphi \big|_{{}_{n}}  |\xi|_i,
\end{equation}
for all $\phi \in \mathscr{E}_1, \varphi \in \mathscr{E}_2$ and $\xi$ in (\ref{E^(q'+1)x ...xE^q}).
The inequality (\ref{BasicInequalityForWickEpsilon->0}) means that
\[
\kappa'_{\epsilon \,\, l, m+q}  \, \otimes|_{{}_{q}}  \,\, \kappa''_{\epsilon \,\, q+\ell, \mathpzc{m}}
\overset{\epsilon \rightarrow 0}{\longrightarrow}
\kappa'_{l, m+q} \, \otimes|_{{}_{q}}  \,\, \kappa''_{q+\ell, \mathpzc{m}}
\]
in
\[
\mathscr{L}\left(\overset{l}{\underset{i=1}{\otimes}} E_{{}_{l_i}}
\overset{m}{\underset{i=1}{\otimes}} E_{{}_{m_i}}
\overset{\ell}{\underset{i=1}{\otimes}} E_{{}_{\ell_i}}
\overset{\mathpzc{m}}{\underset{i=1}{\otimes}} E_{{}_{\mathpzc{m}_i}}, \mathscr{E}_{1}^{*}\otimes \mathscr{E}_{2}^{*}\right),
\]
which proves convergence (\ref{kappaepsilonconvergence}).
\qed

Of course the analogue Wick theorem decomposition holds for the (tensor) product (\ref{Wick(x)Wick(y)}) of more than just two normally ordered
Wick product factors $W_1(x)$ and $W_2(x)$ of the form (\ref{WickMonomialsInGeneralFreeFields}). Repeating the analogous decomposition for more
than just two factors, we easily see that the analogous decomposition holds for the product  
\begin{equation}\label{Wicki(xi)...Wick(xn)NormalKernelDecomposition}
t(x_1, \ldots, x_n) \, {:}W_1(x_1)W_2(x_2) \ldots W_n(x_n){:} \,\,\,\, d(y_1, \ldots, y_k) \, {:}P_1(y_1)W_2(y_2) \ldots P_k(y_k){:}
\end{equation}
where each $W_i(x_i), P_j(y_j)$ are Wick products of free fields (or their derivatives)  at $x_i$ or $y_j$,
and $t(x_1, \ldots, x_n)$, $d(y_1, \ldots, y_k) $, are the translationally invariant scalar $\otimes_q$-contractions (products of pairings) 
of massive or massless free fields, 
and with the kernels $\kappa_{l,m}$ of the Wick decomposition of the operator (\ref{Wicki(xi)...Wick(xn)NormalKernelDecomposition}) equal 
to the contractions
\[
\kappa_{{}_{l,m}}(\phi\otimes \varphi) = \sum\limits_{\kappa'_{{}_{l',m'}},\kappa''_{{}_{l'',m''}}, k}
\kappa'_{{}_{l',m'}}(\phi) \otimes_{k} \kappa''_{{}_{l'',m''}}(\varphi)
\]
\[
\phi \in \mathscr{E}^{\otimes \, n}, \,\, \varphi \in \mathscr{E}^{\otimes \, k}
\]
where in this sum $\kappa'_{{}_{l',m'}},\kappa''_{{}_{l'',m''}}$ range over the kernels respectively
of the operators
\begin{gather*}
t(x_1, \ldots, x_n) \, {:}W_1(x_1)W_2(x_2) \ldots W_n(x_n){:}  
\\
 \textrm{and} 
\\
d(y_1, \ldots, y_k) \, {:}P_1(y_1)W_2(y_2) \ldots P_k(y_k){:}
\end{gather*}
and
\[
l'+l''-k=l, \,\,\, m'+m''-k=m, 
\]
and where the contractios $\kappa'_{{}_{l',m'}}(\phi) \otimes_{k} \kappa''_{{}_{l'',m''}}(\varphi)$ 
are performed upon all $k$ pairs of spin-momenta variables which can be contracted, in which the first variable in the pair
lies among the last $m'$ variables in $\kappa'_{{}_{l',m'}}(\phi)$ and corresponds to an annihilation operator variable 
and the second one lies upon the first $l''$ variables in  $\kappa''_{{}_{l'',m''}}(\varphi)$ and corresponds to the 
creation operator variable of \emph{the same free field}. 
All these contractions are given by absolutely convergent sums/integrals with respect 
to the contracted variables.  After the contraction, 
the kernels should be symmetrized in Boson spin-momentum
variables and antisymmetrized in the Fermion spin-momentum variables
in order to keep one-to-one correspondence between the kernels and operators.

\begin{twr}
If all free fields in the Wick products $W_i(x_i)$ are massive, then the  Wick decomposition of the operator 
product (\ref{Wicki(xi)...Wick(xn)NormalKernelDecomposition}) is equal to finite sum of integral kernel operators 
which belong to
\[
\mathscr{L}\big(\mathscr{E}^{\otimes \,(n+k)}, \,\mathscr{L}((E),(E))\big)
\]
also in case the products of pairings $t(x_1, \ldots, x_n)$, $d(y_1, \ldots, y_k) $,
include the pairings of the massless fields (or their derivatives). 

Namely, in case $t(x_1, \ldots, x_n)$, $d(y_1, \ldots, y_k) $, are products of pairings of possibly massless free fields,
\emph{i.e.} scalar $\otimes_q$-contractions of $\dot{\otimes}$-products of free field kernels, 
and the free fields in $W_i(x)$, $P_j(y)$ are massive, the factors
\begin{gather*}
t(x_1, \ldots, x_n) \, {:}W_1(x_1)W_2(x_2) \ldots W_n(x_n){:} 
\\
\textrm{and} 
\\ 
d(y_1, \ldots, y_k) \, {:}P_1(y_1)P_2(y_2) \ldots P_k(y_k){:}
\end{gather*}
in (\ref{Wicki(xi)...Wick(xn)NormalKernelDecomposition}) 
represent generalized operators which belong, respectively, to 
\[
\mathscr{L}\big(\mathscr{E}^{\otimes \,n}, \,\mathscr{L}((E),(E))\big)
\,\,\, \textrm{and}
\,\,\,
\mathscr{L}\big(\mathscr{E}^{\otimes \,k}, \,\mathscr{L}((E),(E))\big),
\]
so that the product 
\[
t(x_1, \ldots, x_n) \, {:}W_1(x_1)W_2(x_2) \ldots W_n(x_n){:}  \,\,\,
d(y_1, \ldots, y_k) \, {:}P_1(y_1)P_2(y_2) \ldots P_k(y_n){:}
\]  
belongs to
\[
\mathscr{L}\big(\mathscr{E}^{\otimes \,(n+k)}, \,\mathscr{L}((E),(E))\big), 
\]
and for which the above Wick decomposition theorem likewise holds true. 
Here $\mathscr{E} = \mathcal{S}(\mathbb{R}^4; \mathbb{C}^d)$. 
The same holds true if we replace $t,d$ with any translationally invariant tempered distributions, in particular
if we replace $t,d$ with $\textrm{ret} \, t$, $\textrm{ret} \, d$ for any causal translationally invariant tempered distributions
of finite singularity order. 
$t,d$.
\label{ProductTheorem}
\end{twr}

For simplicity of notation
we assumed $W_1, \ldots. P_1, \ldots$, to have the same number $d$ of components -- inessential assumption. 
$ret$ denote the retarded part of a translationally invariant causal
distribution having finite singularity order.

{\bf Proof of theorem \ref{ProductTheorem}}.
Theorem \ref{ProductTheorem} follows from the repeated application of Lemma \ref{rett(x,y):W'(x)W''(y):InSxS->L((E),(E))Contq1>2}
given below.
\qed

In order to simplify notation we introduce the following abbreviation for scalar contractions (of possibly massless kernels) 
\begin{equation}\label{(x)q}
\kappa_q =
\overset{q}{\underset{i=1}{\dot{\otimes}}} \kappa^{{}^{q_i}}_{0,1}
\, \otimes_{{}_{q}} \,\,
\overset{q}{\underset{i=1}{\dot{\otimes}}} \kappa^{{}^{q_i}}_{1,0},
\,\,\,
\kappa_k =
\overset{k}{\underset{i=1}{\dot{\otimes}}} \kappa^{{}^{k_i}}_{0,1}
\, \otimes_{{}_{k}} \,\,
\overset{k}{\underset{i=1}{\dot{\otimes}}} \kappa^{{}^{k_i}}_{1,0},
\,\,\,
\kappa_j =
\overset{j}{\underset{i=1}{\dot{\otimes}}} \kappa^{{}^{j_i}}_{0,1}
\, \otimes_{{}_{j}} \,\,
\overset{k}{\underset{i=1}{\dot{\otimes}}} \kappa^{{}^{j_i}}_{1,0},
\,\,\,
\ldots
\end{equation}
Using thm. 3.13 of \cite{obataJFA} it follows that for the proof of theorem \ref{ProductTheorem} it is sufficient to show that 
(we have indicated the space-time variables $x_r$ with respect to which the corresponding pointwise products $\dot{\otimes}$ are computed)
\begin{gather}
\underbrace{\kappa_q}_{\textrm{funct. of $x_1,x_2$}} \cdot
\underbrace{\overset{\ell}{\underset{i=1}{\dot{\otimes}}} \kappa^{{}^{\ell_i}}_{1,0}
\overset{\mathpzc{m}}{\underset{i=1}{\dot{\otimes}}} \kappa^{{}^{\mathpzc{m}_i}}_{0,1}}_{\textrm{funct. of $x_2$}},
\label{x1(x2)}
\\
\underbrace{\kappa_q}_{\textrm{funct. of $x_1,x_2$}} \cdot
\underbrace{\overset{l}{\underset{i=1}{\dot{\otimes}}} \kappa^{{}^{l_i}}_{1,0} 
\overset{m}{\underset{i=1}{\dot{\otimes}}} \kappa^{{}^{m_i}}_{0,1}}_{\textrm{funct. of $x_1$}}
\otimes
\underbrace{\overset{\ell}{\underset{i=1}{\dot{\otimes}}} \kappa^{{}^{\ell_i}}_{1,0}
\overset{\mathpzc{m}}{\underset{i=1}{\dot{\otimes}}} \kappa^{{}^{\mathpzc{m}_i}}_{0,1}}_{\textrm{funct. of $x_2$}},
\label{x1x2}
\\
\Big[\underbrace{\kappa_q}_{\textrm{$x_1,x_3$}}
\underbrace{\kappa_k}_{\textrm{$x_2,x_3$}}
\Big] \cdot
\underbrace{\overset{l}{\underset{i=1}{\dot{\otimes}}} \kappa^{{}^{l_i}}_{1,0} 
\overset{m}{\underset{i=1}{\dot{\otimes}}} \kappa^{{}^{m_i}}_{0,1}}_{\textrm{$x_1$}}
\,
\otimes
\,
\underbrace{\overset{\ell}{\underset{i=1}{\dot{\otimes}}} \kappa^{{}^{\ell_i}}_{1,0}
\overset{\mathpzc{m}}{\underset{i=1}{\dot{\otimes}}} \kappa^{{}^{\mathpzc{m}_i}}_{0,1}}_{\textrm{$x_2$}}
\,
\otimes
\,
\underbrace{\overset{\mathfrak{l}}{\underset{i=1}{\dot{\otimes}}} \kappa^{{}^{\mathfrak{l}_i}}_{1,0}
\overset{\mathfrak{m}}{\underset{i=1}{\dot{\otimes}}} \kappa^{{}^{\mathfrak{m}_i}}_{0,1}}_{\textrm{$x_3$}},
\label{x1x2x3}
\\
\Big[\underbrace{\kappa_q}_{\textrm{$x_1,x_n$}}
\underbrace{\kappa_k}_{\textrm{$x_2,x_n$}}
\ldots
\underbrace{\kappa_j}_{\textrm{$x_{n-1},x_n$}}
\Big] \cdot
\underbrace{\overset{l}{\underset{i=1}{\dot{\otimes}}} \kappa^{{}^{l_i}}_{1,0} 
\overset{m}{\underset{i=1}{\dot{\otimes}}} \kappa^{{}^{m_i}}_{0,1}}_{\textrm{$x_1$}}
\,
\otimes
\,
\underbrace{\overset{\ell}{\underset{i=1}{\dot{\otimes}}} \kappa^{{}^{\ell_i}}_{1,0}
\overset{\mathpzc{m}}{\underset{i=1}{\dot{\otimes}}} \kappa^{{}^{\mathpzc{m}_i}}_{0,1}}_{\textrm{$x_2$}}
\, 
\otimes
\,
\ldots 
\, \otimes
\,
\underbrace{\overset{\mathfrak{l}}{\underset{i=1}{\dot{\otimes}}} \kappa^{{}^{\mathfrak{l}_i}}_{1,0}
\overset{\mathfrak{m}}{\underset{i=1}{\dot{\otimes}}} \kappa^{{}^{\mathfrak{m}_i}}_{0,1}}_{\textrm{$x_n$}},
\label{x1...xn}
\end{gather}
equal, respectively, to the kernels
\begin{gather*}
\overset{q}{\underset{i=1}{\dot{\otimes}}} \kappa^{{}^{q_i}}_{0,1}
\, \otimes_{{}_{q}} \,\,
\overset{q}{\underset{i=1}{\dot{\otimes}}} \kappa^{{}^{q_i}}_{1,0}
\overset{\ell}{\underset{i=1}{\dot{\otimes}}} \kappa^{{}^{\ell_i}}_{1,0}
\overset{\mathpzc{m}}{\underset{i=1}{\dot{\otimes}}} \kappa^{{}^{\mathpzc{m}_i}}_{0,1}
\\
\overset{l}{\underset{i=1}{\dot{\otimes}}} \kappa^{{}^{l_i}}_{1,0} 
\overset{m}{\underset{i=1}{\dot{\otimes}}} \kappa^{{}^{m_i}}_{0,1}
\overset{q}{\underset{i=1}{\dot{\otimes}}} \kappa^{{}^{q_i}}_{0,1}
\, \otimes_{{}_{q}} \,\,
\overset{q}{\underset{i=1}{\dot{\otimes}}} \kappa^{{}^{q_i}}_{1,0}
\overset{\ell}{\underset{i=1}{\dot{\otimes}}} \kappa^{{}^{\ell_i}}_{1,0}
\overset{\mathpzc{m}}{\underset{i=1}{\dot{\otimes}}} \kappa^{{}^{\mathpzc{m}_i}}_{0,1},
\\
\overset{l}{\underset{i=1}{\dot{\otimes}}} \kappa^{{}^{l_i}}_{1,0} 
\overset{m}{\underset{i=1}{\dot{\otimes}}} \kappa^{{}^{m_i}}_{0,1}
\overset{q}{\underset{i=1}{\dot{\otimes}}} \kappa^{{}^{q_i}}_{0,1}
\, \otimes_{{}_{q}} \,\,
\overset{q}{\underset{i=1}{\dot{\otimes}}} \kappa^{{}^{q_i}}_{1,0}
\overset{\ell}{\underset{i=1}{\dot{\otimes}}} \kappa^{{}^{\ell_i}}_{1,0}
\overset{\mathpzc{m}}{\underset{i=1}{\dot{\otimes}}} \kappa^{{}^{\mathpzc{m}_i}}_{0,1}
\overset{k}{\underset{i=1}{\dot{\otimes}}} \kappa^{{}^{k_i}}_{0,1}
\, \otimes_{{}_{k}} \,\,
\overset{k}{\underset{i=1}{\dot{\otimes}}} \kappa^{{}^{k_i}}_{1,0}
\overset{\mathfrak{l}}{\underset{i=1}{\dot{\otimes}}} \kappa^{{}^{\mathfrak{l}_i}}_{1,0}
\overset{\mathfrak{m}}{\underset{i=1}{\dot{\otimes}}} \kappa^{{}^{\mathfrak{m}_i}}_{0,1},
\\
\overset{l}{\underset{i=1}{\dot{\otimes}}} \kappa^{{}^{l_i}}_{1,0} 
\overset{m}{\underset{i=1}{\dot{\otimes}}} \kappa^{{}^{m_i}}_{0,1}
\overset{q}{\underset{i=1}{\dot{\otimes}}} \kappa^{{}^{q_i}}_{0,1}
\, \otimes_{{}_{q}} \,\,
\overset{q}{\underset{i=1}{\dot{\otimes}}} \kappa^{{}^{q_i}}_{1,0}
\overset{\ell}{\underset{i=1}{\dot{\otimes}}} \kappa^{{}^{\ell_i}}_{1,0}
\overset{\mathpzc{m}}{\underset{i=1}{\dot{\otimes}}} \kappa^{{}^{\mathpzc{m}_i}}_{0,1}
\overset{k}{\underset{i=1}{\dot{\otimes}}} \kappa^{{}^{k_i}}_{0,1}
\, \otimes_{{}_{k}} \,
\ldots
\,
\otimes_{{}_{j}} \,
\overset{j}{\underset{i=1}{\dot{\otimes}}} \kappa^{{}^{j_i}}_{1,0}
\overset{\mathfrak{l}}{\underset{i=1}{\dot{\otimes}}} \kappa^{{}^{\mathfrak{l}_i}}_{1,0}
\overset{\mathfrak{m}}{\underset{i=1}{\dot{\otimes}}} \kappa^{{}^{\mathfrak{m}_i}}_{0,1},
\end{gather*}
with one $\otimes_{{}_{q}}$, two $\otimes_{{}_{q}}, \otimes_{{}_{k}}$, and $n-1$ contracted tensor products 
$\otimes_{{}_{q}}, \otimes_{{}_{k}}, \ldots, \otimes_{{}_{j}}$, belong, respectively, to
\begin{gather}
\mathscr{L}\left(
\left(\overset{\ell}{\underset{i=1}{\otimes}} E_{{}_{\ell_i}}\right)^*
\overset{\mathpzc{m}}{\underset{i=1}{\otimes}} E_{{}_{\mathpzc{m}_i}}, 
\mathscr{E}_{1}^{*}\otimes \mathscr{E}_{2}^{*}\right), \,\,\,
\label{E*xE,S*xS*}
\\
\mathscr{L}\left(
\left(\overset{l}{\underset{i=1}{\otimes}} E_{{}_{l_i}}
\overset{\ell}{\underset{i=1}{\otimes}} E_{{}_{\ell_i}}\right)^*
\overset{m}{\underset{i=1}{\otimes}} E_{{}_{m_i}}
\overset{\mathpzc{m}}{\underset{i=1}{\otimes}} E_{{}_{\mathpzc{m}_i}}, 
\mathscr{E}_{1}^{*}\otimes \mathscr{E}_{2}^{*}\right), \,\,\,
\label{E*..xE..,S*xS*}
\\
\mathscr{L}\left(
\left(\overset{l}{\underset{i=1}{\otimes}} E_{{}_{l_i}}
\overset{\ell}{\underset{i=1}{\otimes}} E_{{}_{\ell_i}}
\overset{\mathfrak{l}}{\underset{i=1}{\otimes}} E_{{}_{\mathfrak{l}_i}}\right)^*
\overset{m}{\underset{i=1}{\otimes}} E_{{}_{m_i}}
\overset{\mathpzc{m}}{\underset{i=1}{\otimes}} E_{{}_{\mathpzc{m}_i}}
\overset{\mathfrak{m}}{\underset{i=1}{\otimes}} E_{{}_{\mathfrak{m}_i}}, \,\,\,
\mathscr{E}_{1}^{*}\otimes \mathscr{E}_{2}^{*}\otimes \mathscr{E}_{3}^{*}\right),
\label{E*..xE..,S*xS*xS*}
\\ 
\mathscr{L}\left(
\left(\overset{l}{\underset{i=1}{\otimes}} E_{{}_{l_i}}
\overset{\ell}{\underset{i=1}{\otimes}} E_{{}_{\ell_i}}
\otimes \ldots
\otimes
\overset{\mathfrak{l}}{\underset{i=1}{\otimes}} E_{{}_{\mathfrak{l}_i}}\right)^*
\overset{m}{\underset{i=1}{\otimes}} E_{{}_{m_i}}
\overset{\mathpzc{m}}{\underset{i=1}{\otimes}} E_{{}_{\mathpzc{m}_i}}
\otimes
\ldots
\otimes
\overset{\mathfrak{m}}{\underset{i=1}{\otimes}} E_{{}_{\mathfrak{m}_i}}, \,\,\,
\mathscr{E}_{1}^{*}\otimes \dots \otimes \mathscr{E}_{n}^{*}\right),
\label{E*..xE..,S*x...xS*}
\end{gather}
provided all non paired free field kernels are massive.

And, moreover, we have to show that it remains true if we replace the scalar contractions
\begin{gather}
t(x_1,x_2) = \kappa_q(x_1,x_2) = \kappa(x_1-x_2)
\label{kappax1x2}
\\
 t(x_1,x_2,x_3) = \kappa_q(x_1,x_3)\kappa_k(x_2,x_3) = \kappa(x_1-x_3,x_2-x_3)
\label{kappax1x2x3}
\\
 t(x_1,x_2, \ldots, x_n) = \kappa_q(x_1,x_n)\kappa_k(x_2,x_n) \ldots \kappa_p(x_{n-1}-x_n) = \kappa(x_1-x_n, x_2-x_{n}, \ldots, x_{n-1}-x_n),
\label{kappax1...xn}
\end{gather}
respectively, in (\ref{x1(x2)}) -- (\ref{x1...xn}),  with any traslationally invariat 
tempered distributions $t$, respectively, in 
\[
\mathscr{E}_{1}^{*}\otimes \mathscr{E}_{2}^{*}, \,\,
\mathscr{E}_{1}^{*}\otimes \mathscr{E}_{2}^{*} \otimes \mathscr{E}_{3}^{*}, \mathscr{E}_{1}^{*}\otimes \ldots \otimes \mathscr{E}_{n}^{*}.
\]
In particular, it is true for $t = \textrm{ret} \, d$ with 
any causal translationally invariant tempered distribution $d$ of finite singularity order.

\begin{lem}
Under the assumption that all non-paired kernels of free fields are massive, respectively, in (\ref{x1(x2)}) -- (\ref{x1...xn}), 
the kernels (\ref{x1(x2)})--(\ref{x1...xn})
belong, respectively, to (\ref{E*xE,S*xS*}) -- (\ref{E*..xE..,S*x...xS*}), 
and the integral kernel operators corresponding, respectively, to the kernels (\ref{x1(x2)}) -- (\ref{x1...xn})
belong, respectively, to
\[
\mathscr{L}\left(\mathscr{E}_{1} \otimes \mathscr{E}_{2}, \mathscr{L}((E), (E))\right), \,\, \ldots, \,\,
\mathscr{L}\left(\mathscr{E}_{1} \otimes \ldots \otimes \mathscr{E}_{n}, \mathscr{L}((E), (E))\right),
\,\,\,\,\,\, \mathscr{E}_i = \mathcal{S}(\mathbb{R}^4) \otimes \mathbb{C}^{d_i}.
\]
This remains true if we replace the products $t$ (\ref{kappax1x2}) -- (\ref{kappax1...xn})  of scalar $\otimes_q$-contractions 
in (\ref{x1(x2)}) -- (\ref{x1...xn}) with any translationally invariant tempered distributions $t$.
\label{rett(x,y):W'(x)W''(y):InSxS->L((E),(E))Contq1>2}
\end{lem}

\qedsymbol \,
We give a proof at once with $t$ equal to any translationally invariant tempered distribution $t$.
Analysis for $n>1$ contracted tensor products is reduced to only minor modification in the analysis of the case $n=1$, 
thus, we start with the case $n=1$, \emph{i.e.} with the kernel (\ref{x1x2}).  Then we extend the proof on 
(\ref{x1...xn}).

Let $\kappa_q(x,y)$ in (\ref{x1x2}) be equal $t(x,y) = \kappa(x-y)$, $\kappa \in \mathscr{E}_{1}^* = \mathcal{S}(\mathbb{R}^4)\otimes \mathbb{C}^{d_1}$.

Without losing generality, we can use the space-time test functions $\chi \in \mathscr{E}_1 \otimes \mathscr{E}_2$ of the form
\[
\chi(x,y) = \phi(x-y)\varphi(y), \,\,\,\, \phi \in \mathscr{E}_1 = \mathcal{S}(\mathbb{R}^4) \otimes \mathbb{C}^{d_1}, \,\,
\varphi \in \mathscr{E}_2= \mathcal{S}(\mathbb{R}^4) \otimes \mathbb{C}^{d_2}.
\]
For such $\chi$, $\langle t, \chi\rangle = \langle \kappa, \phi \rangle \widetilde{\varphi}(0) 
= \langle \widetilde{\kappa}, \widetilde{\phi} \rangle \widetilde{\varphi}(0)$.
 
In particular, there exists $k\in \mathbb{N}$ and finite positive $C_k$, 
such that
\begin{equation}\label{retkappa2Estimation}
|\langle \kappa, \phi \rangle| = \Big|\int\kappa(x)\phi(x) \, \ud^4 x\Big| = \Big|\int\widetilde{\kappa}(p)\widetilde{\phi}(p) \, \ud^4 p\Big| \leq C_{{}_{k}} \big|\phi \big|_{{}_{k}},
\end{equation}
where
\[
\big|\cdot \big|_{{}_{1}}, \big|\cdot \big|_{{}_{2}}, \ldots
\]
is the countable system of Hilbertian norms defining the Schwartz nuclear topology on $\mathscr{E}_1 = \mathcal{S}(\mathbb{R}^4) \otimes \mathbb{C}^{d_1}$.

Let $T_{{}_{w}}\widetilde{\phi}(p) = \widetilde{\phi}(p+w)$.
From  (\ref{retkappa2Estimation}), it follows existence of $k\in \mathbb{N}, 0<c< +\infty$, independent of $\phi$, such that 
\begin{equation}\label{|inttheta.Omega'phi(p',p'',...)| dpk1dpk2<infty}
\left| \left\langle \widetilde{\kappa}, T_{{}_{w}}\widetilde{\phi}  \right\rangle \right| \leq 
c \, \sum\limits_{0=|\gamma| \leq |\alpha| \leq k}|w^\gamma|{\alpha \choose \gamma} \,\, \big|\phi\big|_{{}_{k}},
\end{equation}
for any $w$. In the sequel we will use 
\[
w = 
\Bigg(\,\,\,\, \overbrace{\sum\limits_{i=1}^{l}\boldsymbol{p}_{{}_{l_i}}- \sum\limits_{i=1}^{m}\boldsymbol{p}_{{}_{m_i}}}^{\boldsymbol{w}} \,\,\,\, , 
\,\,\,\,\,\,\,\,\,\
\overbrace{ \sum\limits_{i=1}^{l}p_0(\boldsymbol{p}_{{}_{l_i}}) - \sum\limits_{i=1}^{m}p_0(\boldsymbol{p}_{{}_{m_i}})}^{w_0} \,\,\,\,  \Bigg)
= (w_1,w_2,w_3,w_0),
\]
and $\alpha,\beta, \gamma$, are $4$-component multiindices with
\[
w^{\gamma} = {w}_{0}^{\gamma_0}{w}_{1}^{\gamma_1} {w}_{2}^{\gamma_2} {w}_{3}^{\gamma_3}, \,\,\,
{\alpha \choose \gamma} = {\alpha_0 \choose \gamma_0} {\alpha_1 \choose \gamma_1}{\alpha_2 \choose \gamma_2}{\alpha_3 \choose \gamma_3}.   
\]

Indeed, let
\[
e_{w}(x) = e^{iw\cdot x}.
\]
Using the system of norms 
\[
| \phi |_{{}_{k}} = \underset{|\alpha|,|\beta| \leq k, x\in\mathbb{R}^4}{\textrm{sup}} \, |x^\beta \partial^\alpha \phi(x)|
\]
in (\ref{retkappa2Estimation}), and applying (\ref{retkappa2Estimation})
to $e_{w}\phi$, instead of $\phi$, we get (\ref{|inttheta.Omega'phi(p',p'',...)| dpk1dpk2<infty}). 
From (\ref{|inttheta.Omega'phi(p',p'',...)| dpk1dpk2<infty}) it follows that  
\begin{equation}\label{|inttheta.Omega'phi(p',p'',...)| dpk1dpk2Bounded}
\left| \left\langle \widetilde{\kappa}, T_{{}_{w}}\widetilde{\phi}  \right\rangle \right|
\end{equation}
regarded as the function of the 
variables
\[
\left(\ldots, s_{{}_{l_i}}, \boldsymbol{p}_{{}_{l_i}}, \ldots,s_{{}_{m_i}}, \boldsymbol{p}_{{}_{m_i}}, \ldots  \right) =
\overset{l}{\underset{i=1}{\times}} (s_{{}_{l_i}}, \boldsymbol{p}_{{}_{l_i}}) 
\overset{m}{\underset{i=1}{\times}} (s_{{}_{m_i}}, \boldsymbol{p}_{{}_{m_i}}),
\]
contained in $w$, is bounded by the absolute value of a fixed multiplier of $\overset{l}{\underset{i=1}{\otimes}} E_{{}_{l_i}}
\overset{m}{\underset{i=1}{\otimes}} E_{{}_{m_i}}$ , \emph{i.e.}  by a fixed polynomial in these variables, 
whenever $\phi$ ranges over a bounded set in $\mathscr{E}_1$ (recall that for massive fields $E_i$ are equal to Schwartz test spaces).

We have
\[
\Big \langle \,\, t \, \cdot \, \Big[
\overset{l}{\underset{i=1}{\dot{\otimes}}} \kappa^{{}^{l_i}}_{1,0} 
\overset{m}{\underset{i=1}{\dot{\otimes}}} \kappa^{{}^{m_i}}_{0,1}
\overset{\ell}{\underset{i=1}{\dot{\otimes}}} \kappa^{{}^{\ell_i}}_{1,0}
\overset{\mathpzc{m}}{\underset{i=1}{\dot{\otimes}}} \kappa^{{}^{\mathpzc{m}_i}}_{0,1}
\Big](\chi), \, \xi_1 \otimes \xi_2 \, \Big\rangle
\]
\begin{multline*}
=\sum\limits_{s_{{}_{l_i}}, s_{{}_{\ell_i}}, s_{{}_{\mathpzc{m}_i}}} \bigintsss \Bigg[ 
\prod\limits_{i=1}^{l} v^{{}^{l_i}}_{{}_{s_{{}_{l_i}}}}(\boldsymbol{\p}_{{}_{l_i}})
\prod\limits_{i=1}^{m} u^{{}^{m_i}}_{{}_{s_{{}_{m_i}}}}(\boldsymbol{\p}_{{}_{m_i}})
\prod\limits_{i=1}^{\ell} v^{{}^{\ell_i}}_{{}_{s_{{}_{m_i}}}}(\boldsymbol{\p}_{{}_{\ell_i}})
\prod\limits_{q=1}^{\mathpzc{m}} u^{{}^{\mathpzc{m}_i}}_{{}_{s_{{}_{\mathpzc{m}_i}}}}(\boldsymbol{\p}_{{}_{\mathpzc{m}_i}})
\,
\left\langle \widetilde{\kappa}, 
T_{{}_{w}} \widetilde{\phi} \right\rangle
 \,\, \times
\\
\times \,\, 
\widetilde{\varphi}\left(\sum\limits_{i=1}^{l}\boldsymbol{\p}_{{}_{l_i}} + \sum\limits_{i=1}^{\ell}\boldsymbol{\p}_{{}_{\ell_i}}
-\sum\limits_{i=1}^{m}\boldsymbol{\p}_{{}_{m_i}} 
-\sum\limits_{i=1}^{\mathpzc{m}} \boldsymbol{\p}_{{}_{\mathpzc{m}_i}}, \sum\limits_{i}^{l} p_0(\boldsymbol{\p}_{{}_{l_i}})
+ \sum\limits_{i}^{\ell}p_0(\boldsymbol{\p}_{{}_{\ell_i}})
-\sum\limits_{i=1}^{m}p_0(\boldsymbol{\p}_{{}_{m_i}})
-\sum\limits_{i}^{\mathpzc{m}} p_0(\boldsymbol{\p}_{{}_{\mathpzc{m}_i}})\right) 
\\
\xi_1\left(
\overset{l}{\underset{i=1}{\times}} (s_{{}_{l_i}}, \boldsymbol{p}_{{}_{l_i}} )
\overset{\ell}{\underset{i=1}{\times}} (s_{{}_{\ell_i}}, \boldsymbol{p}_{{}_{\ell_i}}) 
  \right)
\xi_2\left(
\overset{m}{\underset{i=1}{\times}} (s_{{}_{m_i}}, \boldsymbol{p}_{{}_{m_i}})
\overset{\mathpzc{m}}{\underset{i=1}{\times}} (s_{{}_{\mathpzc{m}_i}}, \boldsymbol{p}_{{}_{\mathpzc{m}_i}}) 
\right)
\Bigg] 
\prod\limits_{i}
\ud^3 \boldsymbol{\p}_{{}_{l_i}} 
\ud^3 \boldsymbol{\p}_{{}_{m_i}} 
\ud^3 \boldsymbol{\p}_{{}_{\ell_i}}  
\ud^3 \boldsymbol{\p}_{{}_{\mathpzc{m}_i}}  
\end{multline*}

Because for any fixed tempered distribution, say 
$t$,
the map
\[
\mathcal{O}_M(\mathbb{R}^4\times \mathbb{R}^4) 
\ni \zeta  \longmapsto 
\zeta \cdot t
\in
\mathcal{S}(\mathbb{R}^4\times \mathbb{R}^4)^*
\]
is continuous, then by lemma \ref{DistrProdWickProd}, the kernel 
\begin{equation}\label{tx1x2}
\kappa_{l+\ell,m+\mathpzc{m}} =
t \, \cdot \, \Big[
\overset{l}{\underset{i=1}{\dot{\otimes}}} \kappa^{{}^{l_i}}_{1,0} 
\overset{m}{\underset{i=1}{\dot{\otimes}}} \kappa^{{}^{m_i}}_{0,1}
\overset{\ell}{\underset{i=1}{\dot{\otimes}}} \kappa^{{}^{\ell_i}}_{1,0}
\overset{\mathpzc{m}}{\underset{i=1}{\dot{\otimes}}} \kappa^{{}^{\mathpzc{m}_i}}_{0,1}
\Big]
\end{equation}
 defines a continuous map
\[
\overset{l}{\underset{i=1}{\otimes}} E_{{}_{l_i}}
\overset{\ell}{\underset{i=1}{\otimes}} E_{{}_{\ell_i}}
\overset{m}{\underset{i=1}{\otimes}} E_{{}_{m_i}}
\overset{\mathpzc{m}}{\underset{i=1}{\otimes}} E_{{}_{\mathpzc{m}_i}} \ni
\xi \longmapsto  \kappa_{l+\ell,m+\mathpzc{m}}(\xi) \in \mathscr{E}_{1}^* \otimes \mathscr{E}_{2}^*.
\]

Let us give here the proof that the kernel 
(\ref{tx1x2})
can be extended to a separately continuous map
\[
\left(\overset{l}{\underset{i=1}{\otimes}} E_{{}_{l_i}}^*
\overset{\ell}{\underset{i=1}{\otimes}} E_{{}_{\ell_i}}^*\right)
\times
\left(\overset{m}{\underset{i=1}{\otimes}} E_{{}_{m_i}}
\overset{\mathpzc{m}}{\underset{i=1}{\otimes}} E_{{}_{\mathpzc{m}_i}}\right)
 \longrightarrow \mathscr{E}_{1}^* \otimes \mathscr{E}_{2}^*
\]
over the distributions in $\overset{l}{\underset{i=1}{\otimes}} E_{{}_{l_i}}^*\overset{\ell}{\underset{i=1}{\otimes}} E_{{}_{\ell_i}}^*$
in the variables $\overset{l}{\underset{i=1}{\times}} (s_{{}_{l_i}},\boldsymbol{\p}_{{}_{l_i}}) 
\overset{\ell}{\underset{i=1}{\times}} (s_{{}_{\ell_i}},\boldsymbol{\p}_{{}_{\ell_i}})$.

We note that both $\phi,\varphi$ range over bounded sets in $\mathscr{E}_1,\mathscr{E}_2$,
whenever $\chi$ ranges over a bounded set $B$ in  $\mathscr{E}_1 \otimes \mathscr{E}_2 = \mathcal{S}(\mathbb{R}^4\times \mathbb{R}^4)
\otimes \mathbb{C}^{d_1+d_2}$. Next, we use (\ref{|inttheta.Omega'phi(p',p'',...)| dpk1dpk2<infty}) or, equivalently, 
(\ref{|inttheta.Omega'phi(p',p'',...)| dpk1dpk2Bounded}), in the proof of boundedness of the following sets of functions.
First, we define the following set $B'(S',B)$
of functions
\begin{multline*}
\overset{m}{\underset{i=1}{\times}} (s_{{}_{m_i}}, \boldsymbol{p}_{{}_{m_i}}) 
\overset{\mathpzc{m}}{\underset{i=1}{\times}} (s_{{}_{\mathpzc{m}_i}}, \boldsymbol{p}_{{}_{\mathpzc{m}_i}}) 
\longmapsto
\\
\sum\limits_{s_{{}_{\ell_i}, s_{{}_{\mathpzc{m}_i}}}} \bigintsss \Bigg[ 
\prod\limits_{i=1}^{l} v^{{}^{l_i}}_{{}_{s_{{}_{l_i}}}}(\boldsymbol{\p}_{{}_{l_i}})
\prod\limits_{i=1}^{m} u^{{}^{m_i}}_{{}_{s_{{}_{m_i}}}}(\boldsymbol{\p}_{{}_{m_i}})
\prod\limits_{i=1}^{\ell} v^{{}^{\ell_i}}_{{}_{s_{{}_{m_i}}}}(\boldsymbol{\p}_{{}_{\ell_i}})
\prod\limits_{q=1}^{\mathpzc{m}} u^{{}^{\mathpzc{m}_i}}_{{}_{s_{{}_{\mathpzc{m}_i}}}}(\boldsymbol{\p}_{{}_{\mathpzc{m}_i}})
\,
\left\langle \widetilde{\kappa}, 
T_{{}_{w}} \widetilde{\phi} \right\rangle
 \,\, \times
\\
\times \,\, 
\widetilde{\varphi}\left(\sum\limits_{i=1}^{l}\boldsymbol{\p}_{{}_{l_i}} + \sum\limits_{i=1}^{\ell}\boldsymbol{\p}_{{}_{\ell_i}}
-\sum\limits_{i=1}^{m}\boldsymbol{\p}_{{}_{m_i}} 
-\sum\limits_{i=1}^{\mathpzc{m}} \boldsymbol{\p}_{{}_{\mathpzc{m}_i}}, \sum\limits_{i}^{l} p_0(\boldsymbol{\p}_{{}_{l_i}})
+ \sum\limits_{i}^{\ell}p_0(\boldsymbol{\p}_{{}_{\ell_i}})
-\sum\limits_{i=1}^{m}p_0(\boldsymbol{\p}_{{}_{m_i}})
-\sum\limits_{i}^{\mathpzc{m}} p_0(\boldsymbol{\p}_{{}_{\mathpzc{m}_i}})\right) 
\\
\xi_1\left(
\overset{l}{\underset{i=1}{\times}} (s_{{}_{l_i}}, \boldsymbol{p}_{{}_{l_i}})
\overset{\ell}{\underset{i=1}{\times}} (s_{{}_{\ell_i}}, \boldsymbol{p}_{{}_{\ell_i}}) 
  \right)
\Bigg] 
\prod\limits_{i=1}^{l} \ud^3 \boldsymbol{\p}_{{}_{l_i}} 
\prod\limits_{i=1}^{\ell} \ud^3 \boldsymbol{\p}_{{}_{\ell_i}} 
\end{multline*}
and note that it is bounded in the topology induced from the strong dual topology on 
$\overset{m}{\underset{i=1}{\otimes}} E_{{}_{m_i}}^*
\overset{\mathpzc{m}}{\underset{i=1}{\otimes}} E_{{}_{\mathpzc{m}_i}}^*$
$=\mathcal{S}(\mathbb{R}^3)^{* \otimes \, (m+\mathpzc{m})}\otimes \mathbb{C}^{n_1}$, whenever $\xi_1$ ranges over 
a set $S'$ in $\overset{l}{\underset{i=1}{\otimes}} E_{{}_{l_i}}
\overset{\ell}{\underset{i=1}{\otimes}} E_{{}_{\ell_i}}$
$=\mathcal{S}(\mathbb{R}^3)^{\otimes \, (l+\ell)}\otimes \mathbb{C}^{n_2}$
which is bounded with respect to the strong
dual topology on $\overset{l}{\underset{i=1}{\otimes}} E_{{}_{l_i}}^*
\overset{\ell}{\underset{i=1}{\otimes}} E_{{}_{\ell_i}}^*$
$=\mathcal{S}(\mathbb{R}^3)^{* \otimes \, (l+\ell)}\otimes \mathbb{C}^{n_2}$
and whenever $\chi$ ranges over a bounded set $B$ in $\mathscr{E}_1 \otimes \mathscr{E}_1 = \mathcal{S}(\mathbb{R}^4 \times \mathbb{R}^4)
\otimes \mathbb{C}^{d_1+d_2}$. 
Let us denote the family of subsets in $\overset{l}{\underset{i=1}{\otimes}} E_{{}_{l_i}}
\overset{\ell}{\underset{i=1}{\otimes}} E_{{}_{\ell_i}}$
which are bounded with respect to  the strong dual topology inherited from 
$\overset{l}{\underset{i=1}{\otimes}} E_{{}_{l_i}}^*
\overset{\ell}{\underset{i=1}{\otimes}} E_{{}_{\ell_i}}^*$
by $\mathfrak{S}$.

Because $\overset{m}{\underset{i=1}{\otimes}} E_{{}_{m_i}}
\overset{\mathpzc{m}}{\underset{i=1}{\otimes}} E_{{}_{\mathpzc{m}_i}}$
satisfies the first axiom of countability, then
there exists a zero-neighborhood $V\big(B'(B,S'),\epsilon\big)$  in 
$\overset{m}{\underset{i=1}{\otimes}} E_{{}_{m_i}}
\overset{\mathpzc{m}}{\underset{i=1}{\otimes}} E_{{}_{\mathpzc{m}_i}}$
such that (\cite{GelfandII}, p. 45) 
\[
|\langle \xi_2, f \rangle| <\epsilon, \,\,\, \xi_2 \in V\big(B'(B,S'),\epsilon\big), \,\,\,\,f  \in B'(B,S'). 
\]

Next, using (\ref{|inttheta.Omega'phi(p',p'',...)| dpk1dpk2<infty}), 
or, equivalently, (\ref{|inttheta.Omega'phi(p',p'',...)| dpk1dpk2Bounded}), we note that the following set $B^+(B,S)$ of functions
\begin{multline*}
\overset{l}{\underset{i=1}{\times}} (s_{{}_{l_i}}, \boldsymbol{p}_{{}_{l_i}})
\overset{\ell}{\underset{i=1}{\times}} (s_{{}_{\ell_i}}, \boldsymbol{p}_{{}_{\ell_i}}) 
\longmapsto
\\
\sum\limits_{s_{{}_{m_i}, s_{{}_{\mathpzc{m}_i}}}} \bigintsss \Bigg[ 
\prod\limits_{i=1}^{l} v^{{}^{l_i}}_{{}_{s_{{}_{l_i}}}}(\boldsymbol{\p}_{{}_{l_i}})
\prod\limits_{i=1}^{m} u^{{}^{m_i}}_{{}_{s_{{}_{m_i}}}}(\boldsymbol{\p}_{{}_{m_i}})
\prod\limits_{i=1}^{\ell} v^{{}^{\ell_i}}_{{}_{s_{{}_{m_i}}}}(\boldsymbol{\p}_{{}_{\ell_i}})
\prod\limits_{q=1}^{\mathpzc{m}} u^{{}^{\mathpzc{m}_i}}_{{}_{s_{{}_{\mathpzc{m}_i}}}}(\boldsymbol{\p}_{{}_{\mathpzc{m}_i}})
\,
\left\langle \widetilde{\kappa}, 
T_{{}_{w}} \widetilde{\phi} \right\rangle
 \,\, \times
\\
\times \,\, 
\widetilde{\varphi}\left(\sum\limits_{i=1}^{l}\boldsymbol{\p}_{{}_{l_i}} + \sum\limits_{i=1}^{\ell}\boldsymbol{\p}_{{}_{\ell_i}}
-\sum\limits_{i=1}^{m}\boldsymbol{\p}_{{}_{m_i}} 
-\sum\limits_{i=1}^{\mathpzc{m}} \boldsymbol{\p}_{{}_{\mathpzc{m}_i}}, \sum\limits_{i}^{l} p_0(\boldsymbol{\p}_{{}_{l_i}})
+ \sum\limits_{i}^{\ell}p_0(\boldsymbol{\p}_{{}_{\ell_i}})
-\sum\limits_{i=1}^{m}p_0(\boldsymbol{\p}_{{}_{m_i}})
-\sum\limits_{i}^{\mathpzc{m}} p_0(\boldsymbol{\p}_{{}_{\mathpzc{m}_i}})\right) 
\\
\xi_2\left(
\overset{m}{\underset{i=1}{\times}} (s_{{}_{m_i}}, \boldsymbol{p}_{{}_{m_i}}) 
\overset{\mathpzc{m}}{\underset{i=1}{\times}} (s_{{}_{\mathpzc{m}_i}}, \boldsymbol{p}_{{}_{\mathpzc{m}_i}}) 
  \right)
\Bigg] 
\prod\limits_{i=1}^{m} \ud^3 \boldsymbol{\p}_{{}_{m_i}} 
\prod\limits_{i=1}^{\mathpzc{m}} \ud^3 \boldsymbol{\p}_{{}_{\mathpzc{m}_i}} 
\end{multline*}
is bounded in 
$\overset{l}{\underset{i=1}{\otimes}} E_{{}_{l_i}}
\overset{\ell}{\underset{i=1}{\otimes}} E_{{}_{\ell_i}}$, whenever $\xi_2$ ranges over 
a set $S$ bounded in 
$\overset{m}{\underset{i=1}{\otimes}} E_{{}_{m_i}}
\overset{\mathpzc{m}}{\underset{i=1}{\otimes}} E_{{}_{\mathpzc{m}_i}}$
and whenever $\chi$ ranges over a bounded set $B$ in $\mathscr{E}_1 \otimes \mathscr{E}_2$. 
Let us denote the family of subsets in 
$\overset{m}{\underset{i=1}{\otimes}} E_{{}_{m_i}}
\overset{\mathpzc{m}}{\underset{i=1}{\otimes}} E_{{}_{\mathpzc{m}_i}}$
which are bounded with respect to  the initial topology in 
$\overset{m}{\underset{i=1}{\otimes}} E_{{}_{m_i}}
\overset{\mathpzc{m}}{\underset{i=1}{\otimes}} E_{{}_{\mathpzc{m}_i}}$,
by $\mathfrak{I}$.

Next, for any $S'\in \mathfrak{S}$, $S\in \mathfrak{I}$ and any strong zero-neighborhood $W(B,\epsilon)$
determined by a bounded set $B$ in 
$\mathscr{E}_1 \otimes \mathscr{E}_2$, 
for the zero-neighborhood $V\big(B'(B,S'),\epsilon\big)$ in 
$\overset{m}{\underset{i=1}{\otimes}} E_{{}_{m_i}}
\overset{\mathpzc{m}}{\underset{i=1}{\otimes}} E_{{}_{\mathpzc{m}_i}}$, 
and the strong zero-neighborhood $W\big(B^+(S,B),\epsilon\big)$ in 
\[
\left(\overset{l}{\underset{i=1}{\otimes}} E_{{}_{l_i}}^*
\overset{\ell}{\underset{i=1}{\otimes}} E_{{}_{\ell_i}}^*\right)
\cap
\left(\overset{l}{\underset{i=1}{\otimes}} E_{{}_{l_i}}
\overset{\ell}{\underset{i=1}{\otimes}} E_{{}_{\ell_i}}\right)
=
\left(\mathcal{S}(\mathbb{R}^3)^{* \otimes \, (l+\ell)} \cap \mathcal{S}(\mathbb{R}^3)^{\otimes \, (l+\ell)}\right) \otimes \mathbb{C}^{n_1+n_2}
\]
we have
\[
\Bigg|
\Big \langle \, t \cdot \, \Big[
\overset{l}{\underset{i=1}{\dot{\otimes}}} \kappa^{{}^{l_i}}_{1,0} 
\overset{m}{\underset{i=1}{\dot{\otimes}}} \kappa^{{}^{m_i}}_{0,1}
\overset{\ell}{\underset{i=1}{\dot{\otimes}}} \kappa^{{}^{\ell_i}}_{1,0}
\overset{\mathpzc{m}}{\underset{i=1}{\dot{\otimes}}} \kappa^{{}^{\mathpzc{m}_i}}_{0,1}
\Big](\xi_1 \otimes \xi_2), 
\, \chi \, \Big\rangle
\Bigg| < \epsilon
\]
whenever 
\[
\xi_2 \in S, 
\,\,\,\,\,
\xi_1 \in W\big(B^+(S,B),\epsilon\big)
\]
or whenever
\[
\xi_2 \in V\big(B'(B,S'),\epsilon\big),
\,\,\,\,\,
\xi_1 \in S'.
\]
Put otherwise, 
\[
t \, \cdot \,  \Big[
\overset{l}{\underset{i=1}{\dot{\otimes}}} \kappa^{{}^{l_i}}_{1,0} 
\overset{m}{\underset{i=1}{\dot{\otimes}}} \kappa^{{}^{m_i}}_{0,1}
\overset{\ell}{\underset{i=1}{\dot{\otimes}}} \kappa^{{}^{\ell_i}}_{1,0}
\overset{\mathpzc{m}}{\underset{i=1}{\dot{\otimes}}} \kappa^{{}^{\mathpzc{m}_i}}_{0,1}
\Big](\xi_1 \otimes \xi_2)
\in W(B,\epsilon)
\]
whenever 
\[
\xi_2 \in S, 
\,\,\,\,\,
\xi_1 \in W\big(B^+(S,B),\epsilon\big)
\]
or whenever
\[
\xi_2 \in V(B'(B,S'),\epsilon),
\,\,\,\,\,
\xi_1 \in S'.
\]
Thus, the map
\begin{multline}\label{ret(kappa...)x2(kappa...)hypocont}
\left(\overset{l}{\underset{i=1}{\otimes}} E_{{}_{l_i}}
\overset{\ell}{\underset{i=1}{\otimes}} E_{{}_{\ell_i}}\right)
\times
\left(\overset{m}{\underset{i=1}{\otimes}} E_{{}_{m_i}}
\overset{\mathpzc{m}}{\underset{i=1}{\otimes}} E_{{}_{\mathpzc{m}_i}}\right) \ni
\xi_1 \times \xi_2
\longmapsto
\\
t \, \cdot \, \Big[
\overset{l}{\underset{i=1}{\dot{\otimes}}} \kappa^{{}^{l_i}}_{1,0} 
\overset{m}{\underset{i=1}{\dot{\otimes}}} \kappa^{{}^{m_i}}_{0,1}
\overset{\ell}{\underset{i=1}{\dot{\otimes}}} \kappa^{{}^{\ell_i}}_{1,0}
\overset{\mathpzc{m}}{\underset{i=1}{\dot{\otimes}}} \kappa^{{}^{\mathpzc{m}_i}}_{0,1}
\Big](\xi_1 \otimes \xi_2)
\\
\in \mathscr{E}_{1}^{*} \otimes \mathscr{E}_{2}^{*} = \mathcal{S}(\mathbb{R}^4\times \mathbb{R}^4)^* \otimes \mathbb{C}^{d_1+d_2}
\end{multline}
is $(\mathfrak{S},\mathfrak{I})$-hypocontinuous as a map
\[
\left(\overset{l}{\underset{i=1}{\otimes}} E_{{}_{l_i}}
\overset{\ell}{\underset{i=1}{\otimes}} E_{{}_{\ell_i}}\right)
\times
\left(\overset{m}{\underset{i=1}{\otimes}} E_{{}_{m_i}}
\overset{\mathpzc{m}}{\underset{i=1}{\otimes}} E_{{}_{\mathpzc{m}_i}}\right) \longrightarrow
\mathscr{E}_{1}^{*} \otimes \mathscr{E}_{2}^{*} 
\]
with the topology on 
$\overset{l}{\underset{i=1}{\otimes}} E_{{}_{l_i}}
\overset{\ell}{\underset{i=1}{\otimes}} E_{{}_{\ell_i}}$
induced by the strong dual topology of 
$\overset{l}{\underset{i=1}{\otimes}} E_{{}_{l_i}}^*
\overset{\ell}{\underset{i=1}{\otimes}}^* E_{{}_{\ell_i}}^* \supset \overset{l}{\underset{i=1}{\otimes}} E_{{}_{l_i}}
\overset{\ell}{\underset{i=1}{\otimes}} E_{{}_{\ell_i}}$, and with the initial topology
on $\overset{m}{\underset{i=1}{\otimes}} E_{{}_{m_i}}
\overset{\mathpzc{m}}{\underset{i=1}{\otimes}} E_{{}_{\mathpzc{m}_i}}$
 and the ordinary strong dual topology on $\mathscr{E}_{1}^{*} \otimes \mathscr{E}_{2}^{*}$.

By the Proposition of Chap. III, \S 5.4, p. 90 of \cite{Schaefer}, we see that the map (\ref{ret(kappa...)x2(kappa...)hypocont})
can be uniquely extended to a $(\mathfrak{S}^*,\mathfrak{I})$-hypocontinuous map 
\[
\left(\overset{l}{\underset{i=1}{\otimes}} E_{{}_{l_i}}^*
\overset{\ell}{\underset{i=1}{\otimes}} E_{{}_{\ell_i}}^*\right)
\times
\left(\overset{m}{\underset{i=1}{\otimes}} E_{{}_{m_i}}
\overset{\mathpzc{m}}{\underset{i=1}{\otimes}} E_{{}_{\mathpzc{m}_i}}\right) \longrightarrow
\mathscr{E}_{1}^{*} \otimes \mathscr{E}_{2}^{*},
\]
where $\mathfrak{S}^*$ is the family of all bounded sets in 
$\overset{l}{\underset{i=1}{\otimes}} E_{{}_{l_i}}^*
\overset{\ell}{\underset{i=1}{\otimes}} E_{{}_{\ell_i}}^*$
with respect to the strong dual topology. In particular the map (\ref{ret(kappa...)x2(kappa...)hypocont})
can be uniquely extended to a separately continuous map 
\[
\left(\overset{l}{\underset{i=1}{\otimes}} E_{{}_{l_i}}^*
\overset{\ell}{\underset{i=1}{\otimes}} E_{{}_{\ell_i}}^*\right)
\times
\left(\overset{m}{\underset{i=1}{\otimes}} E_{{}_{m_i}}
\overset{\mathpzc{m}}{\underset{i=1}{\otimes}} E_{{}_{\mathpzc{m}_i}}\right) \longrightarrow
\mathscr{E}_{1}^{*} \otimes \mathscr{E}_{2}^{*},
\]
which was to be proved, because by Theorem 3.13 of \cite{obataJFA} (in the Bose case, or by its Fermi generalization), assertion of our Lemma follows. 
This eds the proof for (\ref{x1x2}).

In passing to (\ref{x1...xn}) with translationally invariant $t(x_1,\ldots,x_n) = \kappa(x_1-x_n,\ldots,x_{n-1}-x_n)$ 
we note that we can analogously, and without losing generality, 
use the space-time test functions $\chi \in \mathscr{E}_1 \otimes \ldots \otimes \mathscr{E}_n$ of the form
\[
\chi(x_1, \ldots, x_n) = \phi_{{}_{1}}(x_1-x_n) \ldots \phi_{{}_{n-1}}(x_{n-1}-x_n)\varphi(x_n), \,\,\,\, \phi_{{}_{i}} \in \mathscr{E}_i 
= \mathcal{S}(\mathbb{R}^4) \otimes \mathbb{C}^{d_i}, \,\,
\varphi \in \mathscr{E}_2= \mathcal{S}(\mathbb{R}^4) \otimes \mathbb{C}^{d_n}.
\]
The above proof remains unchanged, with only minor and obvious replacement 
of $\phi$ with $\phi = \phi_{{}_{1}} \otimes \ldots \otimes \phi_{{}_{n-1}}$ and $y$ with $x_n$. The formula for the pairing of the kernel
(\ref{x1...xn}) with the respective $\chi \otimes \xi_1\otimes \xi_2$ is identical with that given above for the pairing of (\ref{x1x2})
with the obvious replacement of the sum of momenta in the argument of $\widetilde{\varphi}$ enlarged respectively to 
\[
\overset{l}{\underset{i}{\sum}}... + \overset{\ell}{\underset{i}{\sum}}... + \ldots + \overset{\mathfrak{l}}{\underset{i}{\sum}}...
-\overset{m}{\underset{i}{\sum}}... - \overset{\mathpzc{m}}{\underset{i}{\sum}}... - \ldots - \overset{\mathfrak{m}}{\underset{i}{\sum}}...
\]
and with the obvious modification of $w$ with sum of momenta in it respectively enlarged to the sum of momenta corresponding to all
the contracted tensor product factors in (\ref{x1...xn}) except the last one taken with $+$ sign for $1,0$ and $-$ sign for $0,1$
free field kernel factors, and with additional multipliers $u,v$ corresponding to the additional free field kernels. 

Concerning  (\ref{x1(x2)}), the proof remains the same as above provided $\ell,\mathpzc{m} \neq 0$ and in case $\mathpzc{m}=0$ extendibility of (\ref{x1(x2)})
to an element of (\ref{E*xE,S*xS*}) is obvious.   
\qed

We have also the following 
\begin{lem}
Let translationally invariant $t_\epsilon \in \mathscr{E}_{1}^* \otimes \ldots \otimes \mathscr{E}_{n}^*$, $t_\epsilon(x_1,\ldots,x_n) = \kappa_\epsilon(x_1-x_n, \ldots, x_{n-1}-x_n)$, $\kappa_\epsilon \in \mathscr{E}_{1}^* \otimes \ldots \otimes \mathscr{E}_{n-1}^*$, converge to a translationally invariant $t$ in $\mathscr{E}_{1}^* \otimes \ldots \otimes \mathscr{E}_{n}^*$ when $\epsilon \rightarrow 0$. The following  kernel
\[
\kappa_{{}_{\epsilon \,\,\, l+\ell+ \ldots +\mathfrak{l}, \,\, m+ \mathpzc{m} + \ldots + \mathfrak{m}}} =
\big[ t_\epsilon \big] \cdot
\underbrace{\overset{l}{\underset{i=1}{\dot{\otimes}}} \kappa^{{}^{l_i}}_{\epsilon \, 1,0} 
\overset{m}{\underset{i=1}{\dot{\otimes}}} \kappa^{{}^{m_i}}_{\epsilon \, 0,1}}_{\textrm{$x_1$}}
\,
\otimes
\,
\underbrace{\overset{\ell}{\underset{i=1}{\dot{\otimes}}} \kappa^{{}^{\ell_i}}_{\epsilon \, 1,0}
\overset{\mathpzc{m}}{\underset{i=1}{\dot{\otimes}}} \kappa^{{}^{\mathpzc{m}_i}}_{\epsilon \, 0,1}}_{\textrm{$x_2$}}
\, 
\otimes
\,
\ldots 
\, \otimes
\,
\underbrace{\overset{\mathfrak{l}}{\underset{i=1}{\dot{\otimes}}} \kappa^{{}^{\mathfrak{l}_i}}_{\epsilon \, 1,0}
\overset{\mathfrak{m}}{\underset{i=1}{\dot{\otimes}}} \kappa^{{}^{\mathfrak{m}_i}}_{\epsilon \, 0,1}}_{\textrm{$x_n$}},
\]
in which only the non contracted plane wave massless kernels are replaced with their massive counterparts,
converges to
\[
\kappa_{{}_{l+\ell+ \ldots +\mathfrak{l}, \,\, m+ \mathpzc{m} + \ldots + \mathfrak{m}}} =
\big[ t \big] \cdot
\underbrace{\overset{l}{\underset{i=1}{\dot{\otimes}}} \kappa^{{}^{l_i}}_{1,0} 
\overset{m}{\underset{i=1}{\dot{\otimes}}} \kappa^{{}^{m_i}}_{0,1}}_{\textrm{$x_1$}}
\,
\otimes
\,
\underbrace{\overset{\ell}{\underset{i=1}{\dot{\otimes}}} \kappa^{{}^{\ell_i}}_{1,0}
\overset{\mathpzc{m}}{\underset{i=1}{\dot{\otimes}}} \kappa^{{}^{\mathpzc{m}_i}}_{0,1}}_{\textrm{$x_2$}}
\, 
\otimes
\,
\ldots 
\, \otimes
\,
\underbrace{\overset{\mathfrak{l}}{\underset{i=1}{\dot{\otimes}}} \kappa^{{}^{\mathfrak{l}_i}}_{1,0}
\overset{\mathfrak{m}}{\underset{i=1}{\dot{\otimes}}} \kappa^{{}^{\mathfrak{m}_i}}_{0,1}}_{\textrm{$x_n$}},
\]
in
\[
\mathscr{L}\left(
\overset{l}{\underset{i=1}{\otimes}} E_{{}_{l_i}}
\overset{\ell}{\underset{i=1}{\otimes}} E_{{}_{\ell_i}}
\otimes \ldots
\otimes
\overset{\mathfrak{l}}{\underset{i=1}{\otimes}} E_{{}_{\mathfrak{l}_i}}
\overset{m}{\underset{i=1}{\otimes}} E_{{}_{m_i}}
\overset{\mathpzc{m}}{\underset{i=1}{\otimes}} E_{{}_{\mathpzc{m}_i}}
\otimes
\ldots
\otimes
\overset{\mathfrak{m}}{\underset{i=1}{\otimes}} E_{{}_{\mathfrak{m}_i}}, \,\,\,
\mathscr{E}_{1}^{*}\otimes \dots \otimes \mathscr{E}_{n}^{*}\right),
\]
if $\epsilon \rightarrow 0$.
\label{auxiliaryProductLemma}
\end{lem}

\qedsymbol \,
The proof follows from the inequality, which assures existence of finite $c$ and natural $k,n,j$,
such that
\[
\big|
\langle  
\kappa_{{}_{\epsilon \,\,\, l+\ell+ \ldots +\mathfrak{l}, \,\, m+ \mathpzc{m} + \ldots + \mathfrak{m}}}(\phi)
-
\kappa_{{}_{l+\ell+ \ldots +\mathfrak{l}, \,\, m+ \mathpzc{m} + \ldots + \mathfrak{m}}}(\phi), \, \xi
\rangle
\big|
\leq
\epsilon \, c \, \big|\phi  \big|_{{}_{k}}  |\xi|_{{}_{j}}
\]
for all $\phi \in \mathscr{E}_1 \otimes \ldots \otimes \mathscr{E}_n$, $\mathscr{E}_i = \mathcal{S}(\mathbb{R}^4; \mathbb{C}^{d_i})$ and for all 
\[
\xi \in \overset{l}{\underset{i=1}{\otimes}} E_{{}_{l_i}}
\overset{\ell}{\underset{i=1}{\otimes}} E_{{}_{\ell_i}}
\otimes \ldots
\otimes
\overset{\mathfrak{l}}{\underset{i=1}{\otimes}} E_{{}_{\mathfrak{l}_i}}
\overset{m}{\underset{i=1}{\otimes}} E_{{}_{m_i}}
\overset{\mathpzc{m}}{\underset{i=1}{\otimes}} E_{{}_{\mathpzc{m}_i}}
\otimes
\ldots
\otimes
\overset{\mathfrak{m}}{\underset{i=1}{\otimes}} E_{{}_{\mathfrak{m}_i}}.
\]
We do not present derivation of this inequality as it is the same as derivation of (\ref{BasicInequalityForWickEpsilon->0}).
\qed

The above series of Wick decomposition Theorems \ref{WickThmForMassiveFields}-\ref{ProductTheorem} and 
lemma \ref{auxiliaryProductLemma},
gives us a class of generalized operators $\Xi'$, equal to finite sums of integral kernel operators
\[
\Xi' = \sum\limits_{l',m'} \Xi_{l',m'}(\kappa'_{l',m'}) \in \mathscr{L}\big(\mathscr{E}, \mathscr{L}((E),(E)^*) \big), 
\,\,\, \mathscr{E}_{{}_{'}} = \mathcal{S}(\mathbb{R}^{4k'})
\]
for which the product operation is well-defined and the Wick decomposition through the normal ordering is applicable to their products, 
although in general this class includes
operators in 
\[
\mathscr{L}\big(\mathscr{E}_{{}_{'}}, \mathscr{L}((E),(E)^*) \big)
\]
which do not belong to
\[
\mathscr{L}\big(\mathscr{E}_{{}_{'}}, \mathscr{L}((E),(E)) \big).
\]
Namely, we have the following
\begin{twr}
The class of generalized integral kernel operators with $\mathcal{S}(\mathbb{R}^{4k})^*$-valued kernels $\kappa_{l,m}$, which admits the 
operation of (tensor) product, includes the operators  
\begin{equation}\label{ProductClassGenOp}
t(x_1, \ldots, x_n) \, {:}W_1(x_1)W_2(x_2) \ldots W(x_k){:}, \,\,\, t \in  \mathcal{S}(\mathbb{R}^{4k})^* = \mathcal{S}(\mathbb{R}^{4})^{* \, \otimes \, k},
\end{equation}
with translationally invariant $t$, and $W_i$ being Wick products of massless or massive free fields. 
\label{ClassWithProductTheorem}
\end{twr}
\qedsymbol \,
By Theorems \ref{WickThmForMassiveFields}-\ref{ProductTheorem} and 
lemma \ref{auxiliaryProductLemma}, for operators $\Xi', \Xi''$ in this class 
there exist $\epsilon$-approximations
\begin{gather*}
\Xi'_{{}_{\epsilon}} = \sum\limits_{l',m'} \Xi_{l,m}(\kappa'_{\epsilon \, l,m}) \in \mathscr{L}\big(\mathscr{E}_{{}_{'}}, \mathscr{L}((E),(E)) \big), 
\\
\Xi''_{{}_{\epsilon}} = \sum\limits_{l'',m''} \Xi_{l,m}(\kappa''_{\epsilon \, l'',m''}) 
\in \mathscr{L}\big(\mathscr{E}_{{}_{''}}, \mathscr{L}((E),(E)) \big), 
\end{gather*}
(here with $\mathscr{E}_{{}_{'}}$ and $\mathscr{E}_{{}_{''}}$ being some Schwartz spaces of $\mathbb{C}^{d_i}$-valued functions) for which 
\[
\Xi'_{{}_{\epsilon}} \longrightarrow \Xi',
\,\,\,
\Xi''_{{}_{\epsilon}} \longrightarrow \Xi''
\]
in 
\[
\mathscr{L}\big(\mathscr{E}_{{}_{(i)}}, \mathscr{L}((E),(E)^*) \big)
\]
and moreover, the Fock decomposition is naturally applicable
to their operator product $\Xi_{{}_{\epsilon}}$
\[
\Xi_{{}_{\epsilon}} (\phi\otimes\varphi) \overset{\textrm{df}}{=} \Xi'_{{}_{\epsilon}} (\phi) \Xi''_{{}_{\epsilon}}(\varphi) 
\]
and such that the kernels
\[
\kappa'_{\epsilon, l',m'} \otimes_q \kappa''_{\epsilon \, l'',m''}
\]
of the Wick decomposition of the operator product $\Xi_{{}_{\epsilon}}$ converge to some kernels
\[
\kappa'_{l',m'} \otimes_q \kappa''_{l'',m''} \in \mathscr{L}(E^{\otimes(l'+l''+m'+m''-2q)}, \mathscr{E}_{{}_{'}}^{*}\otimes \mathscr{E}_{{}_{''}}^{*})
\]
 representing a well-defined finite sum $\Xi$ of generalized operators. 
Thus, by thm. 3.9 of \cite{obataJFA}, in this class the product operator $\Xi_{{}_{\epsilon}}$ converges
\[
\Xi_{{}_{\epsilon}} \longmapsto \Xi \in \mathscr{L}\big(\mathscr{E}_{{}_{'}}\otimes \mathscr{E}_{{}_{''}}, \mathscr{L}((E),(E)^*) \big)
\]
to an operator $\Xi$ in 
\[
\mathscr{L}\big(\mathscr{E}_{{}_{'}}\otimes \mathscr{E}_{{}_{''}}, \mathscr{L}((E),(E)^*) \big).
\]
In practice, we construct the $\epsilon$-approximation within the class indicated above, just by replacing $|\boldsymbol{\p}|$ by 
$(|\boldsymbol{\p}|^2+\epsilon^2)^{1/2}$ in the
free massless field plane wave kernels.
\qed

Definition of product, given in the above proof, 
can be generalized over a still more general finite sums of integral kernel operators
$\Xi', \Xi''$, respectively in $\mathscr{L}\big(\mathscr{E}_{{}_{(i)}}, \mathscr{L}((E),(E)*) \big)$, $\mathscr{E}_{{}_{(i)}}= \mathcal{S}(\mathbb{R}^{4k_i}$,
provided only they possess the $\epsilon$-approximations 
with the kernels $\kappa'_{\epsilon, l',m'} \otimes_q \kappa''_{\epsilon \, l'',m''}$
of the Wick decompositions of their products
converging in the sense defined above. In fact the class of  generalized operators, on which product operation is well-defined, 
can still be extended over infinite sums 
of integral kernel operators of the class (\ref{ProductClassGenOp}), provided the infinite sums represent Fock expansions convergent in the sense
of \cite{obataJFA}.  In application to causal QFT, where the Lagrangian interaction density operator $\mathcal{L}(x)$
is equal to a Wick polynomial in free fields of finite degree, the class  (\ref{ProductClassGenOp}), allowing
the product operation, is sufficient.

\section{Grassmann valued test functions}\label{Grassmann}

In order to give rigorous formulation of the Bogoliubov causality axioms for the scattering generalized operator with Hida operators
with the generality including the case of Wick monomials $W_{{}_{j}}$ in (\ref{L}) which are of odd degree
in Fermi fields, we need to give, after \cite{Berezin}, a rigorous construction of the Grassmann algebra. As a 
topological linear space, it is equal to the exact inductive limit of a coutable family of Hilbert spaces (strong dual of 
a standard countably Hilbert nuclear space in terminology of \cite{obataJFA}) and, thus, is nuclear. Then we give
definition of the space of Grassmann-valued test functions $g$ of grade $1$ which, 
as a topological vector space, is naturally identified with the Schwartz space $\mathcal{S}(\mathbb{R}^4;\mathbb{C}^d)$ of functions on 
the space-time, and thus equal to
a standard countably Hilbert nuclear space \cite{obataJFA} (projectivie limit of a countable family of Hilbert spaces). Further, we construct
the test space of Grassmann-valued test functions of grade $p$ which, as a t.v.s., is naturally identified with the symmetrized tensor product
$\left(\mathcal{S}(\mathbb{R}^4;\mathbb{C}^d)\right)^{\otimes \, p}_{{}_{\textrm{sym}}}$, and thus also is standard countably Hilbert and nuclear.
All operators, e.g. representations naturally acting in $\mathcal{S}(\mathbb{R}^4;\mathbb{C}^d)$, are, through the tensor
product lifting, trasferred over $\left(\mathcal{S}(\mathbb{R}^4;\mathbb{C}^d)\right)^{\otimes \, p}_{{}_{\textrm{sym}}}$ 
and, thus, over the space of Grassmann-valued test functions
of grade $p$, similarly as for the $p$-powers of tensor products of ordinary $\mathbb{C}^d$-valued test functions.  

It is necessary for example when computing higher order contributions to an interacting Fermi free field. E.g. if one wants to calculate
the said contributions for the spinor field
$\boldsymbol{\psi}$ (and at once to the e.m. potential
field $A$) in spinor QED, we consider the following Krein-self adjoint interaction Lagrangian (\ref{L})
with the switching-off function $g = (g_{{}_{0}}, g_{{}_{1}}, \ldots, g_{{}_{12}}) = (g_{{}_{0}}, h_{a}, h_{b}, j_\mu)$,
$a,b \in \{1,\ldots, 4\}$, $\mu \in\{0,\ldots,3\}$, which is equal
\begin{multline*} 
\sum\limits_{j=0}^{12} g_{{}_{j}}(x) \mathcal{L}_{{}_{j}}(x)
= g_{{}_{0}}(x) \mathcal{L}_{{}_{0}}(x)  + h(x)^\sharp \boldsymbol{\psi}(x) 
+  \boldsymbol{\psi}^\sharp(x) h(x) + j(x)A(x)
\\
=  g_{{}_{0}}(x) \mathcal{L}_{{}_{0}}(x)  + \sum\limits_{a} \overline{h_{a}(x)} \big[\gamma^0 \boldsymbol{\psi}\big]^{a}(x)
 + \sum\limits_{a} h_b(x) \boldsymbol{\psi}^{\sharp \, b}(x) + \sum\limits_{\mu} j_{\mu}(x) A^{\mu}(x),
\end{multline*}
with the four component  bispinor switching-off function 
\begin{equation}\label{h=iota.phi}
h^a(x) = h^a(x)  = \iota_a(x)\phi^a(x) = \iota \cdot \phi^a(x), \,\,\,\, \phi\in \mathcal{S}(\mathbb{R}^4; \mathbb{C}^4),
\end{equation}
whose components are equal to the \emph{generators}
\[
\iota_1(x), \ldots, \iota_4(x), \,\,\,\, x\in\mathbb{R}^4,
\]
\emph{of the Grassmann algebra with inner product and with involution $\overline{\,\, \cdot \,\,}$} in the sense of \cite{Berezin},
multiplied, respectively, by the Schwartz test functions
\[
\phi^1(x), \ldots, \phi^4(x).
\]
Let us construct first the Grassmann algebra without the involution, then we extend construction to include involution. 
The construction remains the same for any number of components of $h$ and $\phi$. The Grassmann algebra without involution
\[
\bigwedge \mathcal{E}^{1*} = \underset{p}{\oplus} \mathcal{E}^{p*} = \underset{p}{\oplus} \mathcal{E}^{1* \wedge \, n}
\]
is constructed as an abstract Gelfand triple 
(or rigged Hilbert space, \cite{GelfandIV}) 
\begin{equation}\label{GammaGelfandTripleGAIP}
(\mathcal{E}^{1}) = \bigwedge \mathcal{E}^{1} \subset \Gamma(\mathcal{H}^1) \subset (\mathcal{E}^{1})^* = \bigwedge \mathcal{E}^{1*}  
\end{equation}
associated to a standard positive operator $\Gamma(A)$ in $\Gamma(\mathcal{H}^1)$, obtained through the application of the Fermi second quantization functor
$\Gamma$ to an abstract Gelfand triple 
\begin{equation}\label{GelfandTripleGAIP}
\mathcal{E}^{1} \subset \mathcal{H}^{1} \subset \mathcal{E}^{1*}
\end{equation}
over a separable Hilbert space $\mathcal{H}^{1}$ and associated to a stanard positive operator $A$ with $\textrm{Inf Spec} \, A >1$,
whose some negative power $A^{-r}$, $r>0$, is of Hilbert Schmidt class. The Grassmann product in $\bigwedge \mathcal{E}^{1*}$
is defined through the wedge product. We construct the generators $\iota^a(x)$, $a=1,\ldots,4$, $x\in \mathbb{R}^4$ 
as in \cite{Berezin} by considering the unitary isomorphism $\iota$ of $L^2(\mathbb{R}^4; \mathbb{C}^4)$ onto
$\mathcal{H}^1$ which is of the form
\begin{equation}\label{BUnitary}
f = \iota(\varphi) = \sum\limits_{a}\int \iota_a(x)\varphi_a(x) \ud^4 x \in \mathcal{H}^1, 
\,\,\,\,\,\, \phi \in L^2(\mathbb{R}^4; \mathbb{C}^4),
\end{equation}
which can be extended to a continuous isomorphism $\mathcal{E}^{1}(\mathbb{R}^4;\mathbb{C}^4) \rightarrow \mathcal{E}^{1}$
and, by duality, to a continuous isomorphism $\mathcal{E}^{1 *}(\mathbb{R}^4;\mathbb{C}^4) \rightarrow \mathcal{E}^{1 *}$, 
where $\mathcal{E}^{1}(\mathbb{R}^4;\mathbb{C}^4)$ is the space of functions corresponding to the elements $f \in \mathcal{E}^{1}$.
Such elements  $\iota_a(x)$ exist and compose a complete system of generalized vectors in the sense of \cite{GelfandIV}, lying in $\mathcal{E}^{1*}$.
Indeed, let us consider the commutative algerba $\mathfrak{A}$  of multiplication operators by smooth bounded $\mathbb{R}^4$-valued functions
in $L^2(\mathbb{R}^4; \mathbb{C}^4)$ with the standard invariant Lebesgue measure on each copy of $\mathbb{R}^4$. 
Let $\iota\mathfrak{A}$ be the algebra corresponding to it through $\iota$. Then $\iota_a(x)$,  
$a=1,\ldots,4$, $x\in \mathbb{R}^4$, compose the generalized eigenvectors common to all operators in  $\iota\mathfrak{A}$. Such
construction posses standard realization, e.g. in the form of the Gelfand triple 
\begin{equation}\label{FunctionGelfandTripleGAIP}
\mathcal{E}^{1}(\mathbb{R}^4;\mathbb{C}^4) = \mathcal{S}(\mathbb{R}^4; \mathbb{C}^4) \,\,\,\, \subset
\,\,\,\,  L^2(\mathbb{R}^4; \mathbb{C}^4)  \,\,\,\, \subset \,\,\,\, \mathcal{E}^{1*}(\mathbb{R}^4; \mathbb{C}^4) = \mathcal{S}(\mathbb{R}^4; \mathbb{C}^4)^* 
\end{equation}
with the standard operator $A = \iota[\oplus H_{{}_{(n)}}]$ equal to the direct sum of four copies of $H_{{}_{(n)}}$, where
$H_{{}_{(n)}}$ is  the Hamiltonian operator of the four dimensional oscillator acting in $L^2(\mathbb{R}^4; \mathbb{C}^4)$.
It has the required properties: the dense nuclear subspace $\mathcal{E}^{1}(\mathbb{R}^4;\mathbb{C}^4) \subset L^2(\mathbb{R}^4; \mathbb{C}^4)$ 
is invariant for $\mathfrak{A}$, the generalized eigenvectors $\iota_a(x)$ are equal to the $\iota$-image of the Dirac delta functionals
$\delta(a,x)$, the function $(a,x) \mapsto \delta(a,x) \in \mathcal{E}^{1*}(\mathbb{R}^4;\mathbb{C}^4)$ is continuous, 
and $(a,x) \mapsto \iota_a(x) \in \mathcal{E}^{1*}$ is continuous. Thus, the integral (\ref{BUnitary}) exists, and by completeness
of the system $\iota^a(x)$, $a=1,\ldots, 4$, $x\in \mathbb{R}^4$, it is unitary. Because $\iota_a(x) \in \mathcal{E}^{1*}$, the inner product
$\left(\cdot, \cdot \right)_{{}_{0}}$ in $\mathcal{H}^1$ is not in general well-defined for $\iota_a(x)$, but still 
$\left(\iota(\varphi), \iota(\varphi) \right)_{{}_{0}}$ is well-defined for $\varphi \in L^2(\mathbb{R}^4; \mathbb{C}^4)$. 
It can be computed as the limit $k \rightarrow \infty$
of the integrals $\left(\iota_{{}_{k}}(\varphi), \iota_{{}_{k}}(\varphi) \right)_{{}_{0}}$, where $\iota_{{}_{k}}(\varphi)$
is understood as the integral (\ref{BUnitary}) in which the functionals $\iota_a(x) \in \mathcal{E}^{1*}$ are replaced with
${\iota_{{}_{k}}}_a(x) \in \mathcal{E}^{1}$ converging to $\iota_a(x)$, when $k \rightarrow \infty$. By completeness of 
the system $\iota_a(x)$, $a=1,\ldots, 4$, $x\in \mathbb{R}^4$, 
$\left({\iota_{{}_{k}}}_a(x), {\iota_{{}_{k}}}_b(x) \right)_{{}_{0}} \overset{k \rightarrow \infty}{\longrightarrow}
\delta_{{}_{a \, b}}\delta(x-y)$, whence the unitarity of (\ref{BUnitary}), as expected of any system of generalized eigenvectors of
the (commuting system) of self adjoint operators acting on the rigged Hilbert space (\ref{GelfandTripleGAIP}), with the nuclear
$\mathcal{E}^1$ as a dense invariant subspace, \cite{GelfandIV}.  
It is easily seen that (\ref{BUnitary}) can be extended
to a continuous isomorphism, as said above, which in fact gives a natural identification of the abstract Gelfand triple (\ref{GelfandTripleGAIP})
with its ``function'' realization (\ref{FunctionGelfandTripleGAIP}) induced by the unitary isomorphism
(\ref{BUnitary}). Extension of (\ref{BUnitary}) to $\varphi\in \mathcal{E}^{1 *}(\mathbb{R}^4;\mathbb{C}^4)$
is understood in the following manner. Let $\varphi_k \in \mathcal{E}^{1}(\mathbb{R}^4;\mathbb{C}^4)$, $k=1,2, \ldots$, converge to $\varphi$ in 
$\mathcal{E}^{1 *}(\mathbb{R}^4;\mathbb{C}^4)$. Each integral  (\ref{BUnitary}) with $\varphi = \varphi_k$ is a well-defined element of 
$\mathcal{E}^{1*}$ naturally identifiable with an element of $\mathcal{E}^{1}$. Each continuous functional $F_{{}_{\iota(\psi)}}$ on $\mathcal{E}^{1*}$ 
is equal to the valuation $\mathcal{E}^{1 *} \ni F \mapsto F(\iota(\psi))$ at an element $\iota(\psi)$ 
of $\mathcal{E}^{1}$, for $\psi \in \mathcal{E}^{1}(\mathbb{R}^4;\mathbb{C}^4)$ (reflexivity
of $\mathcal{E}^{1}$). The value of such a functional $F_{{}_{\iota(\psi)}}$ on (\ref{BUnitary}), with $\varphi=\varphi_k$, is equal to 
$\langle \varphi_k, \psi\rangle$, and for $k$ going to infinity the value converges to $\langle \varphi, \psi\rangle$. Therefore, 
$\langle\iota(\varphi), \iota(\psi)\rangle = \langle \varphi, \psi\rangle$, as expected. In fact, we can put the abstract triple
(\ref{GelfandTripleGAIP}) equal to the ``function'' realization (\ref{FunctionGelfandTripleGAIP}). In this case $\iota^a(x) = \delta(a,x)$,
and (\ref{BUnitary}) becomes nothong else but the natural inclusion $\iota$ of the elements $\varphi \in L^2(\mathbb{R}^4; \mathbb{C}^4)$ (classes 
of functions equal a.e. which are square summable) into $\mathcal{E}^{1*}(\mathbb{R}^4;\mathbb{C}^4)$, i.e. identification of $\varphi$
with the functionals $\mathcal{E}^{1}(\mathbb{R}^4;\mathbb{C}^4) \ni \psi \mapsto \left(\overline{\varphi}, \psi\right)_{{}_{L^2}}$. It induces
natural inclusion $\iota$ of $\varphi \in \mathcal{E}^{1}(\mathbb{R}^4;\mathbb{C}^4)$, and more generally classes $\varphi$ of functions equal a.e. to functions
growing polynomially at infinity,  into function-like functionals $\psi \mapsto \left(\overline{\varphi}, \psi\right)_{{}_{L^2}} = \langle \varphi, \psi\rangle$.      
The unitary map (\ref{BUnitary}) (or its extension $\iota$ to $\mathcal{E}^{1*}(\mathbb{R}^4; \mathbb{C}^4)$) induces a continuous isomporphism 
(Fermi Fock lifting of $\iota$ restricted to $\left(\mathcal{E}^{1*}\right)^{\wedge \, n}(\mathbb{R}^4; \mathbb{C}^4)
= \mathcal{E}^{n*}(\mathbb{R}^4; \mathbb{C}^4)$)
\begin{equation}
\mathcal{E}^{n \, *}(\mathbb{R}^4;\mathbb{C}^4) \ni \varphi^n \longmapsto \iota(\varphi^n) = \sum\limits_{a_1, \ldots, a_n} \int 
\iota^{a_1}(x_1)\ldots \iota^{a_n}(x_n) \varphi^n(a_1,x_1, \ldots, a_n,x_n) \, dx_1 \ldots dx_n \in \mathcal{E}^{n \, *},
\label{iota-n-isomorphism}
\end{equation}
of the space of skew symmetric distributions $\varphi^n  \in \mathcal{E}^{n \, *}(\mathbb{R}^4;\mathbb{C}^4)$
onto $\mathcal{E}^{n \, *}$, compare \cite{Berezin}. This gives the natural identification of the Fermi Fock lifting of the ``function''
Gelfand triple (\ref{FunctionGelfandTripleGAIP})
with the Fermi Fock lifting (\ref{GammaGelfandTripleGAIP}) of the abstract Gelfand triple (\ref{GelfandTripleGAIP}), induced by (\ref{BUnitary}).

Performing summation with respect to $a$ in (\ref{h=iota.phi}) we obtain a most general Grassmann test function of grade $1$ with a fixed
number of components (here we consider four components, but it is arbitrary).
The space $\mathscr{E}^1$ of Grassmann-valued test functions
(\ref{h=iota.phi}), with $\phi \in \mathcal{S}(\mathbb{R}^4, \mathbb{C}^4)$, can be naturally given the topology
of the Schwartz space $\mathcal{S}(\mathbb{R}^4, \mathbb{C}^4)$ on identifying
each $h = \iota \cdot \phi$ in (\ref{h=iota.phi}) with the corresponding $\phi$. 
Similarly, the space $\mathscr{E}^p$ of Grassmann-valued test functions (summation in $a$ is understood in order to obtain most 
general Grassmann test function $h^p$ of grade $p$)
\begin{gather}
h^p(a_1,x_1, \ldots, a_p,x_p)=  \iota^{a_1}(x_1) \ldots \iota^{a_p}(x_p)
\phi(a_1, x_1, \ldots, a_p, x_p)
 = \iota^p \cdot \phi(a_1,x_1, \ldots, a_p,x_p),
\label{h^p}
\\
\phi \in \mathcal{S}(\mathbb{R}^4; \mathbb{C}^4)^{\otimes \, p}_{{}_{\textrm{sym}}}, x_i \in \mathbb{R}^4, a_i \in \{1, \ldots, 4\},
\nonumber
\end{gather}
is naturally endowed with the topology of the symmetrized tensor product $\mathcal{S}(\mathbb{R}^4; \mathbb{C}^4)^{\otimes \, p}_{{}_{\textrm{sym}}}$
through the identification of each $h= \iota^p \cdot \phi$ with the corresponding $\phi$. 
Thus, each skew symmetric distribution $\kappa \in \mathcal{E}^{p \, *}(\mathbb{R}^4;\mathbb{C}^4)$ gives a pairing 
\begin{multline}\label{Grassmann-Berezin-Pairing}
\big\langle\kappa, h^p \big\rangle
\overset{\textrm{df}}{=} 
\sum\limits_{a_1, \ldots, a_p} \int h(a_1, x_1, \ldots, a_p, x_p) \kappa(a_1,x_1, \ldots, a_p,x_p) \, \ud^4x_1 \ldots \ud^4dx_n =
\\
=
\sum\limits_{a_1, \ldots, a_p}\int \iota_{a_1}(x_1)\ldots \iota_{a_p}(x_p) 
\phi(a_1,x_1, \ldots, a_p, x_p) \kappa(a_1,x_1, \ldots, a_p,x_p) \, \ud^4x_1 \ldots \ud^4x_p
\in \mathcal{E}^{n \, *},
\end{multline}
for $h^p$ of the form (\ref{h^p}). Therefore, the value $\kappa(h^p) = \big\langle\kappa, h^p \big\rangle$ of a distribution 
$\kappa \in \mathcal{E}^{p \, *}(\mathbb{R}^4;\mathbb{C}^4)$ on the Grassmann-valued test function $h^p \in \mathscr{E}^p$ is not a complex number
but an element of the abstract Grassmann algebra $\mathcal{E}^{p \, *} \subset \underset{n} \oplus \, \mathcal{E}^{n \, *}$, and, by 
continuity of the map (\ref{iota-n-isomorphism}), gives a continuous map
\begin{equation}\label{ContinuityBerezinPairing}
\mathscr{E}^p \ni h^p \longmapsto \kappa(h^p) = \big\langle\kappa, h^p \big\rangle \in \mathcal{E}^{p \, *}.
\end{equation}
Therefore, each scalar antisymmetric distribution $\kappa \in \mathcal{E}^{p \, *}(\mathbb{R}^4;\mathbb{C}^4)$ 
can also be regarded as a Grassmann-valued 
distribution, when it is regarded as a functional of the Grassmann-valued test functions $h^p \in \mathscr{E}^p$,
\emph{i.e.} $\kappa \in \mathscr{L}(\mathscr{E}^p, \mathcal{E}^{p \, *})$. Indeed, each continuous functional
$F_{{}_{\iota(\psi)}}$ on $\mathcal{E}^{p \, *}$ is equal to the valuation at an element $\iota(\psi) \in \mathcal{E}^{p}$,
for some $\psi \in \mathcal{E}^{p}(\mathbb{R}^4;\mathbb{C}^4)$. The value of the functional  $F_{{}_{\iota(\psi)}}$
on the integral (\ref{Grassmann-Berezin-Pairing}) is equal to the pairing $\langle \phi \kappa, \psi \rangle$, so that
the integral (\ref{Grassmann-Berezin-Pairing}) represents the distribution $\iota(\phi \kappa) \in \mathcal{E}^{p \, *}$.
As explained above, the value is understood as the limit $k\rightarrow \infty$, of the values on the integrals
(\ref{Grassmann-Berezin-Pairing}) with $\kappa$ replaced with $\kappa_k\in \mathcal{E}^{p}(\mathbb{R}^4;\mathbb{C}^4)$ converging to $\kappa$.
Because the multiplication operation 
$\mathcal{S}(\mathbb{R}^4; \mathbb{C}^4)^{\otimes \, p}_{{}_{\textrm{sym}}} \ni \phi \longmapsto \phi \kappa \in \mathcal{E}^{p \, *}(\mathbb{R}^4;\mathbb{C}^4)$ 
is continuous, so is (\ref{ContinuityBerezinPairing}).

More generally, each distribution $\kappa$ with the kernel
\[
\kappa(x_1, \ldots, x_n, a_1,y_1, \ldots, a_p,y_p), \,\,\, x_i,y_j \in \mathbb{R}^4, a_j \in \{1,\ldots, 4\}, 
\]
skew symmetric in $a_j,y_j$, belonging to $\mathscr{E}^{*\otimes \, n} \otimes \mathcal{E}^{p \, *}(\mathbb{R}^4;\mathbb{C}^4)$, 
$\mathscr{E} = \mathcal{S}(\mathbb{R}^4;\mathbb{C})$,
defines also a continuous map
\[
\mathscr{E}^{\otimes \, n} \otimes \mathscr{E}^p \ni \phi \otimes h^p \longmapsto \kappa(\phi \otimes h^p) = 
\big\langle\kappa, \phi \otimes h^p \big\rangle \in \mathcal{E}^{p \, *},
\]
\emph{i.e.} $ \kappa \in \mathscr{L}(\mathscr{E}^{\otimes \, n} \otimes \mathscr{E}^p, \mathcal{E}^{p \, *})$. Therefore,
$\kappa$, when evaluated at $\phi  \otimes h^p$, where $h^p$ is a Grassmann-valued test function, is not scalar valued but
has the values in $\mathcal{E}^{p \, *}$ -- the subspace $\mathcal{E}^{p \, *}$ of $p$-degree of the abstract Grassmann algebra  
$\underset{n} \oplus \, \mathcal{E}^{n \, *}$ with inner product.

In particular each $\mathscr{E}^{*\otimes \, n} \otimes \mathcal{E}^{p \, *}(\mathbb{R}^4;\mathbb{C}^4)$-valued, or equivalently,
each $\mathscr{L}(\mathscr{E}^{\otimes \, n} \otimes \mathcal{E}^{p}(\mathbb{R}^4;\mathbb{C}^4), \mathbb{C})$-valued distribution
\begin{equation}\label{kappa-C-distribution-valued}
\kappa_{\mathpzc{l}, \mathpzc{m}} \in 
\mathscr{L} \big(E^{\otimes (\mathpzc{l}+\mathpzc{m})} , \,  
\mathscr{L}(\mathscr{E}^{\otimes \, n} \otimes \mathcal{E}^{p}(\mathbb{R}^4;\mathbb{C}^4), \mathbb{C})\big)
\cong
\mathscr{L} \big(E^{\otimes (\mathpzc{l}+\mathpzc{m})} , \,  
\mathscr{E}^{*\otimes \, n} \otimes \mathcal{E}^{p \, *}(\mathbb{R}^4;\mathbb{C}^4)\big)
\end{equation}
defines also a vector valued kernel 
\begin{multline}\label{kappa-Grassmann-distribution-valued}
\kappa_{\mathpzc{l}, \mathpzc{m}} \in 
\mathscr{L} \big(E^{\otimes (\mathpzc{l}+\mathpzc{m})} , \, 
\mathscr{L}(\mathscr{E}^{\otimes \, n} \otimes \mathscr{E}^p, \mathcal{E}^{p \, *}) \big)
\cong 
E^{\otimes (* \, \mathpzc{l}+\mathpzc{m})} 
\otimes \mathscr{E}^{* \otimes \, n} \otimes \big[\mathscr{E}^p \big]^* \otimes  \mathcal{E}^{p \, *}
\\
\cong
\mathscr{L} \big(\mathscr{E}^{\otimes \, n} \otimes \mathscr{E}^p, 
\, (E^{\otimes (\mathpzc{l}+\mathpzc{m})})^* \otimes \mathcal{E}^{p \, *} \big)
\end{multline}
of an integral kernel operator 
\begin{multline}\label{classOfXi}
\Xi(\kappa_{\mathpzc{l}, \mathpzc{m}}) \in \mathscr{L} \big((E)\otimes \mathscr{E}^{\otimes \, n} \otimes \mathscr{E}^p, \,  (E)^* \otimes \mathcal{E}^{p \, *}\big)
\\
\cong 
(E)^* \otimes \mathscr{E}^{* \otimes \, n} \otimes  [\mathscr{E}^p]^* \otimes  (E)^* \otimes \mathcal{E}^{p \, *}
\cong 
\mathscr{L}\big((E), (E)^*\big) \otimes \mathscr{L}(\mathscr{E}^{\otimes \, n} \otimes \mathscr{E}^p, \mathcal{E}^{p \, *})
\end{multline}
for which we have the following $\mathcal{E}^{p \, *}$-valued pairing 
\begin{multline*}
\big\langle\kappa_{\mathpzc{l}, \mathpzc{m}}(\eta_{{}_{\Phi,\Psi}}), g^{\otimes \, n} \otimes h^p \big\rangle
\overset{\textrm{df}}{=} 
\sum\limits_{a_1, \ldots, a_p} \int  \Big\{
g(x_1)\ldots g(x_n) h(a_1,y_1)\ldots h(a_p,y_p) \kappa_{\mathpzc{l}, \mathpzc{m}}(\eta_{{}_{\Phi,\Psi}})(x_1, \ldots, x_n, a_1,y_1, \ldots, a_p, y_p)
\\ 
\Big\}
\ud^4x_1 \ldots \ud^4x_n\ud^4y_1 \ldots \ud^4y_n 
\end{multline*}
\begin{multline*}
=
\sum\limits_{a_1, \ldots, a_p} \int  \Big\{ g(x_1)\ldots g(x_n) h(a_1,y_1)\ldots h(a_p,y_p) \, \times
\\
\times \,
 \kappa_{\mathpzc{l}, \mathpzc{m}}(\boldsymbol{\p}_1 \ldots, \boldsymbol{\p}_\mathpzc{l}, \boldsymbol{\q}_1, \ldots, \boldsymbol{\q}_\mathpzc{m}; 
x_1, \ldots, x_n, a_1,y_1, \ldots, a_p, y_p) \eta_{{}_{\Phi,\Psi}}(\boldsymbol{\p}_1 \ldots, \boldsymbol{\p}_\mathpzc{l}, 
\boldsymbol{\q}_1, \ldots, \boldsymbol{\q}_\mathpzc{m})
\\ 
\Big\}
\ud^4x_1 \ldots \ud^4x_n\ud^4y_1 \ldots \ud^4y_n d\boldsymbol{\p}_1 \ldots d\boldsymbol{\p}_\mathpzc{l} d\boldsymbol{\q}_1 \ldots d\boldsymbol{\q}_\mathpzc{m} 
\end{multline*}
\begin{multline}\label{pairingOfkappa}
=
\sum\limits_{a_1, \ldots, a_p} \int  \Big\{ 
 \kappa_{\mathpzc{l}, \mathpzc{m}}(g^{\otimes \, n} \otimes h^p)(\boldsymbol{\p}_1 \ldots, \boldsymbol{\p}_\mathpzc{l}, 
\boldsymbol{\q}_1, \ldots, \boldsymbol{\q}_\mathpzc{m}) 
\eta_{{}_{\Phi,\Psi}}(\boldsymbol{\p}_1 \ldots, \boldsymbol{\p}_\mathpzc{l}, \boldsymbol{\q}_1, \ldots, \boldsymbol{\q}_\mathpzc{m})
\\ 
\Big\}
d\boldsymbol{\p}_1 \ldots d\boldsymbol{\p}_\mathpzc{l} d\boldsymbol{\q}_1 \ldots d\boldsymbol{\q}_\mathpzc{m} 
=
\big\langle\kappa_{\mathpzc{l}, \mathpzc{m}}(g^{\otimes \, n} \otimes h^p ), \eta_{{}_{\Phi,\Psi}} \big\rangle
\in \mathcal{E}^{p \, *}
\end{multline}
for $\kappa_{\mathpzc{l}, \mathpzc{m}}$ understood, respectively, as an element of
\[
\mathscr{L} \big(E^{\otimes (\mathpzc{l}+\mathpzc{m})} , \, 
\mathscr{L}(\mathscr{E}^{\otimes \, n} \otimes \mathscr{E}^p, \mathcal{E}^{p \, *}) \big)
\,\,\,
\textrm{or}
\,\,\,
\mathscr{L} \big(\mathscr{E}^{\otimes \, n} \otimes \mathscr{E}^p, 
\, (E^{\otimes (\mathpzc{l}+\mathpzc{m})})^* \otimes \mathcal{E}^{p \, *} \big),
\]
and for each $\boldsymbol{\p}_i$ and $\boldsymbol{\q}_i$ being the abbreviation for  $(s_i, \boldsymbol{p}_i)$ and for $(\nu_j, \boldsymbol{\q}_j)$, 
including spin component, and with $\int \ldots d\boldsymbol{\p}_i \ldots d\boldsymbol{\q}_j \ldots$ denoting integrations and summations 
$\Sigma_{s_i\ldots \nu_j}\int \ldots \ud^3\boldsymbol{\p}_i \ldots \ud^3\boldsymbol{\q}_j \ldots$
with respect to spin-momentum variables.
Here
\[
\eta_{{}_{\Phi,\Psi}} \in E^{\otimes (\mathpzc{l}+\mathpzc{m})},
\]
for ech pair $\Phi, \Psi$  of elements of the Hida test space $(E)$ and is defined by (with $\boldsymbol{\p}_i$ and $\boldsymbol{\q}_i$
subsuming spin components)
\[
\eta_{{}_{\Phi,\Psi}}(\boldsymbol{\p}_1 \ldots, \boldsymbol{\p}_\mathpzc{l}, 
\boldsymbol{\q}_1, \ldots, \boldsymbol{\q}_\mathpzc{m}) 
= \big\langle\big\langle a(\boldsymbol{\p}_1)^+ \ldots a(\boldsymbol{\p}_\mathpzc{l})^+ 
a(\boldsymbol{\q}_1) \ldots  a(\boldsymbol{\q}_\mathpzc{m}) \Phi, \Psi \big\rangle\big\rangle,
\]
$E_\mathpzc{i} \subset \mathcal{H}_\mathpzc{i} \subset E_{\mathpzc{i}}^{*}$ are the single particle Gelfand triples of the free
fields, and $(E)$ is the test Hida space in the total Fock space of the free fields underlying the actual QFT, equal
to the tensor product of all free fields underlying the theory (tensor product of the Dirac spinor field and the e.m potential field
in case of spinor QED).

Recall, that the integral kernel operator (spin components and summation with respect to them are not explicitly written)
\begin{equation}\label{Xi(kappa_l,m)}
\Xi(\kappa_{\mathpzc{l}, \mathpzc{m}}) = 
\int  \kappa_{\mathpzc{l}, \mathpzc{m}}(\boldsymbol{\p}_1 \ldots, \boldsymbol{\p}_\mathpzc{l}, 
\boldsymbol{\q}_1, \ldots, \boldsymbol{\q}_\mathpzc{m})  \,
a(\boldsymbol{\p}_1)^+ \ldots a(\boldsymbol{\p}_\mathpzc{l})^+ 
a(\boldsymbol{\q}_1) \ldots  a(\boldsymbol{\q}_\mathpzc{m}) \,\,\,\,
d\boldsymbol{\p}_1 \ldots d\boldsymbol{\p}_\mathpzc{l} d\boldsymbol{\q}_1 \ldots d\boldsymbol{\q}_\mathpzc{m} 
\end{equation}
corresponding to the above given vector-valued kernel $\kappa_{\mathpzc{l}, \mathpzc{m}}$ given by (\ref{kappa-C-distribution-valued}),
regarded as an element of (\ref{kappa-Grassmann-distribution-valued}), is uniquely determined by the condition
(for $\mathcal{E}^{p \, *}$-valued pairings)
\begin{equation}\label{X(kappa_l,m)-through-E^*p-valuedPairing}
\big\langle\big\langle \Xi(\kappa_{\mathpzc{l}, \mathpzc{m}}) (\Phi \otimes \phi \otimes h^p), \Psi \big\rangle\big\rangle
= \big\langle\kappa_{\mathpzc{l}, \mathpzc{m}}(\phi \otimes h^p ), \eta_{{}_{\Phi,\Psi}} \big\rangle
= \big\langle\kappa_{\mathpzc{l}, \mathpzc{m}}(\eta_{{}_{\Phi,\Psi}}), \phi \otimes h^p \big\rangle,
\end{equation}
for all
\[
\phi \otimes h^p \in \mathscr{E}^{\otimes \, n} \otimes \mathscr{E}^p, \,\,\,\, \textrm{and} \,\,\,\, \Phi, \Psi \in (E),
\]
or, using scalar pairigs,
\[
\big\langle\big\langle \Xi(\kappa_{\mathpzc{l}, \mathpzc{m}}) (\Phi \otimes \phi \otimes h^p), \Psi \otimes \varphi \big\rangle\big\rangle
= \big\langle\kappa_{\mathpzc{l}, \mathpzc{m}}(\phi \otimes h^p ), \eta_{{}_{\Phi,\Psi}} \otimes \varphi \big\rangle
= \big\langle\kappa_{\mathpzc{l}, \mathpzc{m}}(\eta_{{}_{\Phi,\Psi}}), \phi \otimes h^p \otimes \varphi \big\rangle,
\]
for all
\[ 
\phi \otimes h^p \in \mathscr{E}^{\otimes \, n} \otimes \mathscr{E}^p, \varphi \in \mathcal{E}^p \,\,\,\, \textrm{and} \,\,\,\, \Phi, \Psi \in (E),
\]
compare \cite{obataJFA}.

In order to keep the (Krein-)self-adjointness of (\ref{L}), when considering $W_{{}_{j}}$ of odd degree in charged Fermi fields 
(quantized complex fields) in (\ref{L}), such as the Dirac field, 
we are using not just the abstract Grassmann algebra $\oplus_n \mathcal{E}^{n*}$ with inner product, 
but instead the abstract \emph{Grassman algebra} $\oplus_n \mathcal{E}^{n*}$ \emph{with inner product and with involution} 
$\overline{\,\,\cdot \,\,}$ in the sense of Berezin
\cite{Berezin}, which respects the following conditions
\begin{enumerate}
\item[(1)]
$\overline{\overline{f}} = f$,
\item[(2)]
$\overline{f_1f_2} = \overline{f_2}\overline{f_1}$,
\item[(3)]
$\overline{\alpha f} = \overline{\alpha}\overline{f}$, $\alpha \in \mathbb{C}$,
\item[(4)]
If the inner product $(f,g)$ is well-defined for $f,g$, then it is well-defined for $\overline{f},\overline{g}$
and $(f,g) = (\overline{g}, \overline{f})$,
\item[(5)]
The space $\mathcal{E}^{1*}$ has direct sum decomposition
$\mathcal{E}^{1*} = F \oplus \overline{F}$, such that $F \cap \mathcal{H}^1$ and $\overline{F}\cap \mathcal{H}^1$
are unitarily isomorphic and orthogonal.
\end{enumerate}

It can likewise be realized as the Gelfand triple
\[
\begin{array}{ccccc}
(\mathcal{E}^{1 \, }) & \,\,\,\,\,\,\,\,\, \subset   & \Gamma(\mathcal{H}^1) =\Gamma(\mathcal{H}^{01} \oplus \mathcal{H}^{10})  & \,\,\,\,\,\,\,\,\, \subset & (\mathcal{E}^{1})^*
\\
\parallel & & \parallel & & \parallel
\\
\bigwedge \mathcal{E}^{01} \,\, \widehat{\otimes} \,\, \bigwedge \mathcal{E}^{10} & & \Gamma(\mathcal{H}^{01}) \widehat{\otimes} \Gamma(\mathcal{H}^{10}) &&
\bigwedge \mathcal{E}^{01*} \,\, \widehat{\otimes} \,\, \bigwedge \mathcal{E}^{10*}
\end{array}
\]
arising from the application of the Fermi second quantization functor $\Gamma$ to the Gelfand triple
\begin{equation}\label{GelfandTripleGAIPI}
\mathcal{E}^{1} =\mathcal{E}^{01} \oplus \mathcal{E}^{10} \,\,\,  \subset \,\,\,  \mathcal{H}^{1} = \mathcal{H}^{01} \oplus \mathcal{H}^{10}
\,\,\, \subset \mathcal{E}^{1*} =\mathcal{E}^{01*} \oplus \mathcal{E}^{10*} 
\end{equation}
equal to the direct sum of two copies of the Gelfand triples (\ref{GelfandTripleGAIP}) used above, so that
$\mathcal{H}^{10}=\overline{\mathcal{H}^{01}}$ are both equal to $\mathcal{H}^1$ in (\ref{GelfandTripleGAIP}),
$\mathcal{E}^{10} = \overline{\mathcal{E}^{01}}$ are both equal to $\mathcal{E}^{1}$ in (\ref{GelfandTripleGAIP}),
and  $\mathcal{E}^{10*} = \overline{\mathcal{E}^{01*}}$ are both equal to $\mathcal{E}^{1*}$ in (\ref{GelfandTripleGAIP}),
and with the standard operator $A\oplus A$ of (\ref{GelfandTripleGAIPI}), where $A$ is the standard operator 
of (\ref{GelfandTripleGAIP}). Here $\overline{\mathcal{H}^{01}}$ is the Hilbert space
conjugate to $\mathcal{H}^{01}$, consisting of the same elements as  $\mathcal{H}^{01}$ with the same addition as in 
$\mathcal{H}^{01}$, but with multiplication by a scalar and with inner product defined by
\begin{enumerate}
\item[1)]
\textrm{$\alpha u$ in $\overline{\mathcal{H}^{01}}$ is equal to $\overline{\alpha} u$ in $\mathcal{H}^{01}$}, 
\item[2)]
\textrm{$(u,v)$ in $\overline{\mathcal{H}^{01}}$ is equal to $(v,u)$ in $\mathcal{H}^{01}$},
\end{enumerate}
compare e.g. \cite{Mackey2}, p. 243. Note that any unitary map $U:\mathcal{H}^{01} \longmapsto \mathcal{H}^{01}$, \emph{e.g.} identity map $\textrm{Id}$,
becomes conjugate linear and anti-unitary when regarded as a map  $\overline{U}:\mathcal{H}^{01} \longmapsto \overline{\mathcal{H}^{01}}$
or as a map $\overline{U}:\overline{\mathcal{H}^{01}} \longmapsto \mathcal{H}^{01}$,
and $\overline{\textrm{Id}}$ defines involution in  
\[
\mathcal{H}^{1} = \mathcal{H}^{01} \oplus \mathcal{H}^{10} = \mathcal{H}^{1} = \mathcal{H}^{01} \oplus \overline{\mathcal{H}^{01}}.
\]
The standard operator determining (\ref{GelfandTripleGAIPI}) is equal $\Gamma(A\oplus A)$ and the Grassmann product in the 
abstract Grassmann algebra 
\[
\bigwedge \mathcal{E}^{1 \, *} = \bigwedge \Big(\mathcal{E}^{01*} \oplus \mathcal{E}^{10*}\Big) = 
\bigwedge \mathcal{E}^{01*} \,\,\, \widehat{\otimes} \,\,\, \bigwedge \mathcal{E}^{10*},
\]
is given by the wedge product, as before. By construction
\[
\bigwedge \mathcal{E}^{1 \, *} = \bigwedge \mathcal{E}^{01*} \widehat{\otimes} \bigwedge \mathcal{E}^{10*}
= \underset{n}{\oplus} \mathcal{E}^{n*} = \underset{n}{\oplus}\underset{p+q=n}{\oplus}\mathcal{E}^{pq*},
\,\,\,\,
\mathcal{E}^{pq}= \big[\mathcal{E}^{10}\big]^{\wedge \, p} \wedge \big[\mathcal{E}^{01}\big]^{\wedge \, q}.
\]

By construction, the restriction of the involution $\overline{\textrm{Id}}$
to $\mathcal{E}^{1}$ is continuous on $\mathcal{E}^1$ and can be lifted to a continuous map
$\bigwedge \mathcal{E}^{1} \rightarrow \bigwedge \mathcal{E}^{1}$ by application of the Fermi functor, with the additional modification that in action
on a homogeneous simple tensor 
\[
f_1 \wedge \ldots \wedge f_p \wedge \overline{f_{p+1}} \wedge \overline{f_{p+q}} \in \mathcal{E}^{pq}, \,\,\, f_i,f_j \in \mathcal{E}^{10},
\,\,\, \overline{f_j} \in \mathcal{E}^{01}, 
\]  
it is not merely equal to $\overline{\textrm{Id}}^{\otimes \, (p+q)}$ but in addition we apply the inverse order of the
tensor factors
\[
\mathcal{E}^{pq} \ni
\overline{f_1 \wedge \ldots \wedge f_p \wedge \overline{f_{p+1}} \wedge \ldots \wedge \overline{f_{p+q}}} 
\,\,\,
\overset{\textrm{df}}{=}
\,\,\,
f_{p+1} \wedge \ldots \wedge f_{p+q} \wedge\overline{f_{1}} \wedge \ldots \wedge \overline{f_p} \in \mathcal{E}^{qp}.
\]
This involution is a continuous map
\[
\bigwedge \mathcal{E}^{1} \longrightarrow \bigwedge \mathcal{E}^{1}
\]
which defines, by duality, the continuous involution
\[
\bigwedge \mathcal{E}^{1*} \longrightarrow \bigwedge \mathcal{E}^{1*}
\]
in the Grassmann algebra $\underset{n}{\oplus} \mathcal{E}^{n*}$. 

Let $M = \{1,\ldots,4\}\times \mathbb{R}^4 = \overset{4}{\underset{1}{\sqcup}} \mathbb{R}^4$ with integration $dx$, $x\in M$,
understood as induced by ordinary invariant integration on $\mathbb{R}^4$ on each summand (ordinary integration and summantion 
with respect to the discrete index) 
In the subspace $\mathcal{H}^{01}$ we introduce the generalized basis $\iota(x)$, $x\in M$, as before.
In the subspace $\mathcal{H}^{10} = \overline{\mathcal{H}^{01}}$ we introduce the generalized basis $\overline{\iota(x)}$, $x\in M$,
adjoint to $\iota(x)$, for which
\[
\overline{f} = \int \overline{\iota(x)} \,\,\, \overline{\varphi(x)} \, dx, \,\,\, \overline{f} \in \mathcal{E}^{10*} = \overline{\mathcal{E}^{01*}},
\]
whenever
\[
f = \int \iota(x) \varphi(x) \, dx, \,\,\, f \in \mathcal{E}^{01*}.
\]
The union $\{\overline{\iota(x)},\iota(x), x,y \in M\}$ is the generalized basis of $\mathcal{E}^{1*}$, and provides a system
of \emph{generators} of the Grassmann algebra $\bigwedge \mathcal{E}^{1*} = \underset{n}{\oplus} \mathcal{E}^{n*}$
with involution $\overline{\,\,\cdot \,\,}$. Each element 
$f$ and its adjoint $\overline{f}$ of the Grassmann algebra have the following representation
\begin{gather*}
f = \sum\limits_{p,q} \int \overline{\iota(x_1)} \ldots \overline{\iota(x_p)} \iota(y_1) \ldots \iota(y_q)  
\,\, \varphi^{pq}(x_1, \ldots, x_p, y_1, \ldots, y_q) \, dx_1 \ldots dx_p dy_1 \ldots dy_q \in \mathcal{E}^{pq*},
\\
\overline{f} = \sum\limits_{p,q} \int \overline{\iota(y_q)}\ldots \overline{\iota(y_1)} \iota(x_p) \ldots \iota(x_p)  \,\, 
\overline{\varphi^{pq}(x_1, \ldots, x_p, y_1, \ldots, y_q)} \, dx_1 \ldots dx_p dy_1 \ldots dy_q \in \mathcal{E}^{qp*},
\end{gather*}
with the generalized functions $\varphi^{pq}\in \mathcal{E}^{pq \, *}(M)$ antisymmetric in all variables $x_1, \ldots, y_q \in M$.

We define the space $\mathscr{E}^{pq}$ of Grassmann-valued test functions
\begin{gather}
h^{pq}(a_1,x_1, \ldots, a_p,x_p, b_1,y_1, \ldots, b_q,y_q)
\\
=  
\overline{\iota^{a_1}(x_1)} \ldots \overline{\iota^{a_p}(x_p)} \iota^{b_1}(y_1) \ldots \iota^{b_q}(y_q)
\phi(a_1, x_1, \ldots, a_p, x_p, b_1, y_1, \ldots, b_q, y_q)
\nonumber
\\ = \overline{\iota}^p\iota^q \cdot \phi(a_1,x_1, \ldots, b_q,y_q),
\label{h^pq}
\\
\phi \in \mathcal{S}(\mathbb{R}^4; \mathbb{C}^4)^{\otimes \, (p+q)}_{{}_{\textrm{sym}}}, x_i, y_i \in \mathbb{R}^4, a_i,b_j \in \{1, \ldots, 4\},
\nonumber
\end{gather}
and endow it with the topology of the symmetric tensor product space
$\mathcal{S}(\mathbb{R}^4, \mathbb{C}^4)^{\otimes \, (p+q)}_{{}_{\textrm{sym}}}$ using the identification
of $h^{pq} = \overline{\iota}^p\iota^q \cdot \phi$ with $\phi \in \mathcal{S}(\mathbb{R}^4, \mathbb{C}^4)^{\otimes \, (p+q)}_{{}_{\textrm{sym}}}$. 
We define the pairing of a distribution $\kappa \in \mathcal{E}^{pq \, *}(M)$ with $h^{pq}$ in the same way as before.

Now consider the class (\ref{ProductClassGenOp}) of integral kernel operators in theorem
\ref{ClassWithProductTheorem}. Note, please, that if the distribution $t$ in (\ref{ProductClassGenOp})
is symmetric, say, in $x_1,x_2$, and $W_1$and $W_2$ are odd in Fermi fields, then the kernels $\kappa_{l,m}$
of (\ref{ProductClassGenOp}) have values in $\mathscr{L}\left(\mathscr{E}^n, \mathbb{C}\right)$ 
which are skew symmetric in $x_1,x_2$. And generally, each operator of the class  (\ref{ProductClassGenOp})
with kernels $\kappa_{l,m}$ valued in $\mathscr{L}\left(\mathscr{E}^n, \mathbb{C}\right)$ skew symmetric
in the $p+q$-element subset of space-time variables, can be understood also as the 
$\mathscr{L}\left(\mathscr{E}^{\otimes (n-p-q)} \otimes \mathscr{E}^{pq}, \mathcal{E}^{pq \, *}\right)$-valued 
kernels $\kappa_{l,m}$. The parity of $t$ and $W_i$ is irrelevant, but in case the values 
of $\kappa_{l,m}$ in $\mathscr{L}\left(\mathscr{E}^n, \mathbb{C}\right)$ had zero skew symmetric part in the respective 
space-time variables, the pairing of $\kappa_{l,m}$ with Grassmann test functions would be zero. Therefore, we have
\begin{twr}
The class of generalized integral kernel operators with $\mathscr{L}\left(\mathscr{E}^n, \mathbb{C}\right)$-valued kernels $\kappa_{l,m}$, 
which admits the operation of (tensor) product,
includes all  operators of the class (\ref{ProductClassGenOp}) which can also be regarded
as integral kernel operators with the kernels $\kappa_{l,m}$ with values in 
$\mathscr{L}\left(\mathscr{E}^{\otimes (n-p-q)} \otimes \mathscr{E}^{pq}, \mathcal{E}^{pq \, *}\right)$. 
\label{ClassWithProductTheoremGrassmann}
\end{twr}
\qedsymbol \, 
For the proof we apply the operation of change of the value space in the proof of theorem \ref{ClassWithProductTheorem},
from $\mathscr{L}\left(\mathscr{E}^n, \mathbb{C}\right)$-value to the 
$\mathscr{L}\left(\mathscr{E}^{\otimes (n-p-q)} \otimes \mathscr{E}^{pq}, \mathcal{E}^{pq \, *}\right)$-value
of one and the same kernel $\kappa_{l,m}$ and of its counterpart $\kappa_{\epsilon \, l,m}$, described above in this Section. 
\qed
\begin{rem*}
Note that two operators $\Xi'\left(\kappa'_{\epsilon \, l',m'}(\phi'\otimes h^{'p'q'})\right)$ and 
$\Xi''\left(\kappa''_{\epsilon \, l'',m''}(\phi''\otimes h^{''p''q''})\right)$,
evaluated resp. at $\phi'\otimes h^{'p'q'}$ and $\phi''\otimes h^{''p''q''}$, 
in which the non-paired massless free field kernels are replaced with their massive counterparts, can be multiplied as operators
belonging, respectively, to
\[
\mathscr{L}((E), (E)) \otimes \mathcal{E}^{p'q'*}
\,\,\,\,
\textrm{and}
\,\,\,\,
\mathscr{L}((E), (E)) \otimes \mathcal{E}^{p''q''*}
\]
with the multiplication given by operator composition on the first tensor product factor, and as the Grassmann product
on the second factor, with the product belonging to
\[
\mathscr{L}((E), (E)) \otimes \mathcal{E}^{(p'+q'')(q'+q'')*}.
\]
\qed
\end{rem*}

\section{Axioms for $S$ with Hida operators}\label{axioms}

The axioms (I)-(V) for the scattering operator-valued distribution
$g \rightarrow S(g)$ are the same as in the Bogliubov formulation \cite{Bogoliubov_Shirkov},
\emph{i.e.} Bogliubov's axioms (I)-(IV)  plus the inductive step from $S_1$, ..., $S_{n-1}$ to $S_{n}$,
accompanied by the preservation of the singularity degree axiom (V)
due to Epstein and Glaser \cite{Epstein-Glaser},  are given below in this Section.

The only replacement we are doing is that we are using the free fields as the  (white noise)
integral kernel generalized operators with vector valued (in the distribution space over space-time as the values) kernels.
Put otherwise: we make one refinement, interpreting the creation-annihilation operators at specific momenta as the Hida operators.
We introduce no additions. These white noise free fields are also distributions  with values in $\mathscr{L}((E), (E))$ or in
$\mathscr{L}((E), (E)) \otimes ( \underset{n}{\oplus} \mathcal{E}^{n*} )$,
but, we consider also the more general class of distributions, including all the operators (\ref{ProductClassGenOp}),
with values in $\mathscr{L}((E), (E)^*)$ or in $\mathscr{L}((E), (E)^*) \otimes ( \underset{n}{\oplus} \mathcal{E}^{n*} )$,
and the respective integral kernel operators, which allow the product operation, although they in general belong to
$\mathscr{L}((E), (E)^*)$-valued distributions (resp. $\mathscr{L}((E), (E)^*) \otimes ( \underset{n}{\oplus} \mathcal{E}^{n*} )$-valued distributions) 
but in general not to
$\mathscr{L}((E), (E))$-valued disributions (resp. not $\mathscr{L}((E), (E)) \otimes ( \underset{n}{\oplus} \mathcal{E}^{n*} )$-valued). 
This class includes the interaction Lagrangian $\mathcal{L}$ operator as well as the
the higher order contributions $S_n$ to the scattering operator for all known and realistic QFT.
Because the Wick product operation of the integral
kernel operators within this class is a well-defined integral kernel operator, again within this class,
as well as the ``product'' operation for any two integral kernel operators within this class is a well-defined integral kernel
operator within this class (thms. \ref{WickThmForMassiveFields}-\ref{ClassWithProductTheoremGrassmann}), 
then axioms admit the said mathematical refinement, and are identical 
(except the mathematical refinement in the definition of the allowed class
of generalized operators) as for the Bogoliubov-Epstein-Glaser theory. As we have seen in the previous Subsection, we
have well-defined product and Wick product operations within the generalized integral kernel operators of the class
Including all the operators (\ref{ProductClassGenOp}) --
all we need to formulate the axioms and for computations. This theory uses only these two operations
within the class, including all the operators of the form (\ref{ProductClassGenOp}). Let us consder
the axioms for $S(g)$, using Hida operators, in more detail.

\begin{rem*}
In the axioms below
the adjoint $(\cdot)^+$ acting in  $\mathscr{L}\left((E),(E)^*\right)$, is understood as the linear dual 
preceded and followed by the ordinary complex conjugation operation. Sometimes in literature (\cite{Bogoliubov_Shirkov}, \cite{Scharf}) in case there 
are gauge free fields in (\ref{L}), the symbol $(\cdot)^+$ is used which should be 
translated into ours $\eta(\cdot)^+\eta$, with the left $\eta$ acting in $(E)^*$ and being the 
dual transpose of the right $\eta$ being th Gupta-Bleuler opertor acting in $(E)$.   
\end{rem*}

The scattering operator and its inverse are both constructed as formal power series (\ref{PowerSeriesS}) 
where each $S_n$ is a finite sum (\ref{Sn(g)}) of integral kernel operators $\Xi(\kappa_{\mathpzc{l},\mathpzc{m}})$ 
of the class (\ref{ProductClassGenOp}), with tempered distribution valued kernels $\kappa_{\mathpzc{l},\mathpzc{m}}$ 
in variables $j_1,x_1, \ldots, j_n,x_n$.  Each $j_i$ ranges separately over
the index range $\{0,\ldots, k\}$ of $j$ in (\ref{L}), and each $x_i$ range over the space-time. 
Evaluation $S_n(g^{\otimes \, n}) = \Sigma_{\mathpzc{l},\mathpzc{m}} \Xi\left(\kappa_{\mathpzc{l},\mathpzc{m}}\left(g^{\otimes \, n})\right)\right)$
at $g^{\otimes \, n}$ gives a continuous map
\[
g^{\otimes \, n} \longmapsto \mathscr{L}((E), (E)^*) \otimes ( \overset{n}{\oplus} \mathcal{E}^{p*} )
\,\,\,\,\,\,\,\,\,\,\,\,
\textrm{or}
\,\,\,\,\,\,\,\,\,\,\,\,
g^{\otimes \, n} \longmapsto \mathscr{L}((E), (E)) \otimes ( \overset{n}{\oplus} \mathcal{E}^{p*} ),
\]
depending if there are present or not massless fields in the generalized interaction Lagrangian (\ref{L}), and which is a generalized distribution
\[
S_n(j_1,x_1, \ldots, j_n,x_n) = S(Z)
\]
in the variables $Z= \{j_1,x_1$, $\ldots$, $j_n,x_n\}$. Sometimes, when there is no danger of confusing the $n$-order contribution 
$S_n$ with the entire formal series $S$,  we drop out the subscript $n$ in $S_n$ that can be read off from the number of
variables $Z$ in $S(Z)$.

Bogoliubov's causality axiom (I) reads
\[
\textrm{(I)} \,\,\,\,\,\,\,\,\,\,\,\,\,\,\,\,\,\,\,\,   
              S(g+h) = S(g)S(h), \,\,\, \textrm{whenever} \,\, \textrm{supp} \, g \preceq \textrm{supp} \, h. 
\]
Axiom (I), formulated in terms of generalized integral kernel operators $S_n$
(or generalized distributions) of the said class (\ref{ProductClassGenOp}) reads 
\[
\textrm{(I)} \,\,\,\,\,\,\,\,\,\,\,\,\,\,\,\,\,\,\,\,   
              S_n(j_1,x_1, \ldots, j_n,x_n) = (-1)^{s(X,Y)} \, S_k(j_{r_1},x_{r_1}, \ldots, j_{r_k},x_{r_k})S_{n-k}(j_{r_{k+1}},x_{r_{k+1}}, \ldots, j_{r_n}, x_{r_n}) 
\]
whenever $\{j_{r_{k+1}},x_{r_{k+1}}, \ldots, j_{r_{n}}, x_{r_n} \} \preceq \{j_{r_1},x_{r_1}, \ldots, j_k,x_{r_k}\}$. 

Let $Z$ denote the set $\{j_1,x_1, \ldots, j_n,x_n\}$ of variables. Here we have a partition $Z= X \sqcup Y$ of $Z$ into two disjoint subsets.
In the partition we treat each pair $j_i,x_i$ as a single element. $X= \{j_{r_1},x_{r_1}, \ldots, j_{r_k},x_{r_k} \}$,
$Y = \{j_{r_{k+1}},x_{r_{k+1}}, \ldots, j_{r_n}, x_{r_n}\}$. 
For two subsets $X,Y \subset \mathbb{R}^4$ we say, after Bogoliubov, Epstein and Glaser,
that $Y \preceq X$ if and only if 
\[
X \cap (Y +\overline{V_{-}}) = \emptyset \,\,\, \Longleftrightarrow \,\,\,
Y \cap (X + \overline{V_{+}}) = \emptyset,
\]
or $X$ does not intersect the past causal shadow of $Y$, or equivalently,
$Y$ does not intersect the future causal shadow of $X$, or, equivalently,
$X$ and $Y$ are separated by a space-like hyperplane, and in some Lorentz frame
all points of $Y$ lie in the past of all points of $X$,
compare \cite{Bogoliubov_Shirkov}, \cite{Epstein-Glaser}. We extend this definition over the sets of pairs $j_i,x_i$,
inheriting the relation $\preceq$ from the space-time coefficients $x_i$ of the pairs  $j_i,x_i$.
Symbol $s(X,Y)$ denotes the parity of permutation
of Grassmann variabes in the permutation $Z \rightarrow (X,Y)$.

Covariance axiom (II) reads
\[
\textrm{(II)} \,\,\,\,\,\,\,\,\,\,\,\,\,\,\,\,\,\,\,\,   
              \left(U_{{}_{a,\Lambda}} \otimes 
\boldsymbol{1}
\right)S(g)U_{{}_{a,\Lambda}}^{+} = S(T_{{}_{a,\Lambda}}g), \,\,\, 
\]
or,  in terms of $S_n(j,x_1, ..,j_n,x_n)$,
\[
\textrm{(II)}  \,\,\,\,\,\,\,\,\,\,\,\,\,\,\,\,\,\,\,\,      
 U_{{}_{a,\Lambda}} S_n(j_1,x_1, ..,j_n,x_n) U_{{}_{a,\Lambda}}^{+} 
= \sum\limits_{i_1, \ldots, i_n} V_{{}_{j_1 \, i_1 }} \ldots V_{{}_{j_n \, i_n }} S_n(i_1,\Lambda^{-1}x_1 -a, .., i_n,\Lambda^{-1}x_n -a).
\]
Here $a, \Lambda \mapsto U_{a,\Lambda}$ is the Krein isometric or unitary representation of the  $T_4 \ltimes SL(2, \mathbb{C})$ group
in the Fock space of all free fields present in (\ref{L}).
Here $U_{a,\Lambda}^{+}$ is  the (Krein or Hilbert space adjoint). The left-operator $U_{a,\Lambda}$ is equal to the representor
of the representation acting in the test Hida space $(E)$ or the representation acting in the strong dual $(E)^*$
to the Hida space $(E)$, which is dual to $U_{a,\Lambda}$ acting in the Hida test space, depending on the value space  
$\mathscr{L}((E), (E)) \otimes ( \oplus \mathcal{E}^{p*} )$ or $\mathscr{L}((E), (E)^*) \otimes ( \oplus \mathcal{E}^{p*} )$
of the generalized distribution $S_n$. If no gauge fields are present, then $\eta =1$ and the formal Krein isometricity
is replaced with the formal unitarity of $U_{a,\Lambda}$ and $S(g)$. 
In (II) we have the (finite dimensional) 
matrices $V = V(a,\Lambda)$ which are precisely the matrices in the transformation law of (\ref{L}):
\[
U_{{}_{a,\Lambda}} \mathcal{L}_{{}{j}}(x)  U_{{}_{a,\Lambda}}^{+} = \sum\limits_{i} V_{{}_{j \, i }} \mathcal{L}_{{}_{i}}(\Lambda^{-1}x-a), 
\]
and 
\begin{equation}\label{TestTransformationLaw}
T_{{}_{a,\Lambda}} g_{{}_{j}}(x) = \sum\limits_{i} \iota_{i}(\Lambda(x+a))\big[V^{-1}\big]_{ji} \phi_{{}_{i}}(\Lambda(x+a))),
\,\,\,\,\,\,
T_{{}_{a,\Lambda}} \overline{g_{{}_{j}}(x)} = \sum\limits_{i} \overline{\iota_{i}(\Lambda(x+a))} \, \overline{\big[V^{-1}\big]_{ji}} \overline{\phi_{{}_{i}}(\Lambda(x+a))}
\end{equation}
for $g_{{}_{j}}(x) = \iota_{j}(x)\phi_{{}_{j}}(x)$, $\phi_{{}_{j}} \in \mathcal{S}(\mathbb{R}^4;\mathbb{C})$,
eventually with $\iota_{j}(x)=1$ for the test function $g_{{}_{j}}$ which is not Grassmann but 
$\mathbb{C}$-valued and with $\overline{{}{}{} \cdot{}{}{}}$ acting like ordinary complex conjugation or involution in Grassmann algebra. More generally
\begin{multline*}
T_{{}_{a,\Lambda}} g_{{}_{j_1}}(x_1) \cdots T_{{}_{a,\Lambda}}g_{{}_{j_n}}(x_n) 
\\
= \sum\limits_{i_1,\ldots,i_n} \iota_{i_1}(\Lambda(x_1+a)) \ldots \iota_{i_n}(\Lambda(x_n+a))\big[V^{-1}\big]_{j_1i_1} \ldots \big[V^{-1}\big]_{j_ni_n} \phi_{{}_{i_1}}(\Lambda(x_1+a))
\cdots \phi_{{}_{i_n}}(\Lambda(x_n+a)).
\end{multline*} 
Recall, that whenever $U$ is a representation of the Poincar\'e  group (or its double cover),  then $T$ 
is a represention of the group opposite to it, and \emph{vice versa}.

Let us pass to the unitarity axiom, generalizing to the Krein-isometricity axiom in case including gauge fields with Gupta-Bleuler operator $\eta$ and some 
components $g_{{}_{j}}$ of $g$ being Grassmann test functions, when the corresponding $\mathcal{L}_{{}_{j}}$ in (\ref{L}) are odd in Fermi fields.
The last more general case degenerates to the case when no gauge fields are present, by using $\eta =1$, and with no Grassmann test components in $g$, 
by the replacement of the Grassmann algebra with $\mathbb{C}$. In the
said most general case $S(g)$ is additionally accompanied by the right $\eta$ acting in $(E)$, and with the left $\eta$
equal to the linear transpose of $\eta$ acting in $(E)^*$ tensored with the involution operator $\overline{{}{}{} \cdot {}{}{}}$
acting in the Grassmann algebra:   
\[
\textrm{(III)}  \,\,\,\,\,\,\,\,\,\,\,\,\,\,\,\,\,\,\,\, 
\left(\eta \otimes \left(\overline{{}{}{} \cdot {}{}{}}\right) \right) S(g)^+\eta = S(g)^{-1}.   
\]
In terms of $S_n(j_1,x_1, ..,j_n,x_n)$,
\[
\textrm{(III)}  \,\,\,\,\,\,\,\,\,\,\,\,\,\,\,\,\,\,\,\,        
\overline{S}_n(j_1,x_1, \ldots, j_n,x_n) = \eta S_n(j_1,x_1, \ldots, j_n,x_n)^{+} \eta.
\]

The correspondence principle axiom says that $S_1(g) = \int \mathcal{L}(x) \ud^4x$ where $\mathcal{L}$ is the generalized
interaction Lagrangian (\ref{L}), or 
\[
\textrm{(IV)}  \,\,\,\,\,\,\,\,\,\,\,\,\,\,\,\,\,\,\,\,                    
S_1(j,x) = i \mathcal{L}_{{}_{j}}(x)
\]
where $\mathcal{L}_{{}_{j}}(x)$ are the components in the interaction Lagrangian density operator (\ref{L}), 
equal to a Wick polynomial in free fields, understood as a finite sum of integral kernel operators
with vector-valued kernels (with the values in the distribution space in the variable $j,x$).

Finally, we have the following axiom
\begin{enumerate}
\item[(V)] \,\,\,\,\,\,\,\,\,\,\,\,
The value of the retarded part of a vector valued kernel should coincide with the natural formula given by the multiplication by the step theta 
function on a space-time test function, whenever the natural formula is meaningful for this test function.
\,\,\,\,\,\,\,\,\,\,\,\,\,\,\,\,\,\,\,\,\,\,\,\,\,\,\,\,\,\,\,\,\,\,\,\,\,\,\,\,\,\,\,\,\,\,\,\,\,\,\,\,\,\,\,\,\,\,\,\,
\end{enumerate}

\vspace*{0.5cm}

It means that the singularity degree of the retarded part of a kernel
should coincide with the singularity degree of this kernel, for the kernels of the above defined class of generalized integral kernel operators.
Some authors call axiom (V) ``preservation of the Steinmann scaling degree''.

\begin{twr}
The Bogoliubov causal axioms (I)-(V) for the scattering generalized operator $S$ are consistent and, moreover, 
the higher order constributions $S_n$ to $S$ can be computed by induction with respect to the order $n$.
\label{ConsistencyComputability}
\end{twr}
\qedsymbol \,
Suppose we have given $S_k$ for $k=1, \ldots, n-1$, which respect (I)-(V). 
After \cite{Epstein-Glaser} we apply the following inductive step construction of $S_n$. 
By assumption the following distributions are known
\begin{equation}\label{A'_(n),R'_(n)}
\begin{split}
A'_{(n)}(Z, j_n,x_n) = \sum\limits_{X\sqcup Y=Z, X\neq \emptyset} (-1)^{{}^{s(X,Y,j_n,x_n)}} \overline{S}(X)S(Y,j_nx_n), \\
R'_{(n)}(Z, j_n,x_n) = \sum\limits_{X\sqcup Y=Z, X\neq \emptyset} (-1)^{{}^{s(Y,j_n,x_n,X)}} S(Y,j_nx_n)\overline{S}(X),
\end{split}
\end{equation}
where the sums run over all divisions $X\sqcup Y=Z$ of the set $Z$ of variables $\{j_1,x_1, \ldots, j_{n-1},x_{n-1} \}$
into two disjoint subsets $X$ and $Y$:
\[
\{j_1,x_1, \ldots, j_{n-1},x_{n-1} \} = X \sqcup Y,
\,\,\, \textrm{with} \,\,\,
X \neq \emptyset,
\]
with each pair $j_i,x_i$ treated as a single element in $Z$.
Here $s(X,Y,j_n,x_n)$ is equal to the sign of permutation of the Grassmann variables in the permutation from the order 
$Z,j_n,x_n$ of variables to the order $X,Y,j_n,x_n$. Then we construct
\begin{multline*}
A_{(n)}(Z, j_n,x_n) = \sum \limits_{X\sqcup Y=Z} (-1)^{{}^{s(X,Y,j_n,x_n)}}  \overline{S}(X)S(Y,j_n,x_n)
\\
= \sum \limits_{X\sqcup Y=Z, X\neq \emptyset} (-1)^{{}^{s(X,Y,j_n,x_n)}} \overline{S}(X)S(Y,j_n,x_n) + S(j_1,x_1, \ldots, j_n,x_n), 
\end{multline*}
\begin{multline*}
R_{(n)}(Z, j_n,x_n) = \sum \limits_{X\sqcup Y=Z} (-1)^{{}^{s(Y,j_n,x_n,X)}} S(Y,j_n,x_n)\overline{S}(X)
\\
= \sum \limits_{X\sqcup Y=Z, X\neq \emptyset} (-1)^{{}^{s(Y,j_n,x_n,X)}} S(Y,j_n,x_n)\overline{S}(X) + S(j_1,x_1, \ldots, j_n,x_n),
\end{multline*}
where now, the first summation is extended over all divisions of the set $Z$
into two disjoint subsets $X$ and $Y$, which include the empty set $X= \emptyset$. Note that
\[
D_{(n)} = R'_{(n)} - A'_{(n)} = R_{(n)} - A_{(n)}.
\]
In order to finish presentation of the essential point of the inductive step, 
let us introduce after \cite{Epstein-Glaser} higher dimensional generalization of the backward and forward cones:
\[
\Gamma_{\pm}^{(n)}(y) = \big\{X \in \mathcal{M}^n: x_j - y \in \overline{V_{\pm}} \big\}, \,\,\,\,\,
X = \{x_1, \ldots, x_n \}.
\]
Then, from the axioms (I)-(V), which are assumed to be fulfilled by all $S_k$, $k\leq n-1$, 
it is shown in \cite{Epstein-Glaser} (compare also \cite{Scharf}) that for $n>2$
\[
\begin{split}
\textrm{supp} \, R_{(n)}(j_1,x_1, \ldots, j_{n-1},x_{n-1}, j_nx_n) \subseteq \Gamma_{+}^{(n-1)}(j_n,x_n),
\\
\textrm{supp} \, A_{(n)}(j_1,x_1, \ldots, j_{n-1},x_{n-1}, j_{n},x_n) \subseteq \Gamma_{-}^{(n-1)}(j_n,x_n),
\\
\textrm{supp} \, D_{(n)}(j_1,x_1, \ldots, j_{n-1},x_{n-1}, j_{n},x_n) \subseteq \Gamma_{+}^{(n-1)}(j_n,x_n)
\sqcup \Gamma_{-}^{(n-1)}(j_n,x_n),
\\
\end{split}
\]
and moreover that each $D_{(n)}$ can be 
splitted into sum of operator distributions each having the support, respectively, in
$\Gamma_{+}^{(n-1)}(x_n)$ or in $\Gamma_{-}^{(n-1)}(x_n)$ and that this splitting can be made explicitly
and independently of the axioms (I) -- (V). The splitting is determined up a number of free constants depending on the 
singularity order of the kernels of $D_{(n)}$.  
The essential point is that $R_{(n)}$ and
$A_{(n)}$ can be separately computed as the spitting of $D_{(n)}$ into the advanced $A_{(n)}$ and retarded
$R_{(n)}$ parts, so that
\[
\begin{split}
S_n(j_1,x_1, \ldots, j_{n}, x_n) = A_{(n)}(j_1,x_1, \ldots, j_{n}, x_n) - A'_{(n)}(j_1,x_1, \ldots, j_{n}, x_n)
\\
\,\,\,
\textrm{or equivalently}
\,\,\,
\\
S_n(j_1,x_1, \ldots, j_{n}, x_n) = R_{(n)}(j_1,x_1, \ldots, j_{n}, x_n) - R'_{(n)}(j_1,x_1, \ldots, j_{n}, x_n)
\end{split}
\]
and the inductive step from $n-1$ to $n$ can be computed without encountering any infinities and without any need for renormalization. 
Here we emphasize that the proof \cite{Epstein-Glaser} of the support properties remain completely 
unchanged for $S_n$ understood as finite sums of integral kernel
operators with vector valued kernels. By assumption $S_1$ is equal to a Wick polynomial in free fields, and thus belongs
to the class of thms. \ref{ClassWithProductTheorem} and \ref{ClassWithProductTheoremGrassmann}. By these theorems 
$D_{(2)}(j_1,x_1,j_2,x_2) = S(j_1,x_1)\overline{S}(j_2,x_2) - (-1)^{s(j_2,x_2,j_1,x_1)}\overline{S}(j_2,x_2)S(j_1,x_1)$
makes sense, with the products $S(j_1,x_1)\overline{S}(j_2,x_2)$, $\overline{S}(j_2,x_2)S(j_1,x_1)$ computed 
by application of the Wick theorem \ref{WickThmForMasslessFields}, 
understood in the limit sense when massless fields are in $S_1(j,x) = S(j,x)$. Because the commutation functions
are translationally invariant and have causal support, 
it then follows from the Wick theorem \ref{WickThmForMasslessFields} that $D_{(2)}$ has causal support (compare \cite{Bogoliubov_Shirkov}, \S 17.3) 
and $S_2 = R_{(2)} - R'_{(2)}$ can be computed using the splitting of $D_{(2)}$ into the retarded and advanced part. 
Because (causal combinations of) the contractions
(\ref{(x)q}) (products of pairings) are translationally invariant and have finite singularity order, 
they can be splitted into retarded and advanced part (up to a finite linear combination
of derivatives of the Dirac delta functional). Tensor products of distributions of finite order singularity, 
is again finite order singularity distribution, and the singularity is preserved in the splitting -- axiom (V). From this
it follows that  $S_2 = R_{(2)} - R'_{(2)}$ can be computed (up to a finite set of constants), and moreover, in each step
we apply Wick theorem to the operators of the class of theorems \ref{ClassWithProductTheorem} and \ref{ClassWithProductTheoremGrassmann},
and eventually compute $ret$ or $av$ parts of their causal combinations,  which again leads us into the class of 
theorems \ref{ClassWithProductTheorem} and \ref{ClassWithProductTheoremGrassmann}. Therefore the products in the axiom
(I), as well as in the operators $A'_{(n)},R'_{(n)}$ are well-defined 
by theorems \ref{ClassWithProductTheorem} and \ref{ClassWithProductTheoremGrassmann}. 
\qed

From theorem \ref{ConsistencyComputability} it follows
\begin{cor}
The  $n$-order contributions $W_{{}_{j \,\, \textrm{int}}}^{(n)}$ to interacting 
fields $W_{{}_{j \,\, \textrm{int}}}$ can be computed as finite sums (\ref{Wjint})
of integral kernel operators with vector valued kernels. 
\end{cor}
\qedsymbol \,
Indeed (using the notation of the last proof) it is easily seen that, with the set of variables $Z=\{0,x_1,0,x_2, \ldots, 0,x_n\}$,
the distributional kernels  of  $W_{{}_{j \,\, \textrm{int}}}^{(n)}$ are equal
\[
W_{{}_{j \,\, \textrm{int}}}^{(n)}(x_1, \ldots, x_n,j,x) = \textstyle{\frac{1}{i}} A_{(n+1)}(Z,j,x) 
= \textstyle{\frac{1}{i}} \textrm{av} \, D_{(n+1)}(Z,j,x).
\qed 
\]

We end with several remarks. First, note, please, that axiom (V) is used only to fix the splitting, which makes it
unique up to the number of constants depending on the singularity order. Without this axiom, the splitting would not be fixed even 
up to this freedom and even for distributions of finite order, which we have in QFT.
Next, we note that the singularity order of the contractions (\ref{(x)q}), 
can immediately and easily be computed from the formula 
(\ref{masslesskappa01.masslesskappa10...q'-contraction...masslesskappa10.masslesskappa10}) with 
$|\boldsymbol{p}_i|$ replaced, eventually, by the respective $p_0(\boldsymbol{p}_i)$. 
Further, also the explicit formula for the Fourier transform of the causal combinations of (\ref{(x)q}) can be easily 
computed, by noting the fact that the causal combinations of 
(\ref{masslesskappa01.masslesskappa10...q'-contraction...masslesskappa10.masslesskappa10})
(regarded as acting on space-time test space) is the limit
of the integrals in spin-momenta of the respective ordinary product of the free field kernels, with the range
of integration going to infinity (compare \cite{Scharf}), with 
simple dispersion formula for the retarded part of causal combinations of (\ref{(x)q}) in terms of its Fourier transform. 
Thus, the $\dot{\otimes}$ products of free field kernels, their contractions, and $ret$, $av$ parts of their
causal combinations can be practically computed with a simple dispersion formula, and inductively for
causal combinations of scalar contractions of higher order contributions.

Theory is renormalizable, e.g. QED, if and only if the singularity order $\omega$ of $D_{n}$, equal to the singularity order of $S_n$, 
is bounded, with a bound independent of the order $n$, which, in case of contributions to $D_{n}$ of the form proportional to 
${:}\partial^{\alpha_1}\mathbb{A}^{{}^{1}} \ldots \partial^{\alpha_k}\mathbb{A}^{{}^{k}}{:}$ (\emph{i.e.} with ``external lines''
$\partial^{\alpha_1}\mathbb{A}^{{}^{1}}, \ldots, \partial^{\alpha_k}\mathbb{A}^{{}^{k}}$), the bound is equal to $4$ minus the number
of external lines, counted with the weight $w(i)$ depending on the spin of the field $\mathbb{A}^{{}^{i}}$
in the ``external line'' $\partial^{\alpha_i}\mathbb{A}^{{}^{i}}$, and minus the number of derivatives in external lines.

The presented theory can, in principle, be generalized to include the case of interaction Lagrangian (\ref{L}) with 
the Wick products $W_{{}_{j}}$ replaced by infinite Fock expansions \cite{obataJFA} into integral kernel
operators of the class (\ref{ProductClassGenOp}), involving finite number of different kinds of free fields. Analysis of renormalizability
would be much more difficult in case $\mathcal{L}_{{}_{0}}$ was equal to the said infinite Fock expansion. 
So far, we have rather had no serious physical indications for developing theory in this direction.     

Formal Krein-isometric (and not formal unitary)
scattering operator distribution $S$ is used \cite{Bogoliubov_Shirkov} in case of QED (and this is the case in general
for gauge theory in which unitary gauge is non-local, as in BRST more general case) with the Gupta-Bleuler gauge,
which is non-unitary but Krein-isometric.
This choice of gauge is not a matter of taste, and follows from
the fact that in this gauge we are left with local interaction and local transformation formula.
The unitary gauges (like the Coulomb gauge, or the axial gauge)
are non-local and are in clash with the local implementation of the switching
off multiplication by $g$ mechanism, fundamental for the implementation of the causality axiom.
If one insists on the unitarity of the gauge, then Bogoliubov's causality axioms, which allow avoiding
UV divergence, cannot be applied and one is left with UV-infinities within the renormalization
approach.

\vspace*{0.5cm}

{\bf ACKNOWLEDGEMENTS}

\vspace*{0.3cm}

The author would like to express his deep gratitude to Professor D. Kazakov  and Professor I. Volovich 
for the very helpful discussions. 
He also would like to thank for the excellent conditions for work at JINR, Dubna.
He would like to thank Professor M. Je\.zabek 
for the excellent conditions for work at INP PAS in Krak\'ow, Poland
and would like to thank Professor A. Staruszkiewicz and 
Professor M. Je\.zabek for the warm encouragement.
He also would like to thank the Reviewer for his suggestions.


\begin{thebibliography}{}
\bibitem{Berezin}  Berezin, F. A.: The method of second quantization. Acad. Press, New York, London, 1966.
\bibitem{Bogoliubov_Shirkov} Bogoliubov, N. N., Shirkov, D. V.: Introduction to the Theory of Quantized Fields. New York (1959), second ed. John Wiley \& Sons, Inc., New York, Chichester, Brisbane, Toronto, 1980.
\bibitem{Epstein-Glaser} Epstein, H., Glaser, V.: Ann. Inst. H. Poincar\'e A19, 211 (1973).
\bibitem{GelfandIV} Gelfand, I. M. and Vilenkin, N. Ya.: Applications of Harmonic Analysis: Generalized functions. Vol. 4., Acad. Press, New York, 1964.
\bibitem{GelfandII} Gelfand, I. M., Shilov, G. E.: Generalized Functions. Vol II. Academic Press, New York, San Francisco, London, 1968.
\bibitem{hida} Hida, T, Obata, N., Sait\^o, K.: Nagoya Math. J. 128, 65 (1992).
\bibitem{Mackey2} Mackey, G. W.: The Theory of Unitary Group Representations, The Univeristy of Chicago Press, Chicago, London, 1976.
\bibitem{obata} Obata, N.: J. Math. Soc. Japan 45, 421 (1993).
\bibitem{obataJFA} Obata, N.:  J. of Funct. Anal. 121, 185-232 (1994).
Fourier analysis, self-adjointness, Acad. Press, Inc. 1975.
\bibitem{Rudin} Rudin, W.: Functional Analysis, McGraw-Hill, Inc. 1991.
\bibitem{Schaefer} Schaefer, H. H.: Topological vector spaces, Springer, 2nd ed. 
(rewritten with assistance of M.P. Wolff), New York 1999.
\bibitem{Scharf} Scharf, G: Finite Quantum electrodynamics, Dover Publications, Mineola, New York, 2014.
\bibitem{wig} R. F. Streater and A. S. Wightman, {\em PCT, Spin and Statistics, and All
That}, (W. A. Benjamin, Inc., New York, 1964).  
\bibitem{wawrzycki} Wawrzycki, J: Theoretical and Mathematical Physics {\bf 211}, 775 (2022); ERRATUM: {\bf 212}, 1312 (2022). 
\end{thebibliography}
\end{document}